\documentclass[floatfix,twocolumn,showpacs,preprintnumbers,amsmath,amssymb,pre,superscriptaddress]{revtex4-1}
\usepackage{color}
\usepackage[usenames,dvipsnames,svgnames,table]{xcolor}
\usepackage[colorlinks=true,linkcolor=blue,urlcolor=blue,citecolor=blue]{hyperref}

\usepackage{mathtools}
\usepackage{graphicx}
\usepackage{dcolumn}
\usepackage{array}
\usepackage{lipsum}
\usepackage{bm}
\usepackage{subfigure}
\usepackage{amssymb}
\usepackage{multirow}
\usepackage{tabularx}
\usepackage{amsmath}
\usepackage{braket}
\graphicspath{{plots/}}

\begin{document}
\title{The Strongly Coupled Electron Liquid: \textit{ab initio} Path Integral Monte Carlo Simulations and Dielectric Theories }

\author{Tobias Dornheim}
\email{t.dornheim@hzdr.de}

\affiliation{Center for Advanced Systems Understanding (CASUS), G\"orlitz, Germany}


\author{Travis Sjostrom}

\affiliation{Theoretical Division, Los Alamos National Laboratory, Los Alamos, New Mexico 87545, USA}

\author{Shigenori Tanaka}

\affiliation{Graduate School of System Informatics, Kobe University, Kobe 657-8501, Japan}

\author{Jan Vorberger}

\affiliation{Helmholtz-Zentrum Dresden-Rossendorf, Bautzner Landstra{\ss}e 400, D-01328 Dresden, Germany}


\begin{abstract}
The strongly coupled electron liquid provides a unique opportunity to study the complex interplay of strong coupling with quantum degeneracy effects and thermal excitations. To this end, we carry out extensive \textit{ab initio} path integral Monte Carlo (PIMC) simulations to compute the static structure factor, interaction energy, density response function, and the corresponding static local field correction in the range of $20\leq r_s \leq 100$ and $0.5\leq \theta\leq 4$. We subsequently compare these data to several dielectric approximations, and find that different schemes are capable to reproduce different features of the PIMC results at certain parameters. Moreover, we provide a comprehensive data table of interaction energies and compare those to two recent parametrizations of the exchange-correlation free energy, where they are available. Finally, we briefly touch upon the possibility of a charge-density wave. The present study is complementary to previous investigations of the uniform electron gas in the warm dense matter regime and, thus, further completes our current picture of this fundamental model system at finite temperature. All PIMC data are available online.
\end{abstract}

\maketitle

\section{Introduction}

The uniform electron gas (UEG), often denoted as \textit{jellium} or quantum one-component plasma, is one of the most important and fundamental model systems in physics and beyond~\cite{loos,review}. For example, the UEG has facilitated key insights such as Fermi liquid theory~\cite{quantum_theory}, the quasi-particle picture of collective excitations~\cite{pines}, and the currently prevailing theory of superconductivity~\cite{bcs}. In addition, it offers a plethora of remarkably rich physical effects, such as the emergence of a charge-density wave (CDW)~\cite{iyetomi_cdw,dynamic_ii,schweng} or spin-density wave~\cite{overhauser}, Wigner crystallization at low density~\cite{wigner,gs2,drummond_wigner,ichimaru_wigner,trail_wigner}, and an incipient excitonic mode~\cite{takada2,higuchi,dornheim_dynamic,dynamic_folgepaper}.

Moreover, it was only the availability of accurate parametrizations of the zero-temperature properties of the UEG~\cite{vwn,perdew,perdew_wang,gori-giorgi1,gori-giorgi2,new_param} based on quantum Monte Carlo (QMC) simulations~\cite{gs2,moroni2,spink,ortiz1,ortiz2}, which allowed for the possibly unrivaled success of density functional theory (DFT) regarding the description of real materials~\cite{dft_review,burke_perspective}.

While most of these studies have been limited to the ground state, the recent interest in matter under extreme conditions has led to a spark of activity towards the description of the UEG at finite temperature~\cite{brown_ethan,schoof_prl,dornheim_prl,groth_prl,groth_cpp,malone2,dornheim_pop}. Of particular importance is the so-called warm dense matter (WDM) regime, which is defined in terms of two characteristic parameters that are both of the order of one~\cite{wdm_book,torben_eur}: i) the density parameter (also denoted as quantum coupling parameter or Wigner-Seitz radius) $r_s=\overline{r}/a_\textnormal{B}$, with $\overline{r}$ and $a_\textnormal{B}$ being the average interparticle distance and first Bohr radius, and ii) the degeneracy temperature $\theta=k_\textnormal{B}T/E_\textnormal{F}$, where $E_\textnormal{F}$ denotes the usual Fermi energy~\cite{quantum_theory}. These conditions occur in astrophysical objects like giant planets and brown dwarfs~\cite{militzer1,manuel,saumon1,becker} and are relevant for inertial confinement fusion~\cite{hu_ICF} applications. Moreover, WDM is nowadays routinely realized in experiments using different techniques, see Ref.~\cite{falk_wdm} for a topical review article.

Naturally, the extension of many-body simulation methods like DFT to WDM conditions relies on an accurate description of the UEG in this regime, which was achieved only recently~\cite{groth_prl} on the basis of a combination of novel path integral Monte Carlo (PIMC) methods~\cite{schoof_prl,dornheim,dornheim2,groth,dornheim3,dornheim_prl,dornheim_cpp}. While being an important milestone towards a complete description of the UEG, these studies focused on moderate coupling, i.e., $0\leq r_s \leq 20$.

On the other hand, it is well known that the UEG forms an electron liquid with decreasing density. Despite being unaccessible to current experiments, this regime allows for a theoretical study of the intriguingly intricate interplay of quantum effects such as delocalization with strong Coulomb coupling and thermal excitations. This combination is predicted to give rise to complex phenomena such as an excitonic low-frequency mode~\cite{takada2,higuchi} and a CDW instability. In addition, the availibility of accurate data for these conditions would provide a challenging benchmark for many-body approximations like dielectric theory.

To this end, we carry out extensive \textit{ab initio} path integral Monte Carlo simulations of the strongly coupled electron liquid for $20 \leq r_s \leq 100$ and $0.5 \leq \theta \leq 4$. In particular, we stress that the PIMC approach is capable to treat the complex interplay of all aforementioned effects without any approximations, and therefore constitutes the method of choice. This allows us to present the first unbiased data for the interaction energy $v$ for these parameters, which are complementary to previous studies~\cite{groth_prl,ksdt} and can be used to further complete our current picture of the UEG at finite temperature~\cite{review,karasiev_status}.

The second key objective of this paper is the study of the response of the electron liquid to an external harmonic perturbation. This question is typically investigated within the purview of dielectric theories, which give an approximate description of the local field correction and, thus, the density response function. More specifically, we consider the finite-temperature versions~\cite{stls,stls2} of the methods by Singwi-Tosi-Land-Sj\"olander (STLS)~\cite{stls_original} and Vashista and Singwi (VS)~\cite{vs_original}, and a recently introduced improved approach by Tanaka~\cite{tanaka_hnc}.
These data are then compared to our new PIMC results, which allows us to unambiguously characterize the strengths and weaknesses of the different dielectric approximations for different quantities.

Finally, we discuss the possibility of a CDW instability, which, however, does not occur under the present conditions.

The paper is organized as follows: In Sec.~\ref{sec:theory}, we introduce the relevant theoretical background, starting with the PIMC method and its application to fermions (\ref{sec:PIMC}). In Sec.~\ref{sec:dielectric}, we give an overview of linear response theory and introduce the relevant concepts and quantities, explain the three dielectric approximations studied in this work, and explain how the density response can be studied using PIMC. Sec.~\ref{sec:results} is devoted to our simulation results, beginning with a discussion of PIMC data for different quantities (\ref{sec:PIMC_results}) and the extrapolation to the thermodynamic limit (TDL).
We then compare our PIMC results to dielectric approximations regarding the static structure factor (\ref{sec:S}), the interaction energy (\ref{sec:vvv}), and the density response (\ref{sec:response_and_LFC}). Finally, we investigate the possibility of a charge-density wave (\ref{sec:CDW}) and the paper is concluded with a brief summary and outlook (\ref{sec:summary}).

\section{Theory\label{sec:theory}}

\subsection{Path Integral Monte Carlo\label{sec:PIMC}}

Let us consider a system of $N$ electrons in a cubic box with volume $V=L^3$ at an inverse temperature $\beta=1/T$, i.e., in thermodynamic equilibrium. Note that we restrict ourselves to the unpolarized case (i.e., an equal number of spin-up and -down electrons, $N/2=N_\uparrow=N_\downarrow$), and we assume Hartree atomic units throughout this work. All thermodynamic expectation values of such a system are fully defined by the canonical partition function~\cite{review}
\begin{eqnarray}\label{eq:Z}
 Z &=& \frac{1}{N_\uparrow!N_\downarrow!} \sum_{\sigma_\uparrow\in S_{N_{\uparrow}}}\sum_{\sigma_\downarrow\in S_{N_{\downarrow}}} \textnormal{sgn}^\textnormal{f}(\sigma_\uparrow,\sigma_\downarrow) \\ \nonumber
 & &
 \int \textnormal{d}\mathbf{R}\ \bra{ \mathbf{R} } e^{-\beta\hat{H}} \ket{ \hat{\pi}_{\sigma_\uparrow}\hat{\pi}_{\sigma_\downarrow}\mathbf{R}} \quad ,
\end{eqnarray}
where the double sum is carried out over all possible permutations $\sigma_k$ from the respective permutation group $S_{N_k}$, and $\hat\pi_{\sigma_k}$ being the corresponding permutation operators. Of particular importance is the sign function $\textnormal{sgn}^\textnormal{f}(\sigma_\uparrow,\sigma_\downarrow)$, which is positive (negative) for an even (odd) number of pair permutations of both spin-up and -down electrons. It is well known that the matrix elements of the density operator $\hat\rho = e^{-\beta\hat H}$ are not suitable for a straightforward evaluation as the operators for the potential and kinetic energies, $\hat V$ and $\hat K$, do not commute, 
\begin{eqnarray}\label{eq:primitive}
 e^{-\beta\hat H} =  e^{-\beta\hat K} e^{-\beta\hat V} + \mathcal O\left(\beta^2 \right) \quad .
\end{eqnarray}
To overcome this obstacle, one can exploit a semi-group property of the density operator,
\begin{eqnarray}\label{eq:group}
 e^{-\beta\hat H} = \prod_{\alpha=0}^{P-1}e^{-\epsilon\hat H} \quad ,
\end{eqnarray}
with $\epsilon=\beta/P$, which allows us to transform Eq.~(\ref{eq:Z}) into the sum over $P$ sets of particle coordinates, but at $P$ times the temperature,
\begin{eqnarray}\label{eq:Z_PIMC}
 Z &=& \frac{1}{N_\uparrow!N_\downarrow!} \sum_{\sigma_\uparrow\in S_{N_{\uparrow}}}\sum_{\sigma_\downarrow\in S_{N_{\downarrow}}} \textnormal{sgn}^\textnormal{f}(\sigma_\uparrow,\sigma_\downarrow) \\ \nonumber
 & &
 \int \textnormal{d}\mathbf{R}_0 \dots \textnormal{d}\mathbf{R}_{P-1} \prod_{\alpha=0}^{P-1} \bra{ \mathbf{R}_\alpha } e^{-\epsilon\hat{H}} \ket{ \hat\pi_{P}\mathbf{R}_{\alpha+1}}  \quad .
\end{eqnarray}
Note that $\hat\pi_{P}$ combines $\hat\pi_{\sigma_\uparrow}$ and $\hat\pi_{\sigma_\downarrow}$, and only acts on the final set of coordinates $\mathbf{R}_P$.
The crucial advantage of Eq.~(\ref{eq:Z_PIMC}) is that the factorization error from Eq.~(\ref{eq:primitive}) can be made arbitrarily small by increasing $P$, and the partition function can finally be written as
\begin{eqnarray}\label{eq:def:W}
 Z = \int \textnormal{d}\mathbf{X}\ W(\mathbf{X}) \quad ,
\end{eqnarray}
with the meta-variable $\mathbf{X}=(\mathbf{R}_0,\dots,\mathbf{R}_{P-1})^T$ denoting a so-called \textit{configuration}, and $W(\mathbf{X})$ being the corresponding configuration weight, which is a function that can be readily evaluated.

\begin{figure}
\includegraphics[width=0.4147\textwidth]{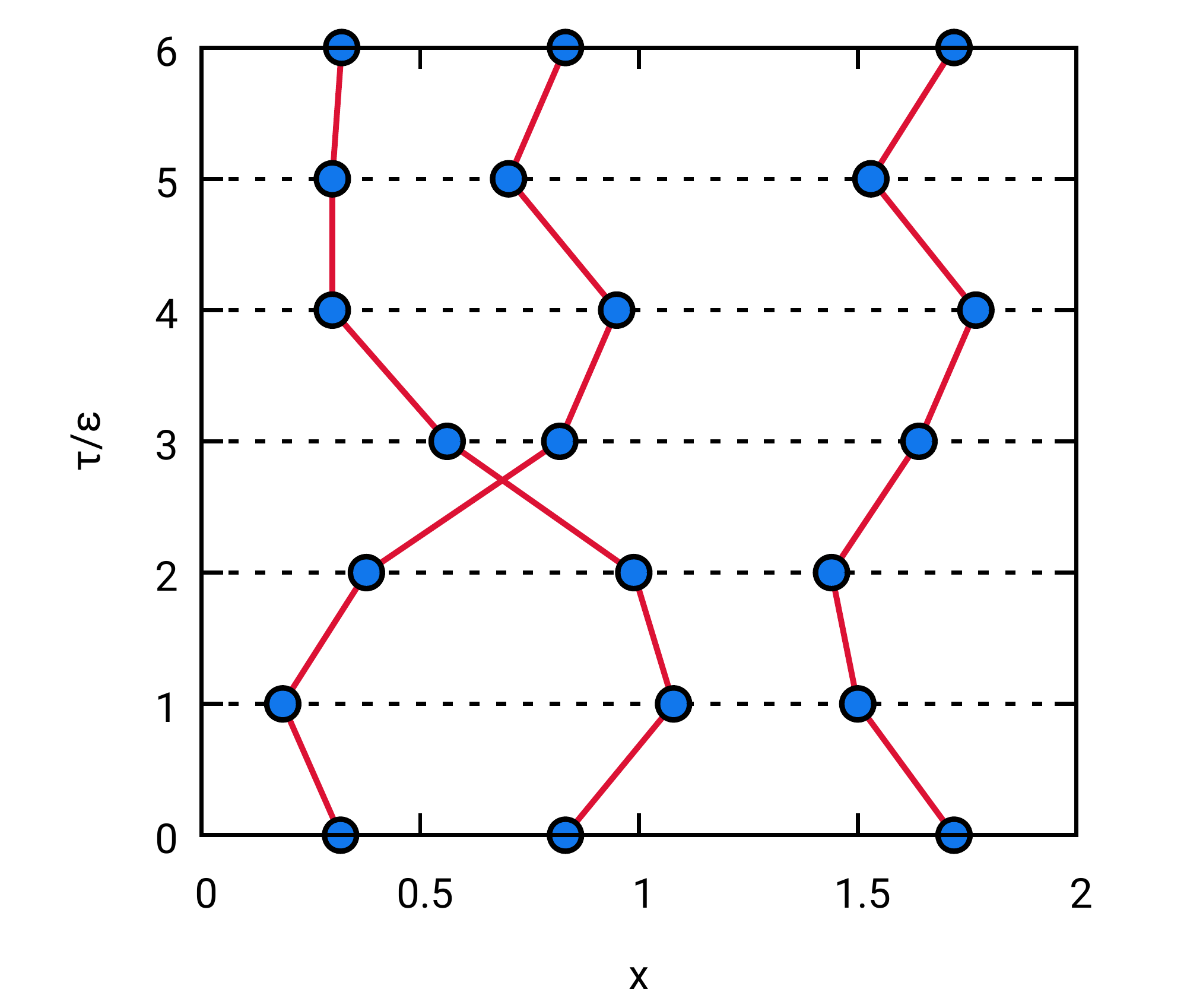}
\caption{\label{fig:PIMC}
Schematic illustration of Path Integral Monte Carlo---Shown is a configuration of $N=3$ electrons with $P=6$ imaginary--time propagators in the $x$-$\tau$ plane. Due to the single pair-exchange, the corresponding configuration weight $W(\mathbf{X})$ [cf.~Eq.~(\ref{eq:def:W})] is negative. Taken from Ref.~\cite{dornheim_permutation_cycles} with the permission of the authors.
}
\end{figure}  

This is illustrated in Fig.~\ref{fig:PIMC}, where we show a configuration of $N=3$ particles in the $x$-$\tau$-plane for $P=6$. First and foremost, we note that each particle is now represented by an entire path of $P$ particle coordinates in the imaginary time $\tau\in[0,\beta]$, with $\epsilon$ being the time-step. Moreover, the paths are closed, i.e., $\mathbf{R}_0=\mathbf{R}_P$, which reflects the definition of $Z$ as the sum over diagonal elements. Finally, we mention the presence of a single pair permutation in this example, which results in a path comprised of two particles, and the sign function is negative, $W(\mathbf{X})<0$.

The basic idea of the path integral Monte Carlo method~\cite{berne3,imada,pollock,cep,review} is to randomly generate a Markov chain of configurations $\mathbf{X}$ distributed according to $P(\mathbf{X})=W(\mathbf{X})/Z$ by using the celebrated Metropolis algorithm~\cite{metropolis}. Unfortunately, this is not directly possible in the case of fermions, as a probability must not be negative. To circumvent this issue, we switch to the modified configuration space
\begin{eqnarray}\label{eq:Z_prime}
 Z' = \int \textnormal{d}\mathbf{X}\ |W(\mathbf{X})| \quad ,
\end{eqnarray}
where the paths are generated proportionally to the absolute value of the weight function. The exact fermionic expectation value of an arbitrary observable $\hat A$ can then be obtained from
\begin{eqnarray}\label{eq:fermionic_expectation_value}
 \braket{\hat A} = \frac{\braket{\hat A\hat S}'}{\braket{\hat S}'} \quad ,
\end{eqnarray}
with $\hat S=W(\mathbf{X})/|W(\mathbf{X})|$ being the corresponding sign function.

The partial cancellation of positive and negative terms in Eq.~(\ref{eq:fermionic_expectation_value}) is the origin of the notorious fermion sign problem~\cite{loh,troyer} (FSP), which results in an exponential scaling with increasing system size $N$ and inverse temperature $\beta$, see Ref.~\cite{dornheim_sign_problem} for an extensive topical discussion. 
More specifically, the denominator in Eq.~(\ref{eq:fermionic_expectation_value}), which is typically referred to simply as the \textit{average sign} $S$, constitutes a direct measure for the degree of severity of the FSP, and simulations become unfeasible when $S\sim10^{-3}$.

Regarding physical parameters, fermionic PIMC simulations break down when quantum degeneracy effects start to dominate, i.e., towards high density and low temperature. Consequently, there have recently appeared a number of new approaches to mitigate this issue, see Refs.~\cite{schoof_prl,malone1,malone2,dornheim,dornheim2,groth,dornheim3,dornheim_prl,groth_prl} for details, and Refs.~\cite{review,dornheim_pop} for topical overviews and comparisons.

Since the present work is devoted to the study of the strongly correlated electron liquid, the standard PIMC method as introduced above is fully sufficient to obtain accurate results down to half the Fermi temperature. Still, even here we observe the exponential scaling of computation time, and we find average signs between $S=0.01$ ($N=34$, $r_s=20$, and $\theta=0.5$) and $S\approx1$ at $r_s=100$ and $\theta=4$.

For completeness, we mention that all PIMC results have been obtained using a canonical adaption~\cite{mezza} of the worm algorithm introduced by Boninsegni \textit{et al.}~\cite{boninsegni1,boninsegni2}.

\subsection{Linear Response Theory\label{sec:dielectric}}

The basic idea of linear response theory is to apply a harmonic perturbation to the UEG and subsequently measure its response, i.e., the deviation to the unperturbed case. In particular, the perturbation amplitude $A$ must be sufficiently small such that all terms beyond a linear treatment in $A$ can be neglected. Since detailed and accessible introductions to LRT have been presented elsewhere~\cite{quantum_theory,dynamic_folgepaper,review}, here will only repeat the most important relations.

\subsubsection{Density response and structure factors}

The central quantity of LRT is the density response function
\begin{eqnarray}\label{eq:chi_tilde}
\tilde \chi({q}, \overline{t}) = - i\braket{[\hat\rho({q},t),\hat\rho(-{q},t')]} \quad ,
\end{eqnarray}
with $\overline{t}=t-t'$, which fully describes the effects of an, in general, dynamic perturbation on the total density of the system. Note that $\hat\rho({q},t)$ denotes the Fourier component of the density evaluated at a time $t$, and the expectation value in Eq.~(\ref{eq:chi_tilde}) has to be computed with respect to the unperturbed system. In addition, we mention that the density response only depends on the modulus on the wave vector, $q=|\mathbf{q}|$ due to the homogeneity of the UEG. It is typically more convenient to work in frequency space, which leads to~\cite{quantum_theory}
\begin{eqnarray}\label{eq:chi_omega}
\chi({q},\omega) = \lim_{\eta\to 0} \int_{-\infty}^\infty \textnormal{d}\overline{t}\ e^{(i\omega-\eta)\overline{t}}\tilde\chi({q},\overline{t}) \quad .
\end{eqnarray}
It is important to note that the static density response function $\chi({q})$ is defined as the static limit of Eq.~(\ref{eq:chi_omega}),
\begin{eqnarray}\label{eq:chi_static}
\chi({q}) = \lim_{\omega\to0} \chi({q},\omega) \quad ,
\end{eqnarray}
and, thus, describes the response to a time-independent perturbation, which is in contrast to the static structure factor $S({q})$, see Eq.~(\ref{eq:static_structure_factor}) below.

Remarkably, $\chi({q},\omega)$ provides the complete information about all thermodynamic properties of the unperturbed system of interest. This can be seen from the well-known fluctuation--dissipation theorem,
\begin{eqnarray}\label{eq:FDT}
S({q},\omega) = - \frac{ \textnormal{Im}\chi({q},\omega)  }{ \pi n (1-e^{-\beta\omega})} \quad .
\end{eqnarray}
which gives a straightforward relation to the dynamic structure factor $S({q},\omega)$. The latter is defined as the Fourier transform of the intermediate scattering function $F$,
\begin{eqnarray}\label{eq:scattering_function}
F({q},t) &=& \frac{1}{N} \braket{\rho({q},t)\rho(-{q},0)} \\
\Rightarrow S({q},\omega) &=& \frac{1}{2\pi} \int_{-\infty}^\infty \textnormal{d}t\ F({q},t)\ e^{i\omega t} \quad ,\label{eq:dyn_def}
\end{eqnarray}
and is directly accessible in X-ray Thomson scattering experiments~\cite{siegfried_review}. The normalization of Eq.~(\ref{eq:dyn_def}) is commonly known as the \textit{static} structure factor
\begin{eqnarray}\label{eq:static_structure_factor}
S({q}) = \int_{-\infty}^\infty \textnormal{d}\omega\ S({q},\omega) \quad ,
\end{eqnarray}
and is directly connected to the pair correlation function $g(\mathbf{r})$ via a Fourier transform with respect to the wave vector $\mathbf{q}$.
Hence, Eq.~(\ref{eq:static_structure_factor}) can be used to compute the interaction energy of the system via
\begin{eqnarray}\label{eq:v}
v = \frac{1}{\pi} \int_0^\infty \textnormal{d}q\ \left( S(q)-1 \right) \quad ,
\end{eqnarray}
which, in turn, can be used to compute the exchange--correlation part of the free energy via the adiabatic connection formula~\cite{review,groth_prl},
\begin{eqnarray}\label{eq:ACF}
f_\textnormal{xc}(r_s) = \frac{1}{r_s^2}\int_0^{r_s} \textnormal{d}\overline{r}_s\ v(\overline{r}_s) \overline{r}_s \quad .
\end{eqnarray}
Since the free energy is equivalent to the partition function $Z$, Eq.~(\ref{eq:chi_omega}) entails the full thermodynamic information about the system of interest.

\subsubsection{Dielectric Theory\label{sec:DielectricTheory}}

A particularly important concept of LRT is the dynamic local field correction $G$, which is defined by~\cite{kugler1}
\begin{eqnarray}\label{eq:define_LFC}
\chi({q},\omega) = \frac{ \chi_0({q},\omega) }{ 1 - 4\pi/q^2\big[1-G({q},\omega)\big]\chi_0({q},\omega)} \quad ,
\end{eqnarray}
with $\chi_0$ being the density response function of the ideal Fermi gas. Note that setting $G=0$ in Eq.~(\ref{eq:define_LFC}) gives the widely used random phase approximation, which entails a mean-field description of the density response. Hence, $G(q,\omega)$ provides a wave-number and frequency resolved description of all exchange--correlation effects, and plays a similar role as the self energy in Green function methods~\cite{vanLeeuwen}.
Therefore, the exact local field correction is a-priori unknown, but can be reasonably approximated within the framework of dielectric theory as we shall see in the following section.

In this work, we follow Tanaka and Ichimaru~\cite{stls} and introduce the density-response function for complex frequencies as
\begin{eqnarray}
\tilde\chi(q,z) = \int_{-\infty}^\infty \frac{\textnormal{d}\nu}{\pi} \frac{\textnormal{Im}\chi(q,\nu)}{\nu-z} \quad .
\end{eqnarray}
This allows to express the static structure factor as 
\begin{eqnarray}\label{eq:ancilla}
S(q) = - \frac{T}{n} \sum_{l=-\infty}^\infty \tilde \chi(q,z_l) \quad ,
\end{eqnarray}
with $z_l=2\pi i l T$ being the so-called Matsubara frequencies, see Ref.~\cite{review} for a derivation.

The basic idea of the dielectric approximations is to express the unknown LFC as a functional of the static structure factor $S(q)$, i.e., $G(q)=G[S(q)]$. This results in a closed set of equation, which can be solved iteratively in the following way:
\begin{enumerate}
    \item Compute the ideal response function $\tilde\chi_0(q,z_l)$ for sufficiently large $l$.
    \item Compute $\tilde\chi(q,z_l)$ via Eq.~(\ref{eq:define_LFC}), use $G(q)=0$ for the first iteration.
    \item Compute $S(q)$ via Eq.~(\ref{eq:ancilla}).
    \item Use the new $S(q)$ to compute the next iteration of $G[S(q)]$.
\end{enumerate}
These steps are then repeated until convergence is reached.

A particularly successful scheme was introduced by Singwi, Tosi, Land, and Sj\"olander (STLS)~\cite{stls_original} by approximating the classical dynamic two-particle distribution function by a product ansatz of the form
\begin{eqnarray}\label{eq:STLS_derivation}
f_2(r_1,p_1,r_2,p_2) \approx f_1(r_1,p_1,t) f_1(r_2,p_2,t) g_\textnormal{eq}(r_1-r_2) \quad ,
\end{eqnarray}
with $g_\textnormal{eq}(r)$ being the pair distribution function in thermodynamic equilibrium, and $r_k$ ($p_k$) denoting particle coordinates (momenta).
This idea has subsequently been extended to finite temperature by Tanaka and co-workers~\cite{stls,stls_1985},
and one finds the following expression for the LFC:
\begin{eqnarray}\label{eq:STLS}
G_\textnormal{STLS}(q) &=& - \frac{1}{n} \int_0^\infty \frac{\textnormal{d}k}{(2\pi)^2} k^2 \left[ S(k) - 1 \right] \\ & & \nonumber
\left[ \frac{q^2 - k^2}{4kq} \textnormal{log}\left( \frac{(q+k)^2}{(q-k)^2}\right) +1 \right] \quad .
\end{eqnarray}
Note that Eq.~(\ref{eq:STLS}) [like Eqs.~(\ref{eq:VS}) and (\ref{eq:HNC}) below] does not depend on $\omega$ and, thus, constitutes a \textit{static approximation}, which is a direct consequence of the classical ansatz for $f_2$ in Eq.~(\ref{eq:STLS_derivation}). A somewhat more sophisticated quantum mechanical derivation gives the \textit{quantum} version of STLS (qSTLS) explored in Refs.~\cite{dynamic_i,dynamic_ii,dynamic_iii,schweng,arora}.

The STLS approximation has been successfully applied both in the ground state and at finite temperature and is known to give remarkably accurate results for the interaction energy $v$~\cite{review}. On the other hand, it strongly violates the compressibility sum-rule~\cite{stls2}, which gives an exact relation between the long-wavelength limit of $G(q,0)$ and the partial derivative of $f_\textnormal{xc}$ with respect to the density $n$,
\begin{eqnarray}\label{eq:CSR}
\lim_{q\to0} G(q,0) = - \frac{q^2}{4\pi} \frac{\partial^2}{\partial n^2} \left( n f_\textnormal{xc} \right) \quad .
\end{eqnarray}
More specifically, the lhs.~of Eq.~(\ref{eq:CSR}) as obtained from Eq.~(\ref{eq:STLS}) does not agree with the rhs.~computed via the adiabatic connection formula Eq.~(\ref{eq:ACF}).

To overcome this shortcoming, Vashista and Singwi (VS)~\cite{vs_original} proposed a modified expression for $G(q,0)$,
\begin{eqnarray}\label{eq:VS}
G_\textnormal{VS}(q,0) = \left( 1 + a n \frac{\partial}{\partial n}\right) G_\textnormal{STLS}(q,0) \quad ,
\end{eqnarray}
where the free parameter $a$ is chosen such that Eq.~(\ref{eq:CSR}) is exactly satisfied. This idea was later extended to finite temperature by Stolzmann and R\"osler~\cite{stolzmann} and, more rigorously, by Sjostrom and Dufty~\cite{stls2}, who found that the free parameter must depend both on density and temperature, i.e., $a=a(r_s,\theta)$.
Interestingly, it was found that the incorporation of the exact physical relation Eq.~(\ref{eq:CSR}) leads to an overall decreased accuracy in other thermodynamic properties like the static structure factor, pair distribution function, or interaction energy in the warm dense matter regime~\cite{review}.

The last dielectric approximation to be considered in this work is the recent scheme by Tanaka~\cite{tanaka_hnc}, which was derived from the hypernetted-chain (HNC) equations. More specifically, the HNC method is well known to accurately reproduce exact molecular dynamics results even for strongly coupled classical systems~\cite{lkw_fahrer}. Therefore, this new relation for the LFC,
\begin{eqnarray}\label{eq:HNC}
G_\textnormal{HNC}(q,0) &=& G_\textnormal{STLS}(q,0) + \frac{1}{n}\int \frac{\textnormal{d}\mathbf{k}}{(2\pi)^3} \frac{\mathbf{k}\cdot\mathbf{q}}{k^2} \\ \nonumber & & \left[ S(\mathbf{q}-\mathbf{k})-1 \right] \left[ G(\mathbf{k},0)-1\right]\left[ S(\mathbf{k})-1\right] \quad ,
\end{eqnarray}
is expected to constitute a significant improvement over STLS in particular in the electron liquid regime that is considered in the present work.




\subsubsection{PIMC approach to the static density response}

The first quantum Monte Carlo results for the static density response function and local field correction in the ground state have been obtained by simulating a harmonically perturbed, inhomogeneous electron gas and subsequently measuring the effect of the perturbation on an observable like the total energy~\cite{bowen,moroni,moroni2,bowen2}. This idea was recently extended by Dornheim, Groth, and co-workers to the finite-temperature permutation blocking PIMC~\cite{dornheim_pre} and configuration PIMC~\cite{groth_jcp} methods, which has allowed for the first reliable benchmarks of the static LFC in the warm dense matter regime. 
While being in principle exact, this approach is computationally very involved as one has to carry out multiple simulations with different perturbation amplitudes $A$ for each single wave number $q$.

In contrast, the standard PIMC method introduced above allows to obtain the entire $q$-dependence of both $\chi(q)$ and $G(q)$ from a single simulation of the unperturbed electron gas when the sign problem is not too severe.
In this context, the key quantity is given by imaginary-time density--density correlation function
\begin{eqnarray}\label{eq:F}
F(\mathbf{q},\tau) = \frac{1}{N} \braket{\rho(\mathbf{q},\tau)\rho(-\mathbf{q},0)} \quad ,
\end{eqnarray}
which corresponds to the intermediate scattering function Eq.~(\ref{eq:scattering_function}) evaluated at an argument $\tau\in[0,\beta]$.
Eq.~(\ref{eq:F}) can be straightforwardly evaluated within a PIMC simulation~\cite{berne1,berne2} and is of high importance for different applications. For example, it is directly connected to the dynamic structure factor $S(q,\omega)$ via a Laplace transform,
\begin{eqnarray}\label{eq:FS}
F(\mathbf{q},\tau) = \int_{-\infty}^\infty \textnormal{d}\omega\ S(\mathbf{q},\omega) e^{-\tau\omega} \quad ,
\end{eqnarray}
which can be used as the starting point for a so-called analytic continuation. While being notoriously difficult~\cite{jarrell}, such a reconstruction of a dynamic property based on PIMC data obtained in the thermodynamic equilibrium constitutes a valuable alternative to a direct propagation in real time, which typically rely on an uncontrolled, perturbative treatment of exchange-correlation effects. In particular, the numerical inversion of Eq.~(\ref{eq:FS}) has recently allowed to obtain the first accurate data for $S(q,\omega)$ for the UEG in the warm dense matter regime~\cite{dornheim_dynamic,dynamic_folgepaper}.

For the present purposes, we use Eq.~(\ref{eq:F}) as input for the imaginary-time version of the fluctuation--dissipation theorem~\cite{bowen}, 
\begin{eqnarray}\label{eq:static_chi}
\chi(\mathbf{q}) = -n\int_0^\beta \textnormal{d}\tau\ F(\mathbf{q},\tau) \quad ,
\end{eqnarray}
which states that the density response of the perturbed system  can be obtained as a simple one-dimensional integral over the correlation function computed for the unperturbed case. The corresponding PIMC results for the static LFC are subsequently computed by solving Eq.~(\ref{eq:define_LFC}) for $G$, i.e.,
\begin{eqnarray}\label{eq:Get_G}
G(q) = 1 - \frac{q^2}{4\pi}\left( \frac{1}{\chi_0(q)} - \frac{1}{\chi(q)} \right).
\end{eqnarray}

\section{Results\label{sec:results}}

\subsection{PIMC data and finite-size effects\label{sec:PIMC_results}}

In the following section, we give some practical remarks regarding our PIMC simulations of the electron liquid, with a particular focus on the finite system size. A more detailed discussion of the physical properties is postponed until Secs.~\ref{sec:S}-\ref{sec:CDW}.

\begin{figure}
\includegraphics[width=0.48\textwidth]{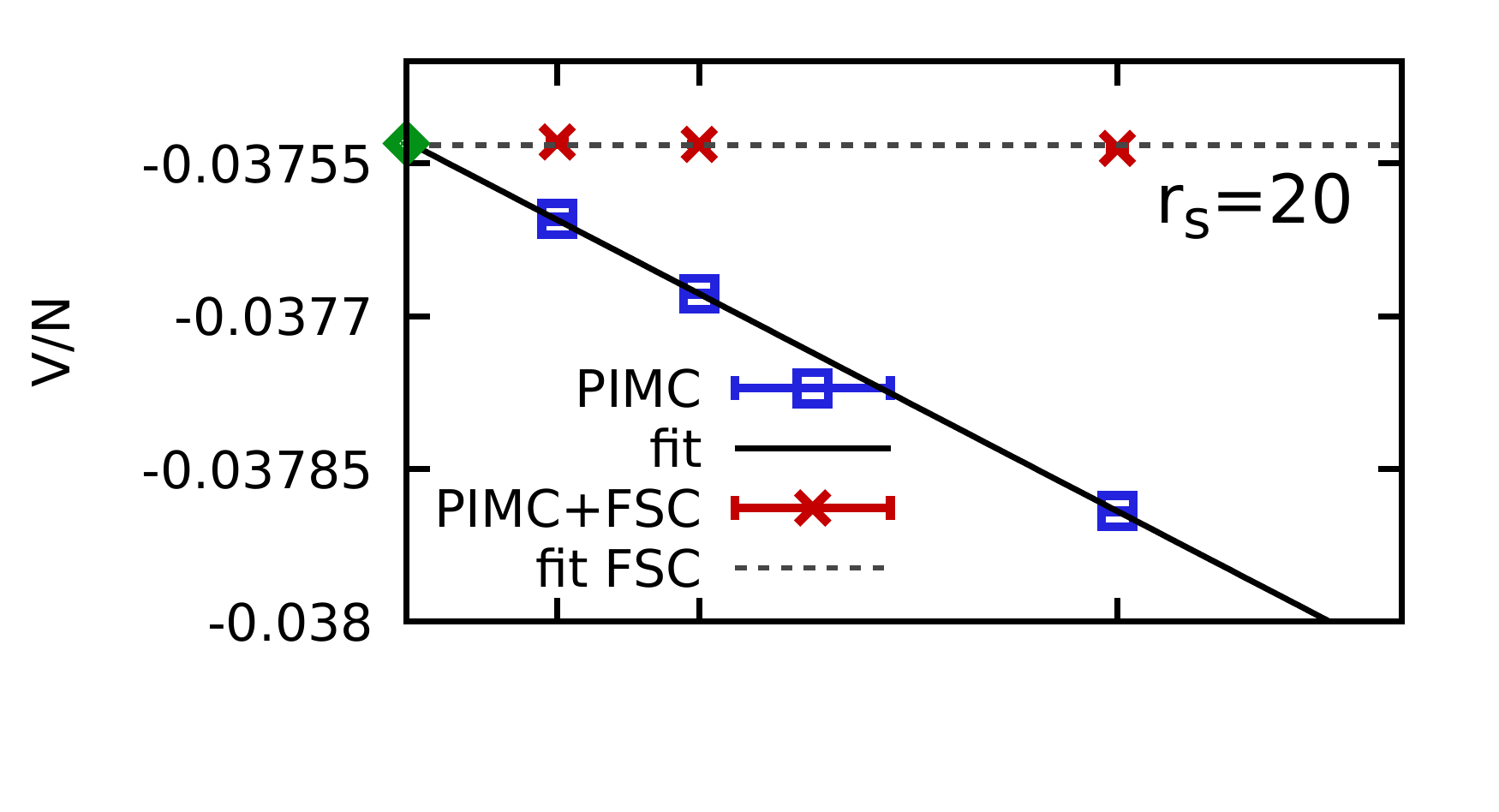} \\ \vspace*{-1cm}
\includegraphics[width=0.48\textwidth]{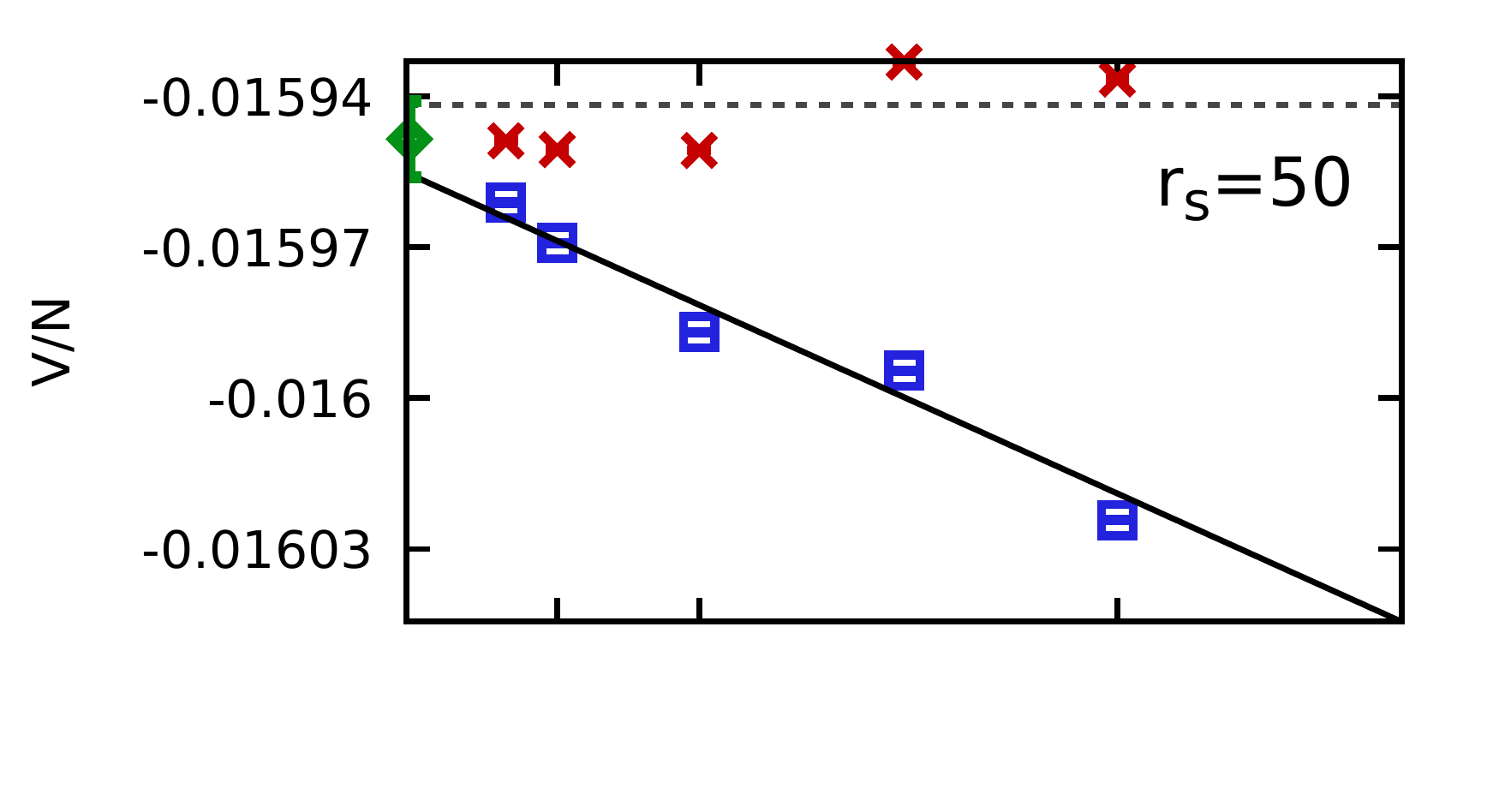} \\ \vspace*{-1cm}
\includegraphics[width=0.48\textwidth]{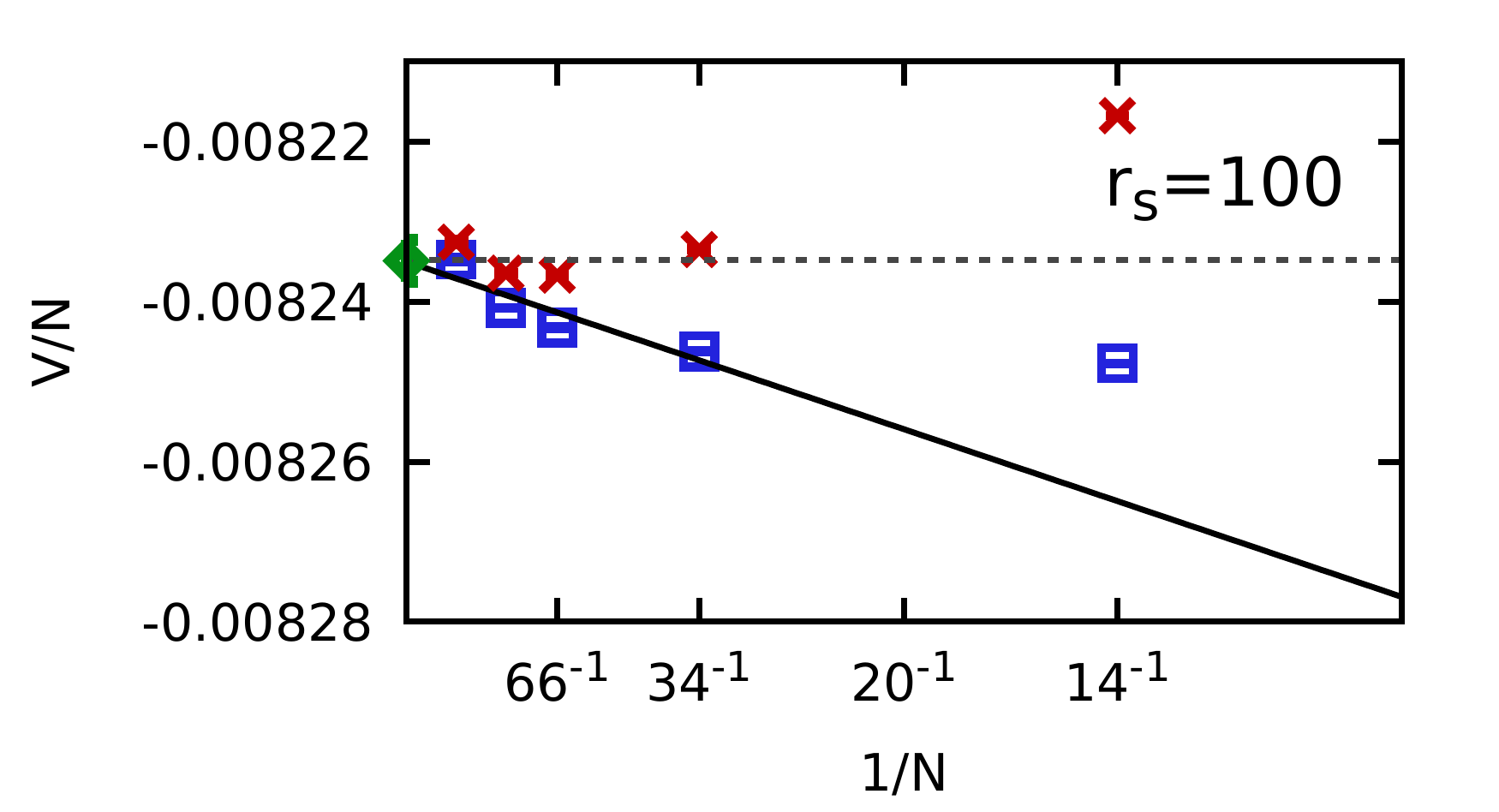}
\caption{\label{fig:FSC_v}
System-size dependence of the interaction energy per particle $v$ of the UEG at $\theta=1$ for $r_s=20$ (top), $r_s=50$ (center), and $r_s=100$ (bottom). The blue squares and solid black lines depict the raw PIMC data and a corresponding linear fit. The red squares have been obtained by adding to the PIMC data the finite-size correction from Eq.~(\ref{eq:FSC_V}), and the dashed grey lines depict the corresponding average values. Finally, the green diamonds show our extrapolated result for $v$ in the thermodynamic limit, which is computed as the mean of the dashed grey and solid black line at $N^{-1}=0$, and the errorbar takes into account all remaining uncertainties. All extrapolated results for $v(r_s,\theta)$ are given in Tab.~\ref{tab:interaction}.
}
\end{figure}

\begin{figure}\centering
\includegraphics[width=0.448\textwidth]{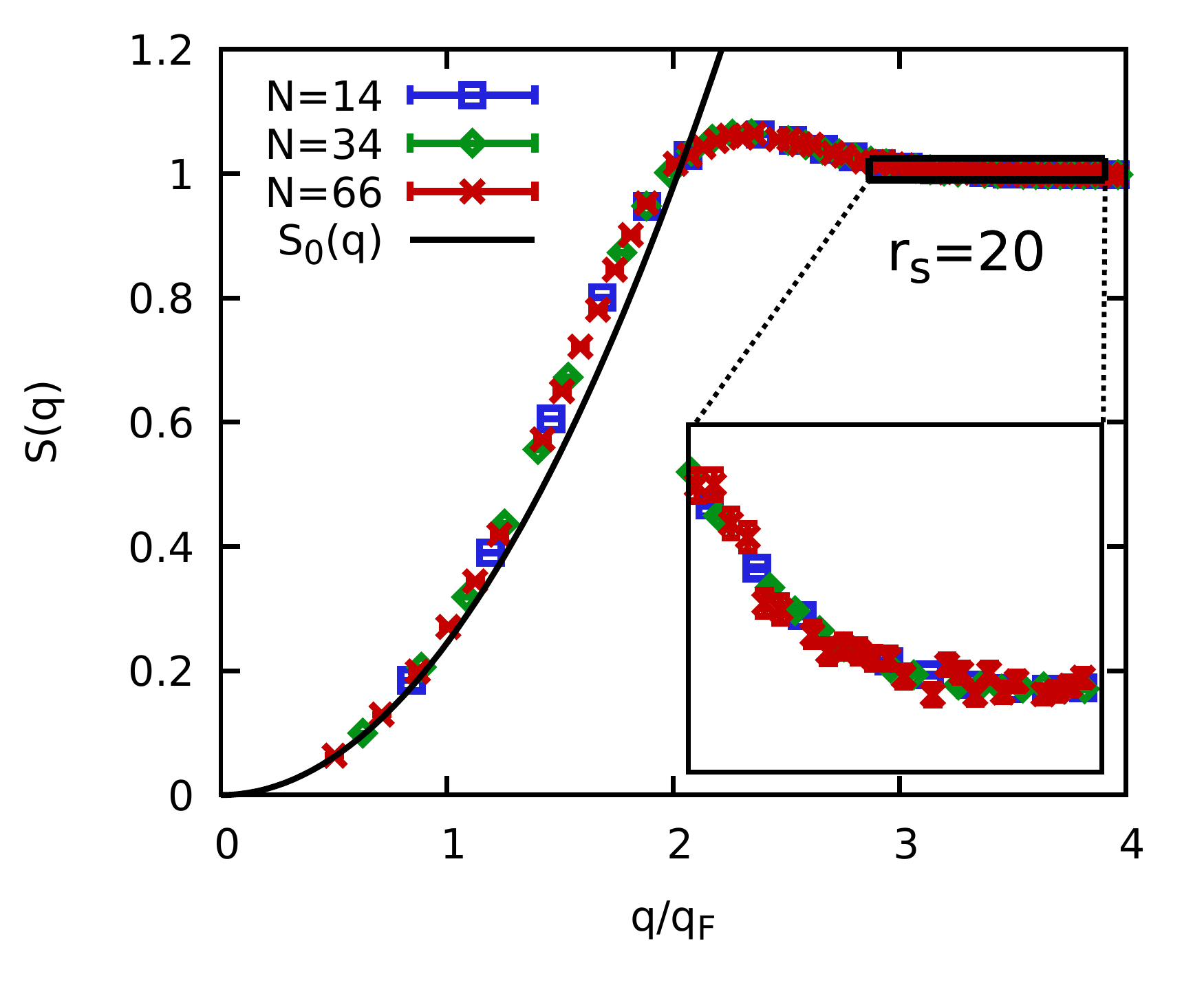}
\includegraphics[width=0.448\textwidth]{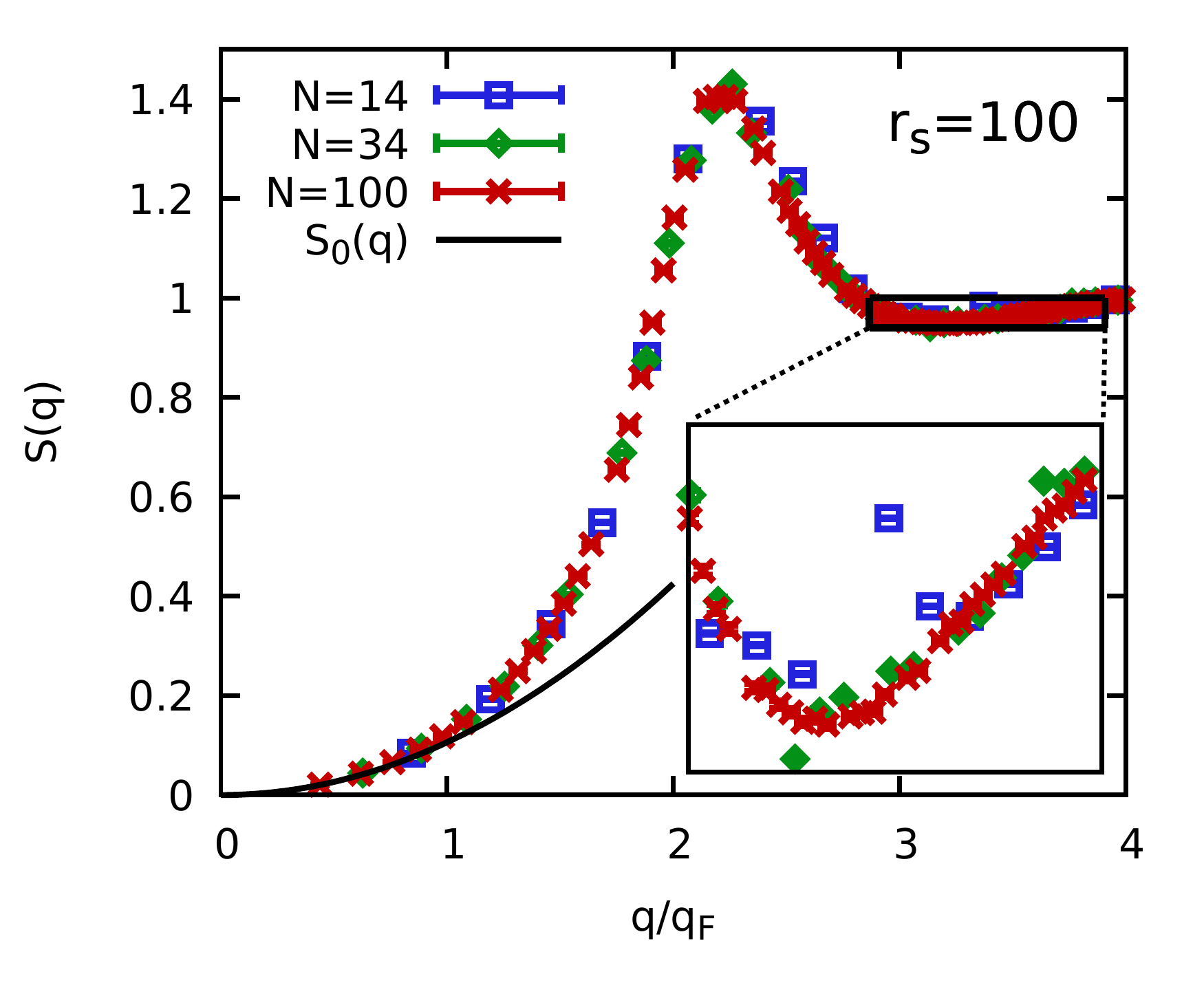}
\caption{\label{fig:FSC_S}
System-size dependence of the static structure factor of the UEG at $\theta=1$ for $r_s=20$ (top) and $r_s=100$ (bottom). The symbols depict the raw PIMC data for different particle numbers and the solid black line the long-wave length expansion of RPA, cf.~Eq.~(\ref{eq:S0}). The insets show magnified segments of the curves.
}
\end{figure}

The PIMC method as introduced in Sec.~\ref{sec:PIMC} is formulated in the canonical ensemble and, consequently, allows for simulations with a finite system size $N$. For realistic applications, however, we are interested in the properties of the electron gas in the thermodynamic limit, i.e., for the limit of an infinite number of electrons, but at a constant density $n$. For example, the corresponding TDL of the interaction energy is defined as
\begin{eqnarray}\label{eq:TDL}
v = \lim_{N\to\infty} \left. \frac{V_N}{N}\right|_{r_s} \quad ,
\end{eqnarray}
with $V_N$ being the total interaction energy obtained from a PIMC simulation, and the deviation between $v$ and $V_N/N$ is commonly known as the \textit{finite-size error}~\cite{fraser,chiesa,drummond,holzmann,dornheim_prl}.
This issue is investigated in Fig.~\ref{fig:FSC_v} at the Fermi temperature for $r_s=20$ (top), $r_s=50$ (center), and $r_s=100$ (bottom).
At $r_s=20$, the raw PIMC data (blue squares) exhibit finite-size errors of the order of $\Delta V/V\sim10^{-3}$, and $V_N/N$ is exactly reproduced by a linear fit (solid black line). The reason for this behavior was first reported by Chiesa \textit{et al.}~\cite{chiesa} in the context of ground-state QMC simulations, and recently studied for the finite temperature case as well~\cite{brown_ethan,dornheim_prl}: for a finite number of particles $N$, the continuous integral in Eq.~(\ref{eq:v}) has to be replaced by a discrete sum over reciprocal lattice vectors. Since the static structure factor $S_N(q)$ is known to only weakly depend on $N$ (cf.~Fig.~\ref{fig:FSC_S}), the finite-size error can be viewed as a simple discretization error, and the dominant contribution comes from the $q=0$ term, which is completely omitted for finite $N$.
Since the exact $q\to0$ limit is known exactly from the perfect screening sum-rule~\cite{kugler_bounds},
\begin{eqnarray}\label{eq:S0}
S_0(q) = \frac{q^2}{2\omega_\textnormal{p}} \textnormal{coth}\left( \frac{\beta\omega_\textnormal{p}}{2}
\right) \quad ,
\end{eqnarray}
one can derive a corresponding \textit{finite-size correction} as~\cite{brown_ethan}
\begin{eqnarray} \label{eq:FSC_V}
\Delta V(N) = \frac{\omega_\textnormal{p}}{4N} \textnormal{coth}\left( \frac{\beta\omega_\textnormal{p}}{2}
\right) \quad .
\end{eqnarray}
Indeed, Eq.~(\ref{eq:FSC_V}) does predict a linear dependence of $V_N/N$ on $1/N$, and adding the correction to our PIMC data (red crosses) completely removes the finite-size errors, and the resulting data points are perfectly reproduced by a constant fit (dashed grey line). To further verify the validity of the theory behind Eq.~(\ref{eq:FSC_V}), we investigate the system-size dependence of $S_N(q)$ in Fig.~\ref{fig:FSC_S}.
The top panel again shows results for $r_s=20$ for $N=14$ (blue squares), $N=34$ (green diamonds), and $N=66$ (red crosses) unpolarized electrons. While the PIMC data are available on different $q$-points, which is due to the momentum quantization in a finite simulation cell, $S_N(q)$ itself is already converged within the given statistical uncertainty even for the smallest considered system size. In addition, the solid black line shows the exact parabolic long-wave length expansion of $S$ from Eq.~(\ref{eq:S0}), which smoothly connects to the smallest wave number $q_\textnormal{min}=2\pi/L$ where our PIMC data are available. In summary, these findings do indeed confirm the validity of Eq.~(\ref{eq:FSC_V}), which then holds both in theory and in practice for $r_s=20$ and $\theta=1$.

Let us next consider the center (bottom) panel of Fig.~\ref{fig:FSC_v}, where we analyze the system size dependence for $r_s=50$ ($r_s=100$).
In both cases, we again find finite-size errors of the order of $\Delta V/V\sim 10^{-3}$, which are, overall, somewhat smaller than for $r_s=20$, but less regular. More specifically, our $V_N/N$ data cannot be exactly reproduced by a linear fit, and the $N=14$ result at $r_s=100$ seems to constitute an outlier. Consequently, Eq.~(\ref{eq:FSC_V}) does not cover the entire finite-size error, and the corrected points fluctuate around the constant fit (excluding the aforementioned outlier at $r_s=100$). To understand this nontrivial behaviour, we might again consider our PIMC data for $S(q)$, which are shown in the bottom panel of Fig.~\ref{fig:FSC_S} for $r_s=100$. First and foremost, we note that the long-wave length expansion from Eq.~(\ref{eq:S0}) again perfectly connects to our PIMC data, which means that the finite-size correction from Eq.~(\ref{eq:FSC_V}) should indeed fully correct the previously discussed contribution for $q=0$. The reason for the remaining $N$-dependence in the red crosses in Fig.~\ref{fig:FSC_v} is the functional form of $S_N(q)$ itself, which exhibits some unsmooth fluctuations (cf.~the inset) that are particularly pronounced for small particle number. More specifically, this somewhat peculiar behaviour is a commensurability effect: with increasing coupling strength, the electron liquid starts to exhibit a long-range order in coordinate space, and adding or subtracting a few particles can significantly shape the dominant packing structure.

To remove the residual system-size dependence in $V_N/N$, we average over the constant fits to the corrected data and the linear fits to the raw PIMC results. Our thus determined final results for $v$ are shown as the green diamonds in all three panels. In addition, we obtain a reliable measure for the remaining uncertainty taking into account both the statistical (Monte Carlo) errors of all data points and the respective fitting errors from the linear and constant fits. All extrapolated PIMC results listed in Tab.~\ref{tab:interaction} have been obtained in this way.

\begin{figure}\centering
\includegraphics[width=0.5\textwidth]{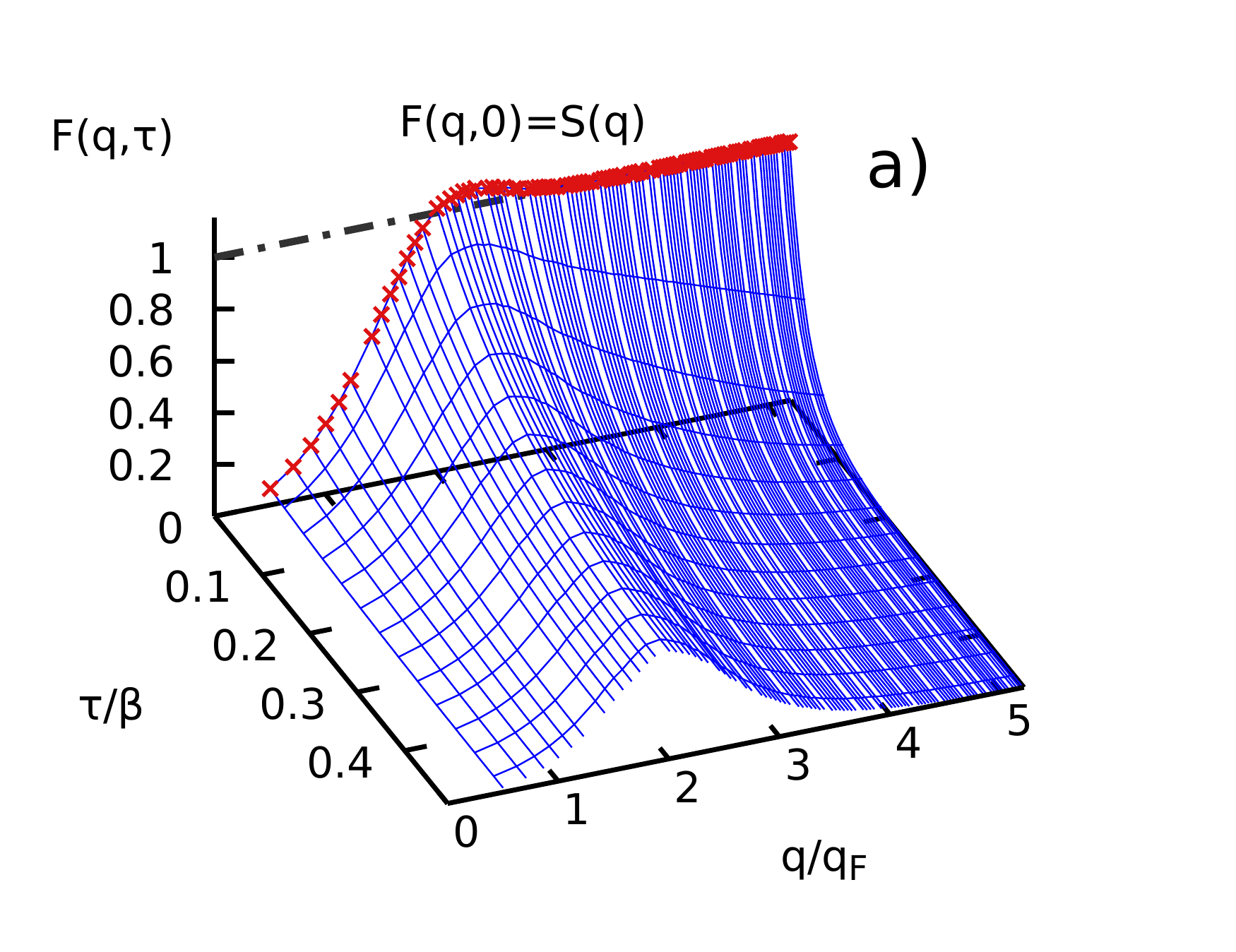} \\ \vspace*{-0.75cm}
\includegraphics[width=0.5\textwidth]{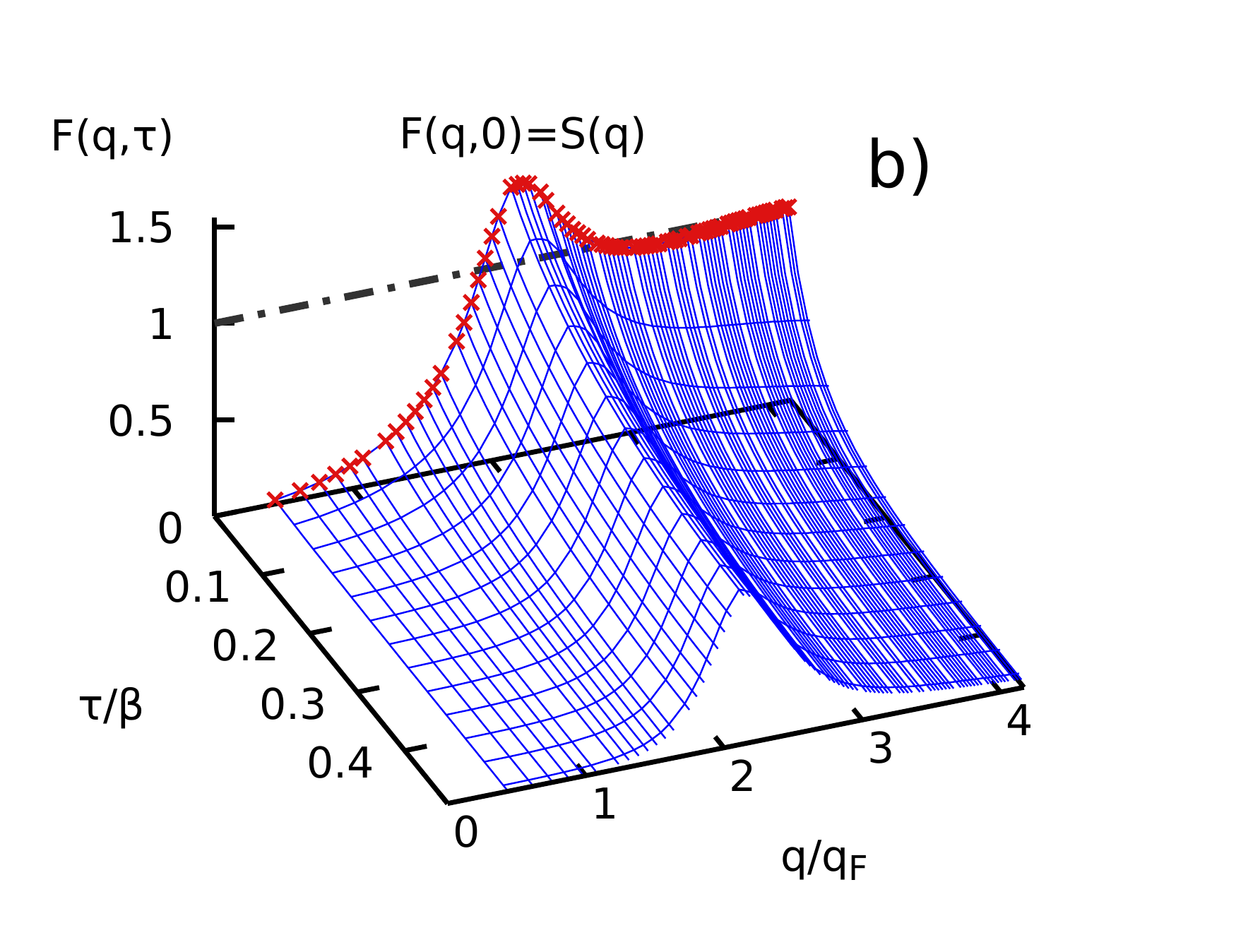}

\caption{\label{fig:FPLOT}
PIMC results for the imaginary-time density--density correlation function $F(q,\tau)$ [cf.~Eq.~(\ref{eq:F})] of the UEG at $\theta=1$ for $N=66$ and $r_s=20$ (panel a) and $N=100$ and $r_s=100$ (panel b). $F(q,\tau)$ is symmetric in $\tau$ with respect to $\tau=\beta/2$ and approaches the static structure factor $S(q)$ [cf.~Eq.~(\ref{eq:static_structure_factor})] for $\tau\to0$ (red crosses).
}
\end{figure}

Another important goal of the present paper is the comparison of the static density response of the electron liquid computed within different dielectric methods to the exact results from PIMC. For this purpose, we compute the imaginary-time density--density correlation function $F(q,\tau)$ [cf.~Eq.~(\ref{eq:F})], which is shown in the relevant $\tau$-$q$-plane at the Fermi temperature in Fig.~\ref{fig:FPLOT} for two different values of the density parameter. Firstly, we note that the wave-number grid is non-equidistant and directly follows from the momentum quantization as explained in the discussion of $S(q)$, cf.~Fig.~\ref{fig:FSC_S}. In contrast, the $\tau$-grid is determined by the number of high-temperature factors $P$ as introduced in Sec.~\ref{sec:PIMC} and can, in principle, be made arbitrarily fine. Although a vivid physical interpretation of $F(q,\tau)$ is rather difficult, we mention that its $\tau\to 0$ limit is given by the static structure factor (see the red crosses), and that it is symmetric in $\tau$ with respect to $\tau=\beta/2$. Moreover, we find a single distinct maximum in $F$ around twice the Fermi wave number $q_\textnormal{F}$, which somewhat decreases in magnitude with increasing $\tau$ and is significantly more pronounced for $r_s=100$. In fact, there appears a second maximum around $q/q_\textnormal{F}\approx5$ in the latter case, which is of the order of $10^{-3}$.

\begin{figure}\centering
\includegraphics[width=0.448\textwidth]{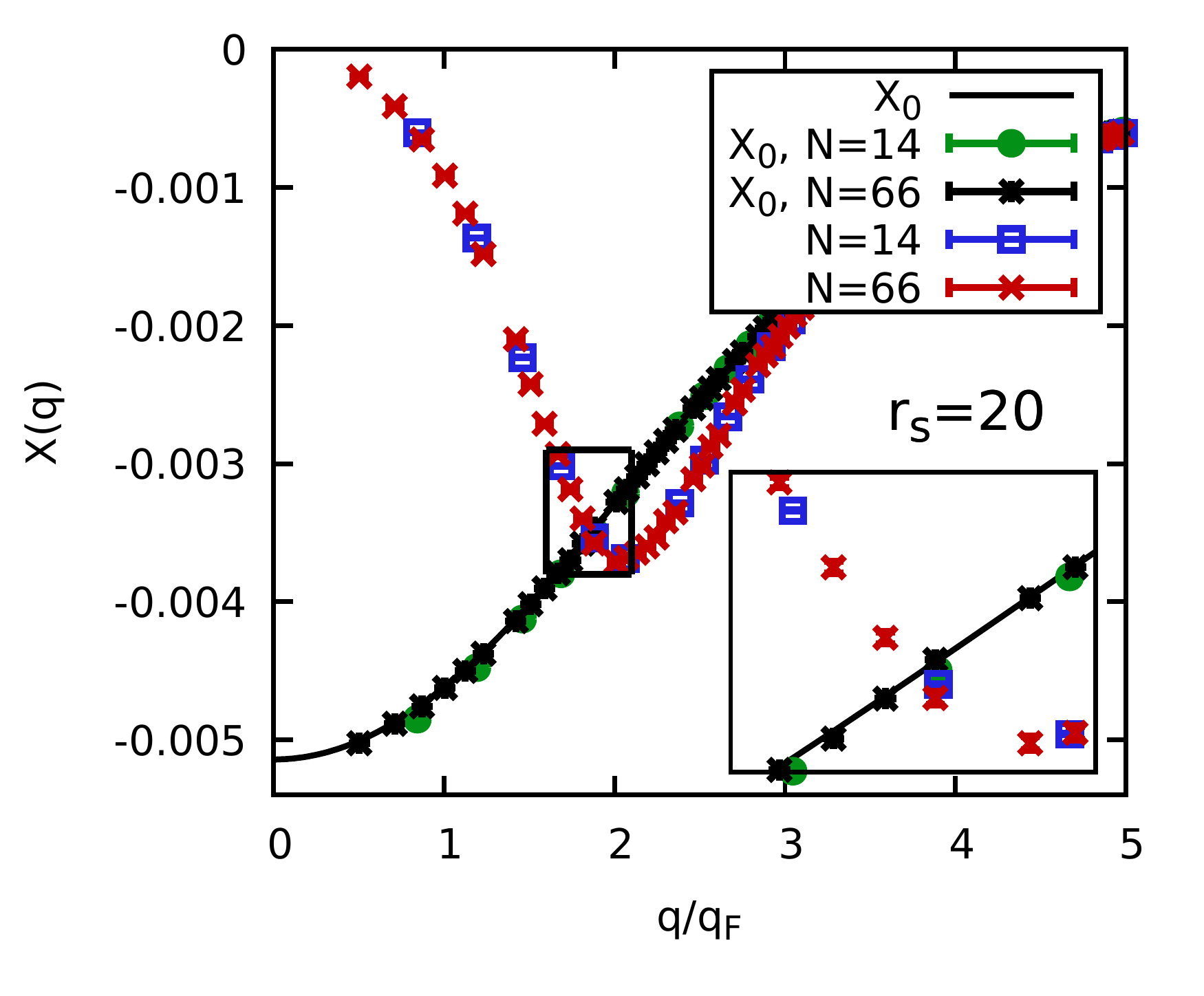}
\includegraphics[width=0.448\textwidth]{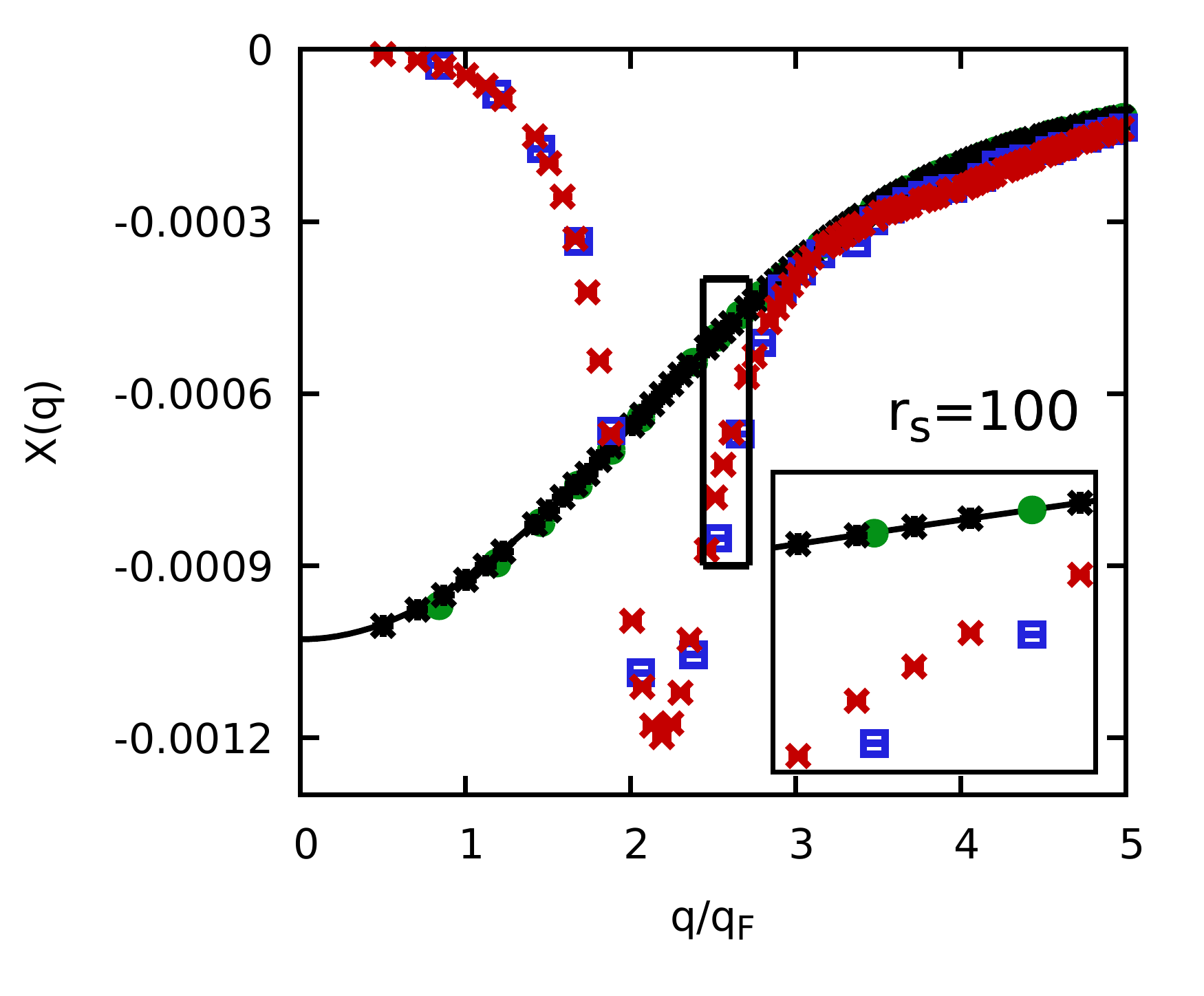}
\caption{\label{fig:FSC_CHI}
System-size dependence of the static density response function of the UEG at $\theta=1$ for $r_s=20$ (top) and $r_s=100$ (bottom).
The green circles and black stars show CPIMC data for the ideal response function $\chi_0(q)$ for $N=14$ and $N=66$, and the solid black line the corresponding thermodynamic limit. The blue squares and red crosses show PIMC data for the interacting UEG, again for $N=14$ and $N=66$, respectively. The insets show magnified segments of the curves.
}
\end{figure}

The main utility of $F(q,\tau)$ in the context of the present work is Eq.~(\ref{eq:static_chi}), which implies that the static density response function $\chi(q)$ can be obtained from a simple one-dimensional integration along the $\tau$-axis. The results from this procedure are shown in the top panel of Fig.~\ref{fig:FSC_CHI} for $r_s=20$ and $\theta=1$. Let us first consider our PIMC data for the interacting UEG, which are depicted by the blue squares (red crosses) for $N=14$ ($N=66$). Similarly as for the static structure factor, no system-size dependence in $\chi_N$ can be resolved within the given statistical uncertainty. Further, the bottom panel shows the same information for strong coupling, $r_s=100$, and we find similar commensurability effects as in $S_N(q)$.

\begin{figure}\centering
\includegraphics[width=0.448\textwidth]{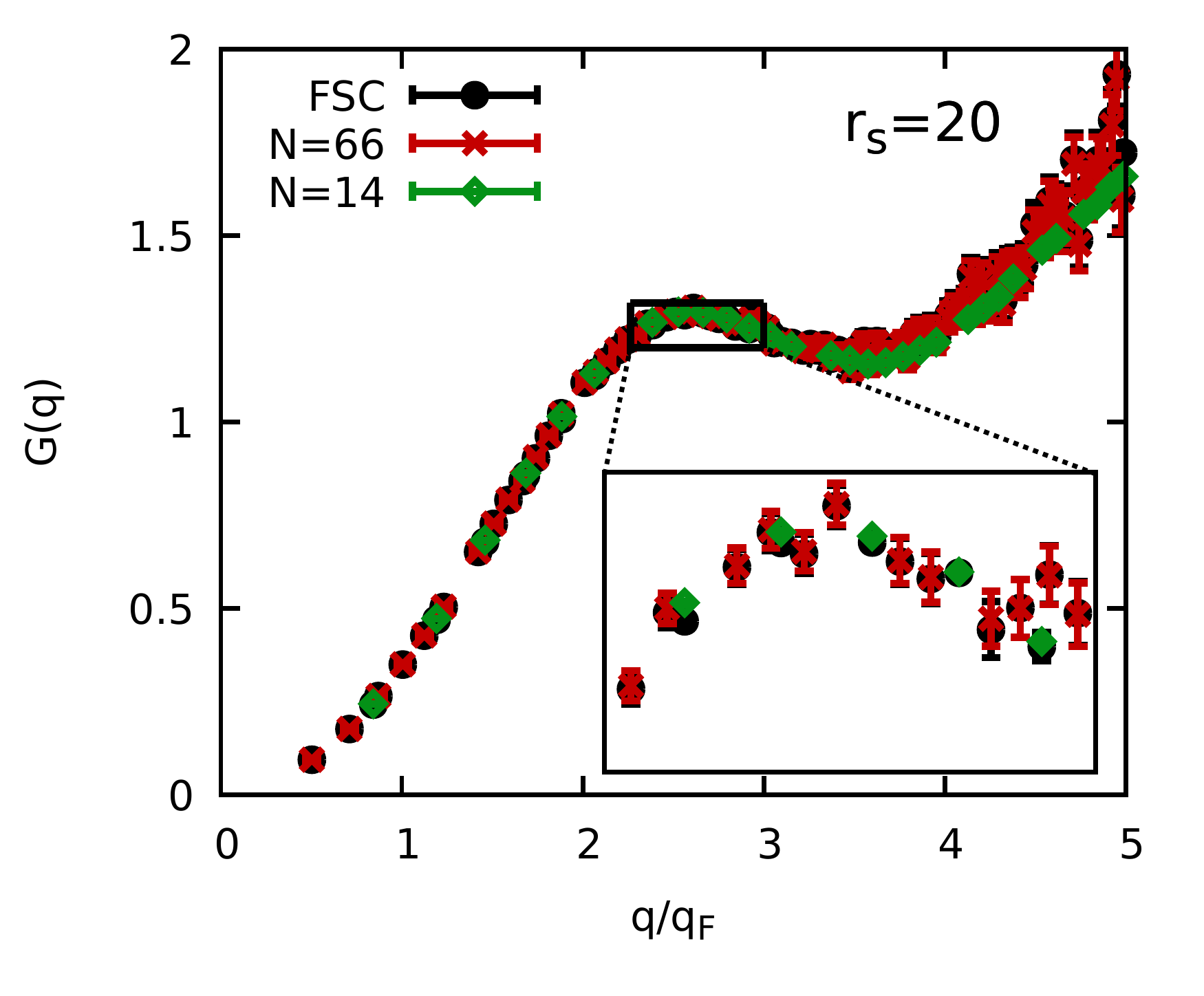}
\includegraphics[width=0.448\textwidth]{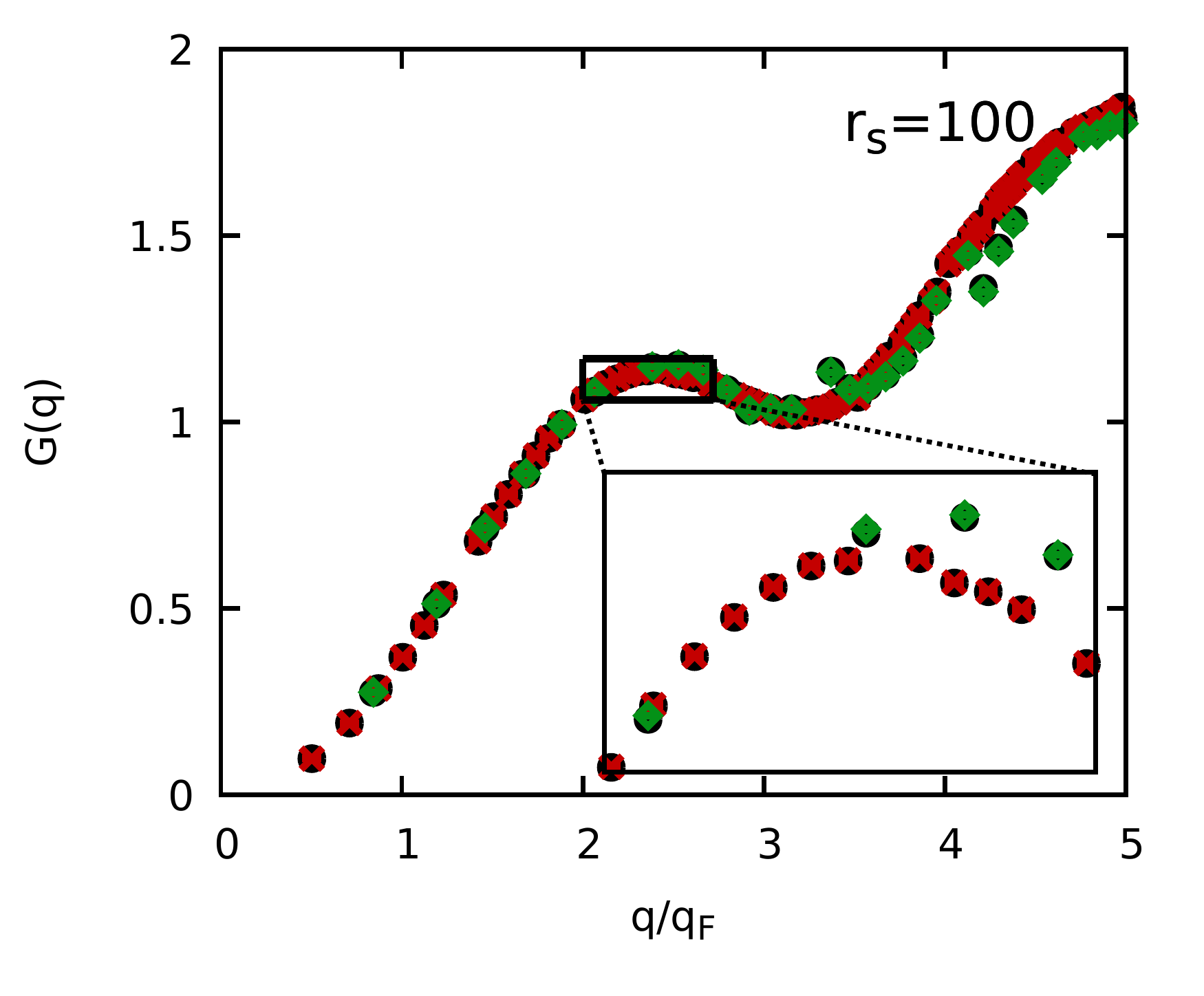}
\caption{\label{fig:FSC_G}
System-size dependence of the static local field correction of the UEG at $\theta=1$ for $r_s=20$ (top) and $r_s=100$ (bottom). The red crosses and green diamonds show the raw PIMC data for $N=66$ and $N=14$, and the black dots the corresponding finite-size corrected values.
}
\end{figure}

Finally, Fig.~\ref{fig:FSC_G} shows the corresponding PIMC results for the static local field correction $G_N(q)$ as obtained from Eq.~(\ref{eq:Get_G}) for the same conditions. Unsurprisingly, here, too, no finite-size effects can clearly be resolved for $r_s=20$ (top), whereas the $N=14$ data at $r_s=100$ (bottom) exhibit pronounced fluctuations and systematically lower values than the $N=66$ points for large $q$. It is well known that the system-size dependence in $G_N(q)$ can often be effectively removed by replacing in Eq.~(\ref{eq:Get_G}) the ideal response function $\chi_0(q)$ by its finite-size pendant $\chi_0^N(q)$, which can be obtained using the configuration PIMC (CPIMC) method~\cite{groth_jcp,dornheim_ML}. The thus corrected data are shown as the black dots in Fig.~\ref{fig:FSC_G} for both $N$- and $r_s$-values. Interestingly, they do not remove the $N$-dependence in our data and can hardly be distinguished from the raw PIMC values. To understand this behaviour, we must return to the static density response function $\chi(q)$ shown in Fig.~\ref{fig:FSC_CHI}. In particular, the solid black line shows $\chi_0(q)$ in the TDL, and the green dots (black crosses) correspond to the CPIMC data for $N=14$ ($N=66$).
Evidently, the ideal response function exhibits almost no system-size dependence for both densities, which is in stark contrast to the warm dense matter regime studied in Refs.~\cite{groth_jcp,dornheim_ML}. Therefore, the finite-size effects in the interacting response function and local field correction are not due to finite-size effects in $\chi_0^N(q)$, but are intrinsic to the local field correction $G_N(q)$ itself.
Still, it should be noted that these effects do hardly diminish the value of the PIMC data as a benchmark for dielectric theory, as the system-size dependence is much smaller compared to the systematic errors of these approximations, cf.~Sec.~\ref{sec:response_and_LFC}.

\subsection{Static structure factor\label{sec:S}}

\begin{figure*}\centering
\hspace*{-0.4cm}\includegraphics[width=0.36\textwidth]{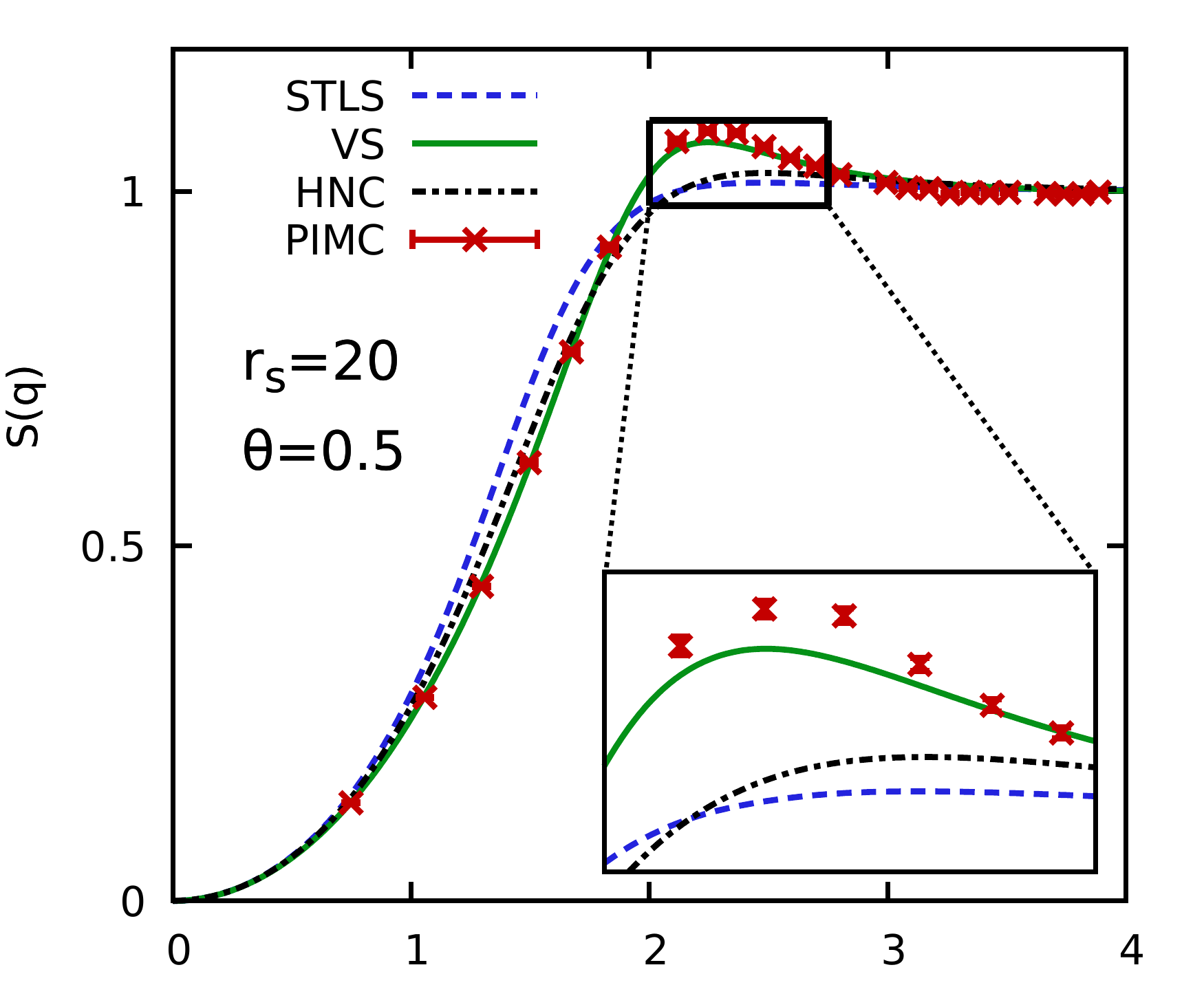}\hspace*{-0.4cm}\includegraphics[width=0.36\textwidth]{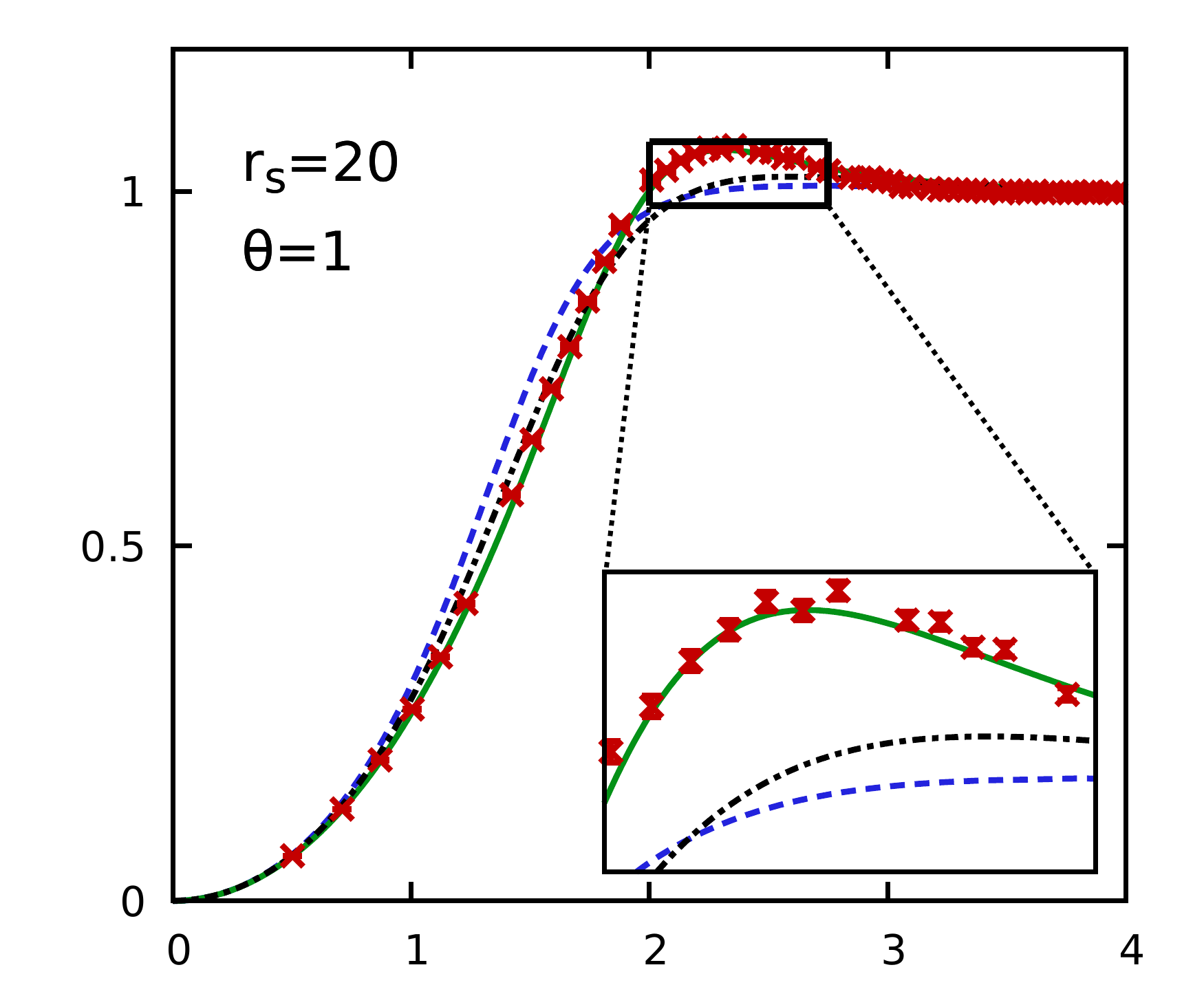}\hspace*{-0.4cm}\includegraphics[width=0.36\textwidth]{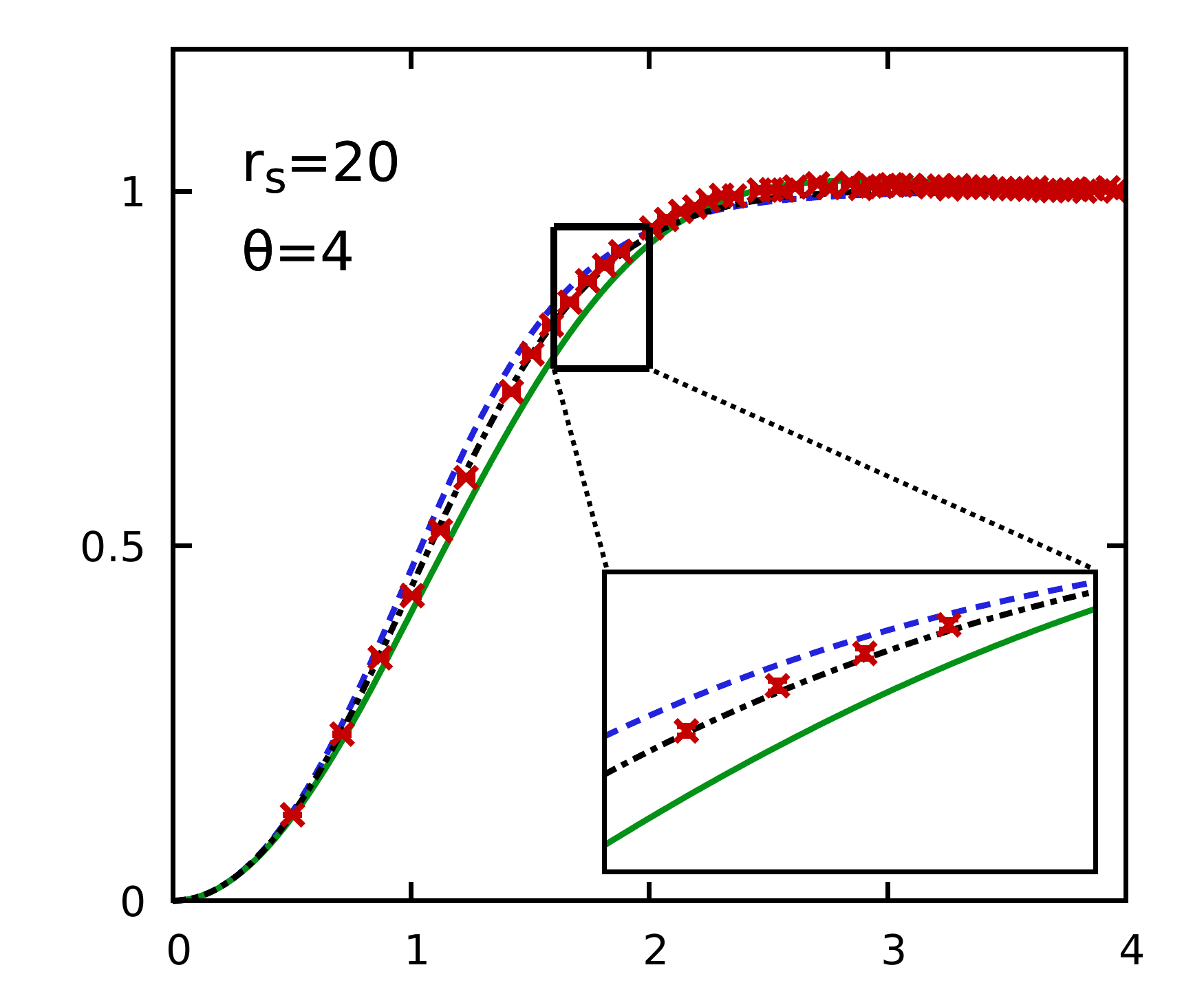}\\ 
\hspace*{-0.4cm}\includegraphics[width=0.36\textwidth]{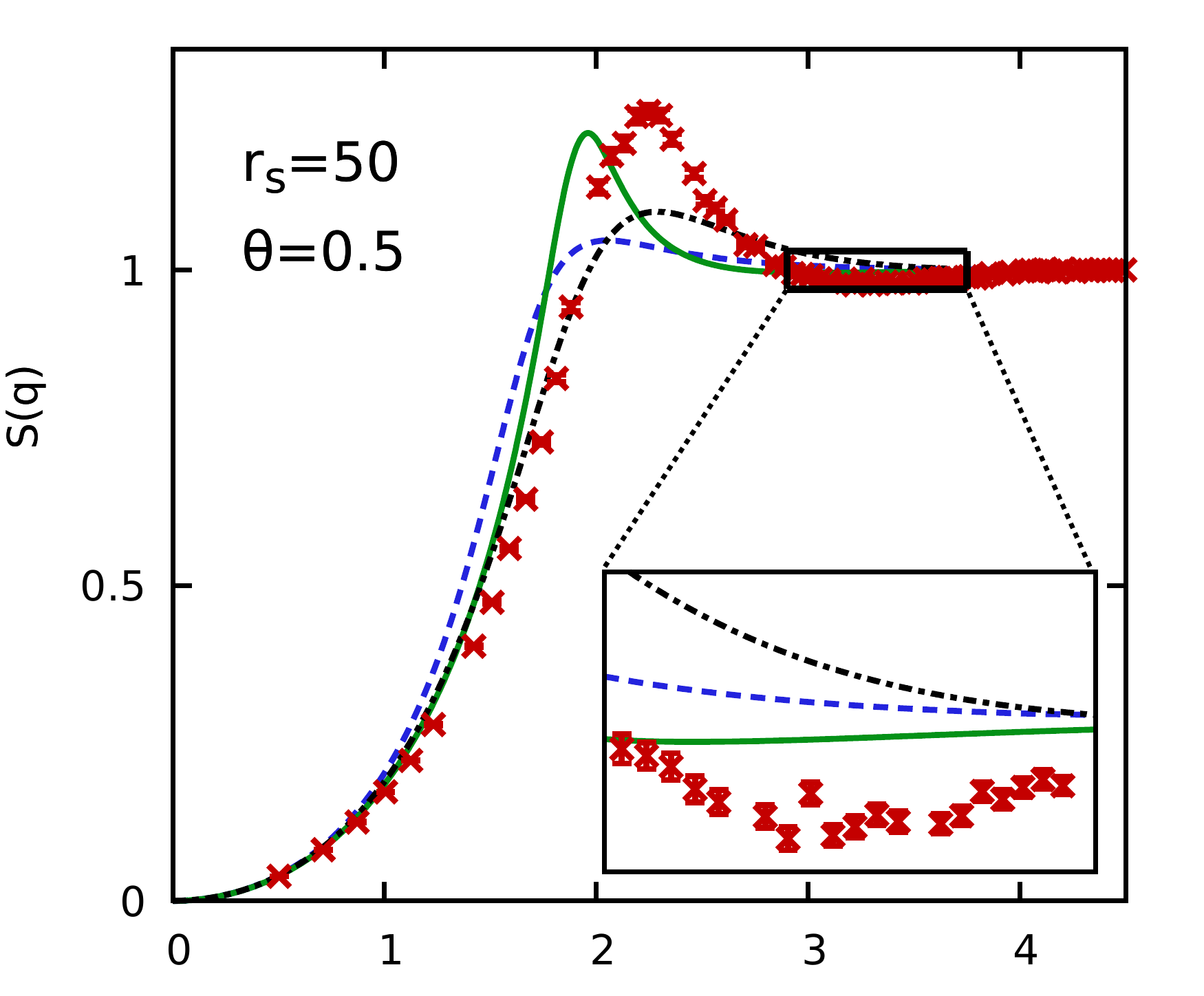}\hspace*{-0.4cm}\includegraphics[width=0.36\textwidth]{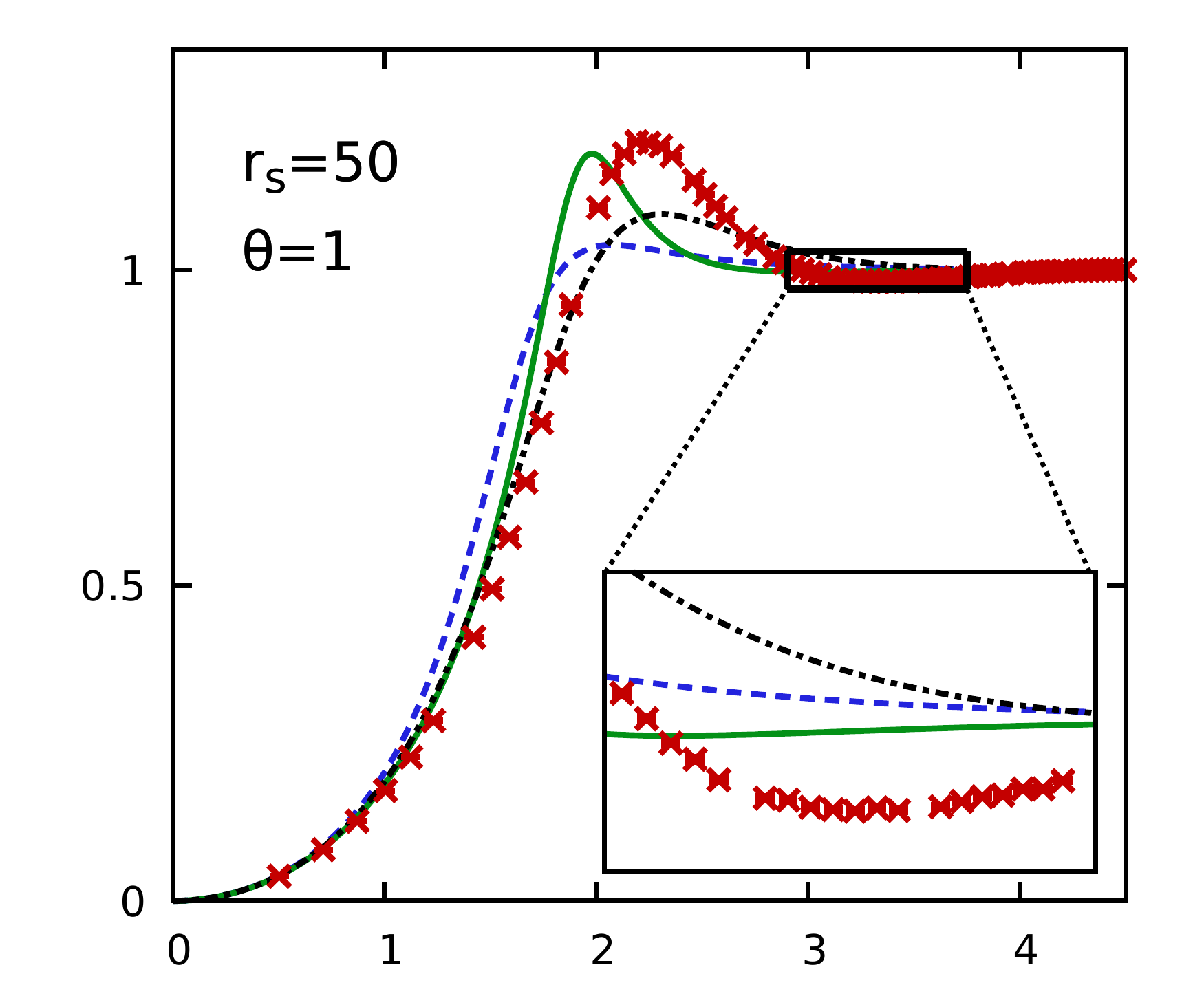}\hspace*{-0.4cm}\includegraphics[width=0.36\textwidth]{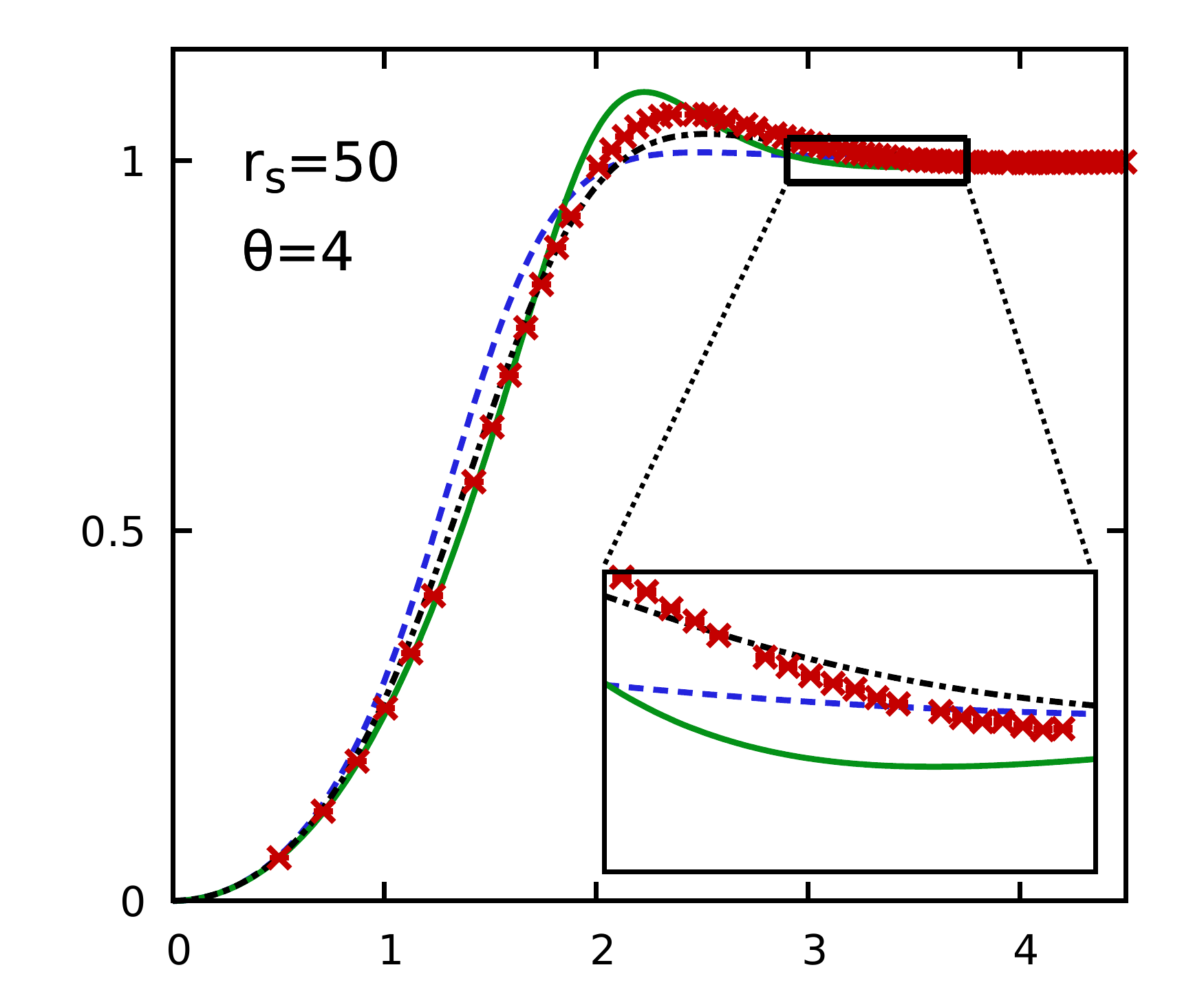}\\ 
\hspace*{-0.4cm}\includegraphics[width=0.36\textwidth]{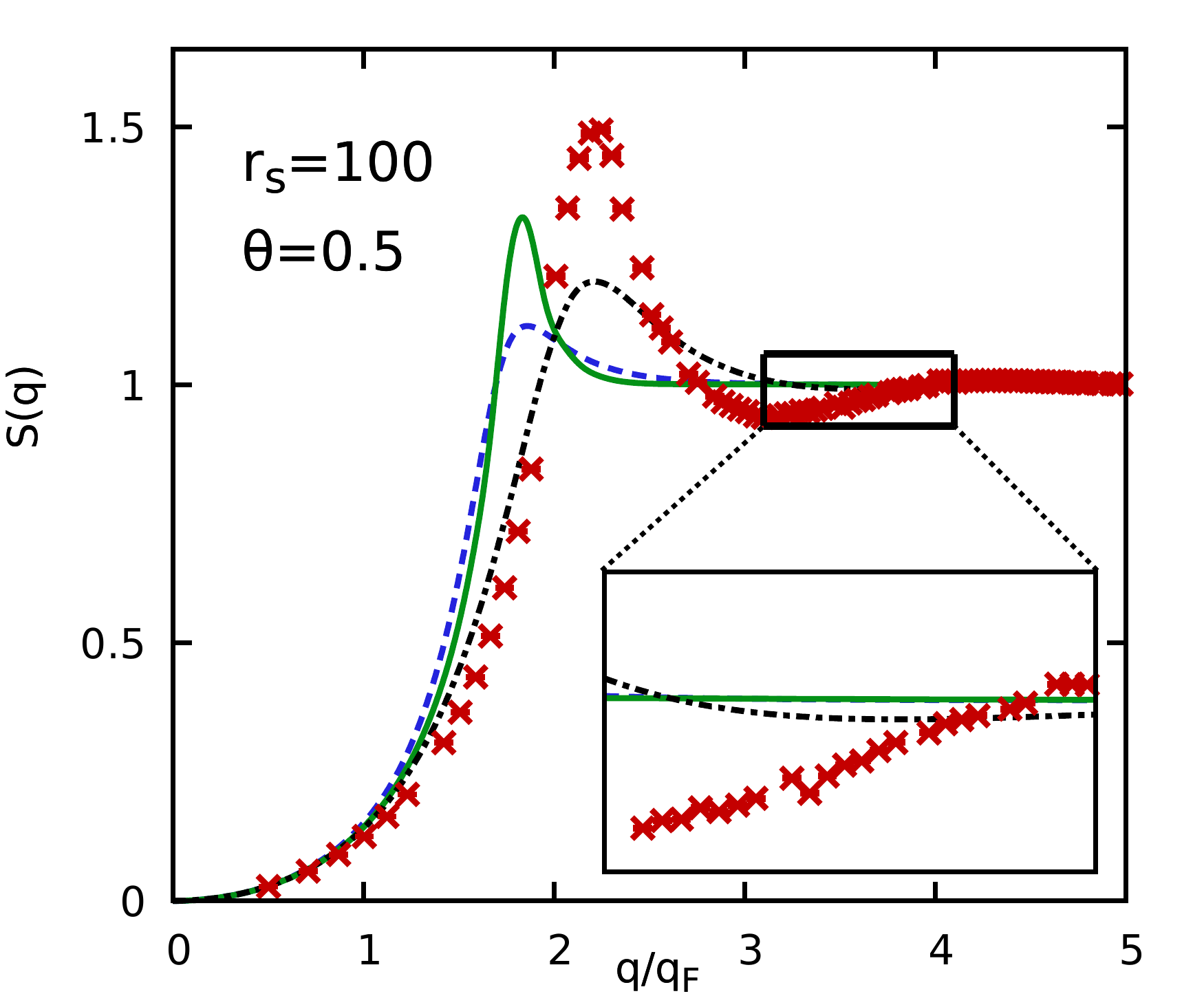}\hspace*{-0.4cm}\includegraphics[width=0.36\textwidth]{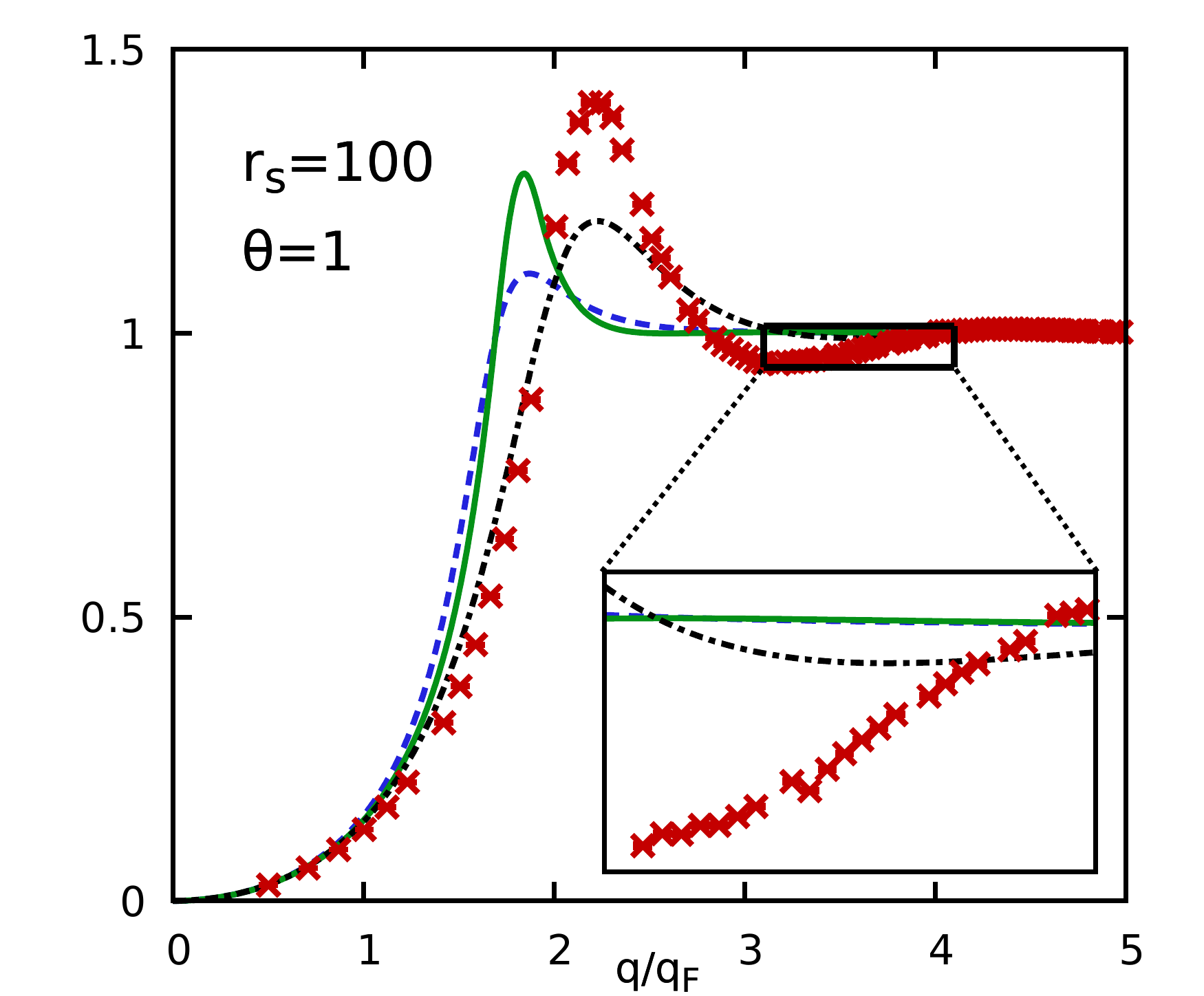}\hspace*{-0.4cm}\includegraphics[width=0.36\textwidth]{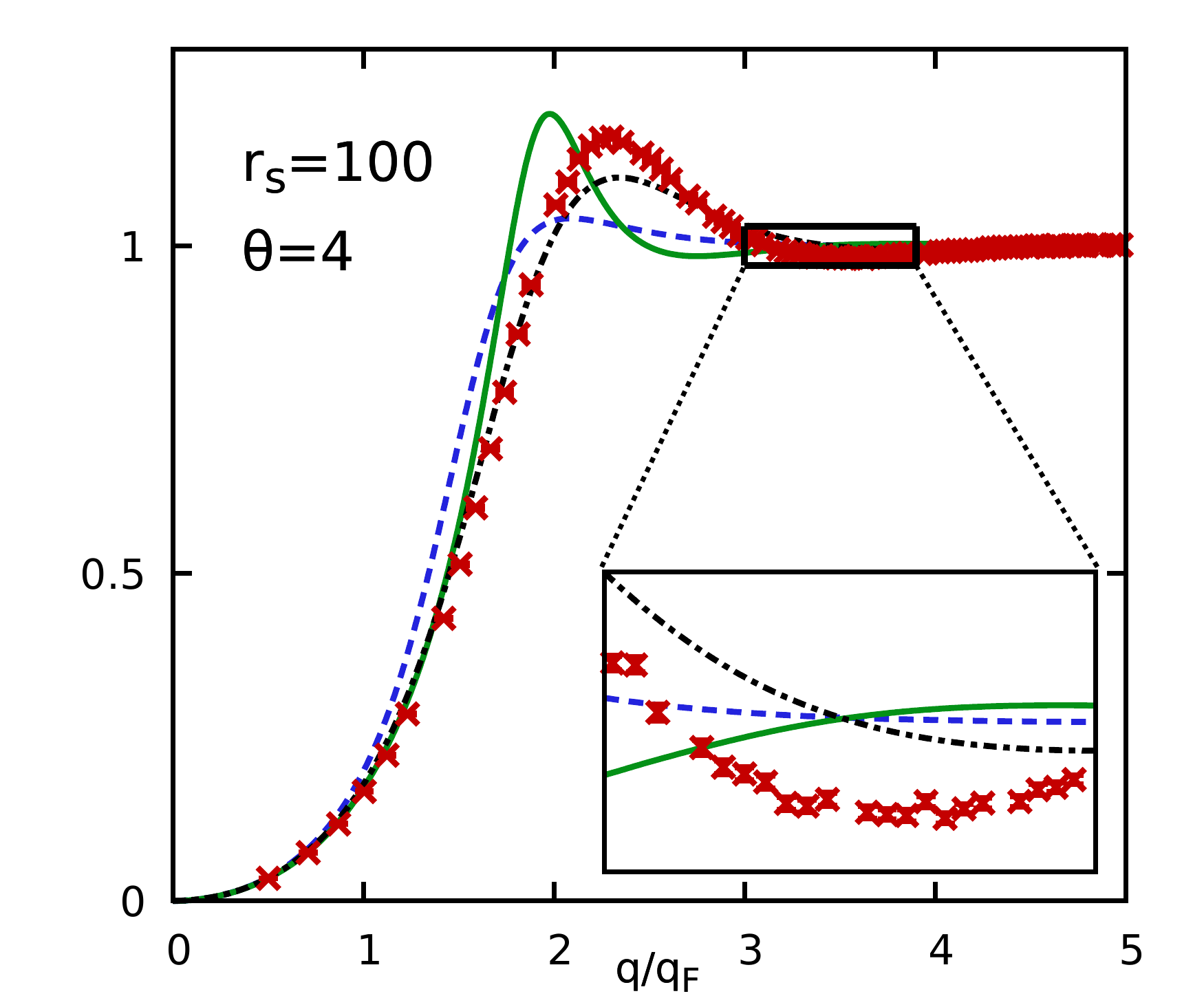}
\caption{\label{fig:S}
Comparison of the static structure factor of the unpolarized electron liquid. Shown are PIMC data (red crosses, $N=66$ electrons, except for $r_s=20$ and $\theta=0.5$ where we show $N=20$) and the results from various dielectric approximations, in particular STLS (dashed blue), VS (solid green), and HNC (dash-dotted black). All PIMC data are available online~\cite{github_link}.
}
\end{figure*}

Let us start our discussion of the physics of the unpolarized electron liquid by examining the static structure factor $S(q)$, which is shown in Fig.~\ref{fig:S} for nine different $r_s$-$\theta$-combinations.
The red crosses correspond to our new PIMC results and have been obtained for $N=66$ electrons, except for $r_s=20$ and $\theta=0.5$, where we are restricted to $N=20$ due to the fermion sign problem. In particular, $S(q)$ provides the complete wave-number resolved information about coupling effects in the system, and is equivalent to the pair-correlation function $g(r)$.

The top row shows results for the highest considered density in this work, $r_s=20$, and three values of the temperature parameter, namely $\theta=0.5$ (left), $\theta=1$ (center), and $\theta=4$ (right). At the lowest two temperatures, we observe a significant maximum in $S(q)$ around $q_\textnormal{max}\approx2.3q_\textnormal{F}$, which is an indication of an incipient short-range order. Consequently, this feature decreases with increasing $\theta$ and vanishes for $\theta=4$.

The middle row shows results for $S(q)$ for stronger coupling, $r_s=50$, and the same three values of the temperature parameter. Naturally, the system becomes more ordered, which manifests in a larger amplitude of the main peak, which even persists at $\theta=4$. Moreover, this feature is followed by a significant minimum around $q\approx3q_\textnormal{F}$ both for $\theta=0.5$ and $\theta=1$, see also the insets showing a magnified segment.

In the bottom row, we show results for $r_s=100$, which is the largest value of the coupling parameter considered in this work. Under these extreme conditions, the electrons exhibit a more pronounced liquid behavior, with the first peak exceeding (almost reaching) $S(q_\textnormal{max})=1.5$ at $\theta=0.5$ ($\theta=1$). In addition, we can clearly resolve a small second peak around $q=4q_\textnormal{F}$ in both cases.

Let us next consider the results for $S(q)$ from dielectric theory (cf.~Sec.~\ref{sec:DielectricTheory}), which are depicted as the dashed blue (STLS), solid green (VS), and dash-dotted black (HNC) lines. First and foremost, we note that all dielectric results exactly reproduce the exact parabolic long wave-length limit of $S(q)$ given by Eq.~(\ref{eq:S0}) for all density-temperature combinations by design, approach unity for large $q$, and deviate in between.

At $r_s=20$ and $\theta=0.5,1$, both STLS and the HNC-based scheme do not qualitatively reproduce the peak in our PIMC data, although HNC does constitute an improvement over STLS over the entire $q$-range. In contrast, the VS approach is in spectacular agreement to the red crosses and reproduces the maximum in $S(q)$ regarding position, width, and height. Interestingly, this behavior changes significantly for $\theta=4$, where STLS (VS) is significantly too high (low) for $q\lesssim2$, wheras the HNC approach is in good agreement to the PIMC data everywhere.

Upon increasing the coupling strength to $r_s=50$, no dielectric approach is capable to reproduce all qualitative aspects of the PIMC results for $\theta=0.5,1$: The HNC scheme constitutes the most accurate approximation for $q\leq q_\textnormal{max}$ and correctly predicts the position of the first peak, but considerably underestimates the height. In contrast, the VS curve is superior regarding peak height, but gives the wrong position. Lastly, the STLS approach combines the shortcomings of both and gives a too low peak at the wrong position. In addition, we note that no dielectric method is capable to resolve the shallow minimum in $S(q)$ around $q\approx3q_\textnormal{F}$, cf.~the insets showing a magnified segment around this feature. At $\theta=4$, the respective accuracy of the dielectric methods somewhat changes and HNC and VS provide a similarly accurate description of $S(q)$.

Let us conclude this section with an assessment of dielectric theory at strong coupling, $r_s=100$. While there appear significant deviations between the former and the PIMC data for the entire relevant $q$-range, the HNC-scheme still constitutes a remarkable improvement over STLS everywhere. More specifically, HNC gives the correct peak position, though again with an underestimated height, and even predicts a subsequent minimum in $S(q)$, see the inset. 
Furthermore, VS provides the most accurate description of the peak height, but at a too small $q$.

\subsection{Interaction energy\label{sec:vvv}}

\begin{table*}
\caption{\label{tab:interaction}Interaction energy of the unpolarized electron liquid. Shown are finite-size corrected PIMC data and the corresponding extrapolation error ($\Delta$PIMC) as discussed in Sec.~\ref{sec:PIMC_results}, as well as dielectric results within STLS~\cite{stls,stls2}, VS~\cite{stls2}, and HNC~\cite{tanaka_hnc} computed from the static structure factor $S(q)$ via Eq.~(\ref{eq:v}). All data points are shown in Fig.~\ref{fig:v}.
}
\begin{ruledtabular}
\begin{tabular}{lllllll}
$r_s$ & $\theta$ & PIMC & $\Delta$PIMC & STLS & VS & HNC \\ \hline 
$20$ & $0.5$ & $-0.037812$ & $3.6\cdot10^{-5}$ & $-0.0368208$ & $-0.0377668$ & $-0.0379232$ \\
$ $ & $0.75$ & $-0.037723$ &$1.8\cdot10^{-5}$ & $-0.0368128$ & $-0.0378456$ & $-0.0379253$ \\
$ $ & $1$    & $-0.0375304$ & $1.9\cdot10^{-6}$ & $-0.0366796$ & $-0.0378388$ & $-0.0377905$ \\
$ $ & $2$    & $-0.0363461$ & $6.1\cdot10^{-6}$ & $-0.0356344$ & $-0.0374242$ & $-0.0366536$ \\
$ $ & $4$    & $-0.0338924$ & $4.2\cdot10^{-6}$ & $-0.0333573$ & $-0.035110$  & $-0.034162$ \\ \hline 
$30$& $0.5$  & $-0.0259156$ & $6.3\cdot10^{-6}$ & $-0.025017$  & $-0.025534$  & $-0.025868$ \\
$ $ & $0.75$ & $-0.0258671$ & $4.7\cdot10^{-6}$ & $-0.0250256$ & $-0.025567$  & $-0.0258855$ \\
$ $ & $1$    & $-0.0257792$ & $2.0\cdot10^{-6}$ & $-0.024979$  & $-0.025605$  & $-0.0258388$ \\
$ $ & $2$    & $-0.0252204$ & $2.3\cdot10^{-6}$ & $-0.0245047$ & $-0.0255615$ & $-0.0253119$ \\
$ $ & $4$    & $-0.0239098$ & $2.5\cdot10^{-6}$ & $-0.0233338$ & $-0.0250387$ & $-0.0239991$ \\ \hline 
$50$& $0.5$  & $-0.016007$  & $1.4\cdot10^{-5}$ & $-0.0152889$ & $-0.0154739$ & $-0.0158967$ \\
$ $ & $0.75$ & $-0.0159838$ & $7.4\cdot10^{-6}$ & $-0.0152991$ & $-0.0154985$ & $-0.0159106$ \\
$ $ & $1$    & $-0.0159485$ & $7.6\cdot10^{-6}$ & $-0.0152884$ & $-0.0155096$ & $-0.015902$ \\
$ $ & $2$    & $-0.0157336$ & $5.6\cdot10^{-6}$ & $-0.0151287$ & $-0.0154601$ & $-0.0157178$ \\
$ $ & $4$    & $-0.0151703$ & $1.3\cdot10^{-6}$ & $-0.0146535$ & $-0.0151513$ & $-0.0151599$ \\ \hline 
$100$&$0.5$  & $-0.008255$  & $1.2\cdot10^{-5}$ & $-0.00778221$& $-0.00782947$& $-0.00815748$ \\
$ $  &$0.75$ & $-0.0082457$ & $7.5\cdot10^{-6}$ & $-0.00778623$& $-0.00783455$& $-0.0081637$ \\
$ $  &$1$    & $-0.0082349$ & $2.6\cdot10^{-6}$ & $-0.00778611$& $-0.00783633$& $-0.00816496$ \\
$ $  &$2$    & $-0.0081765$ & $2.9\cdot10^{-6}$ & $-0.0077569$ & $-0.0078318$ & $-0.00812802$ \\
$ $  &$4$    & $-0.00800623$& $8.4\cdot10^{-7}$ & $-0.00763381$& $-0.00777045$& $-0.00796771$
 
\end{tabular}
\end{ruledtabular}
\end{table*}

\begin{figure*}\centering
\includegraphics[width=0.448\textwidth]{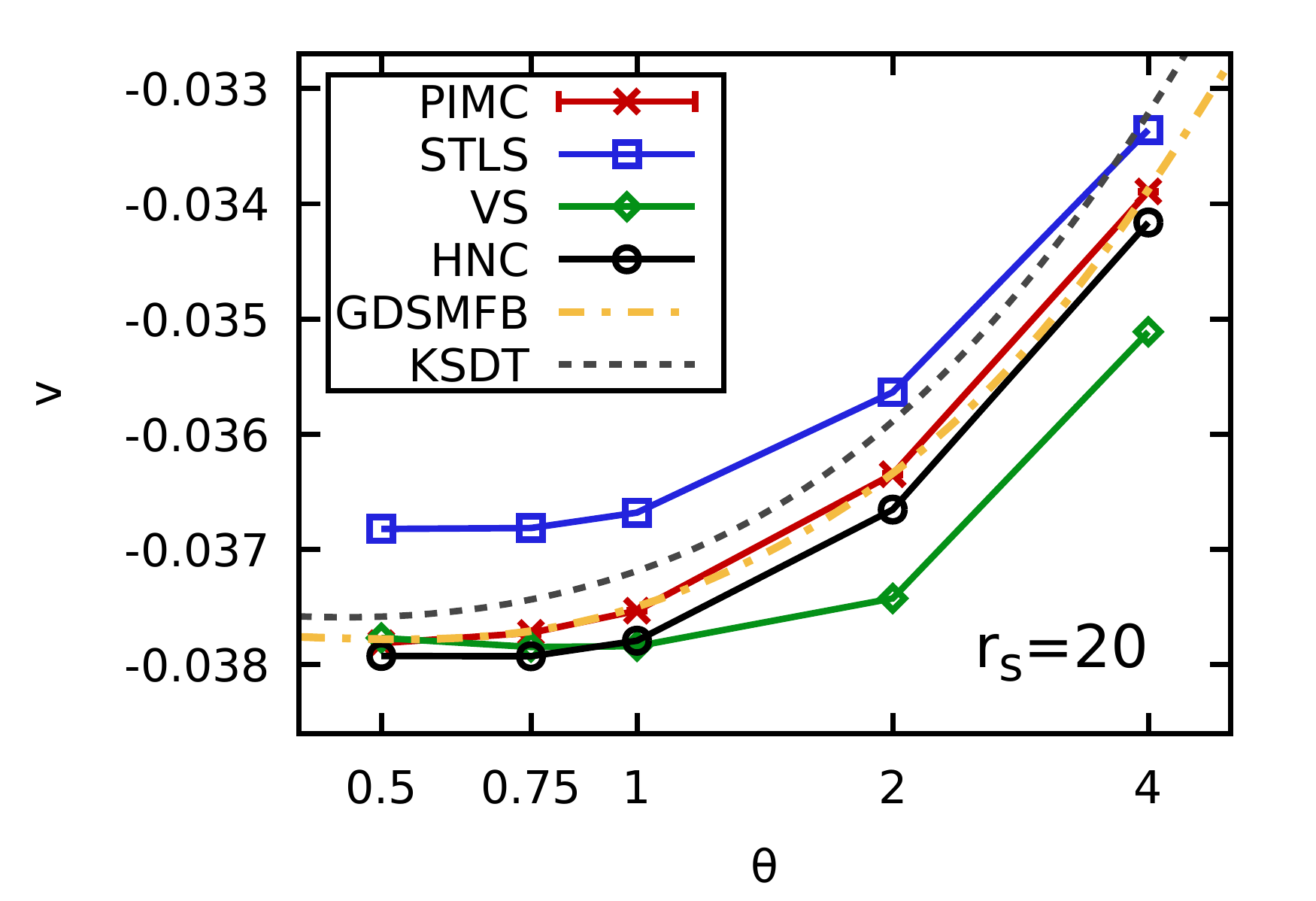}\includegraphics[width=0.448\textwidth]{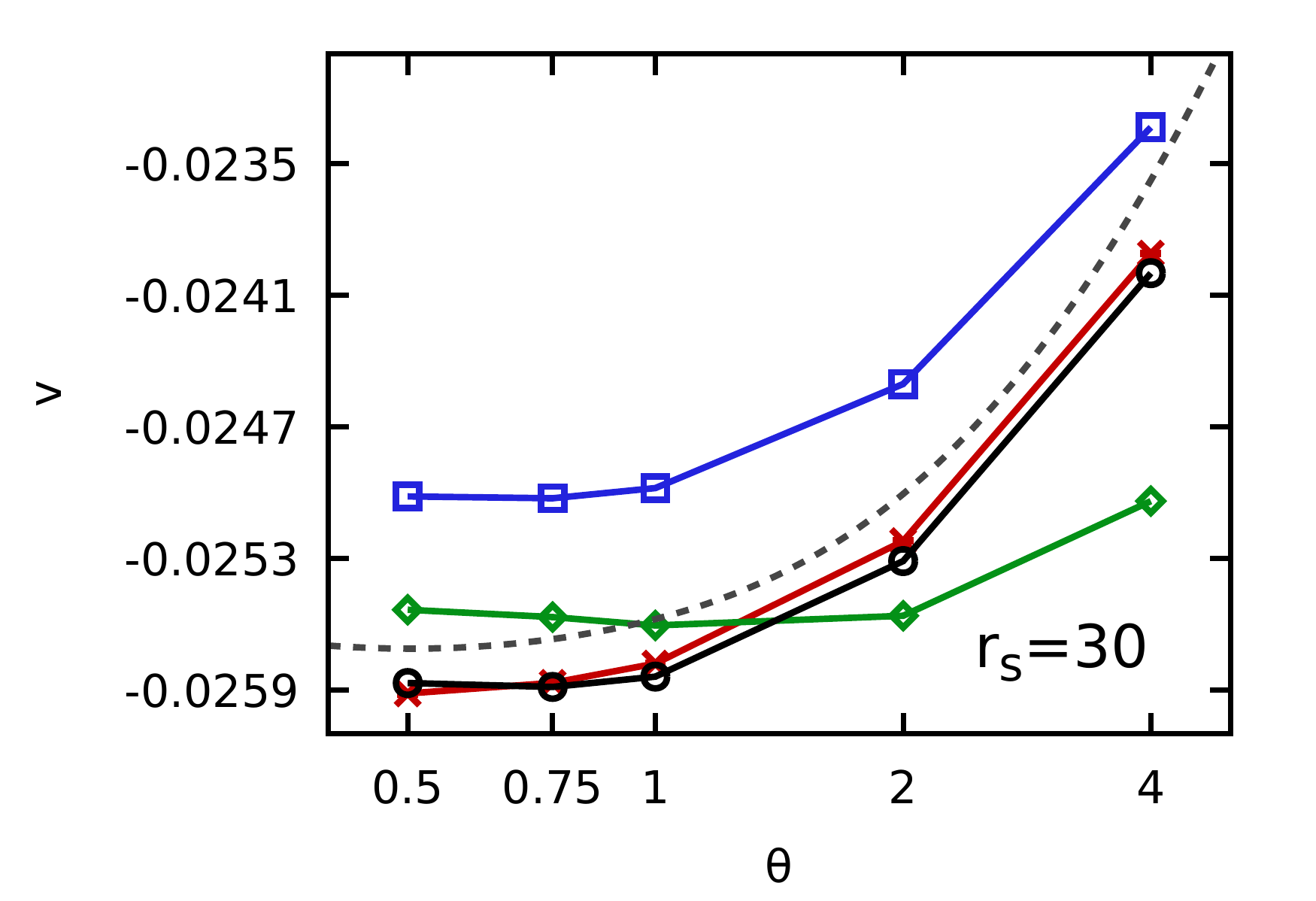} \\
\includegraphics[width=0.448\textwidth]{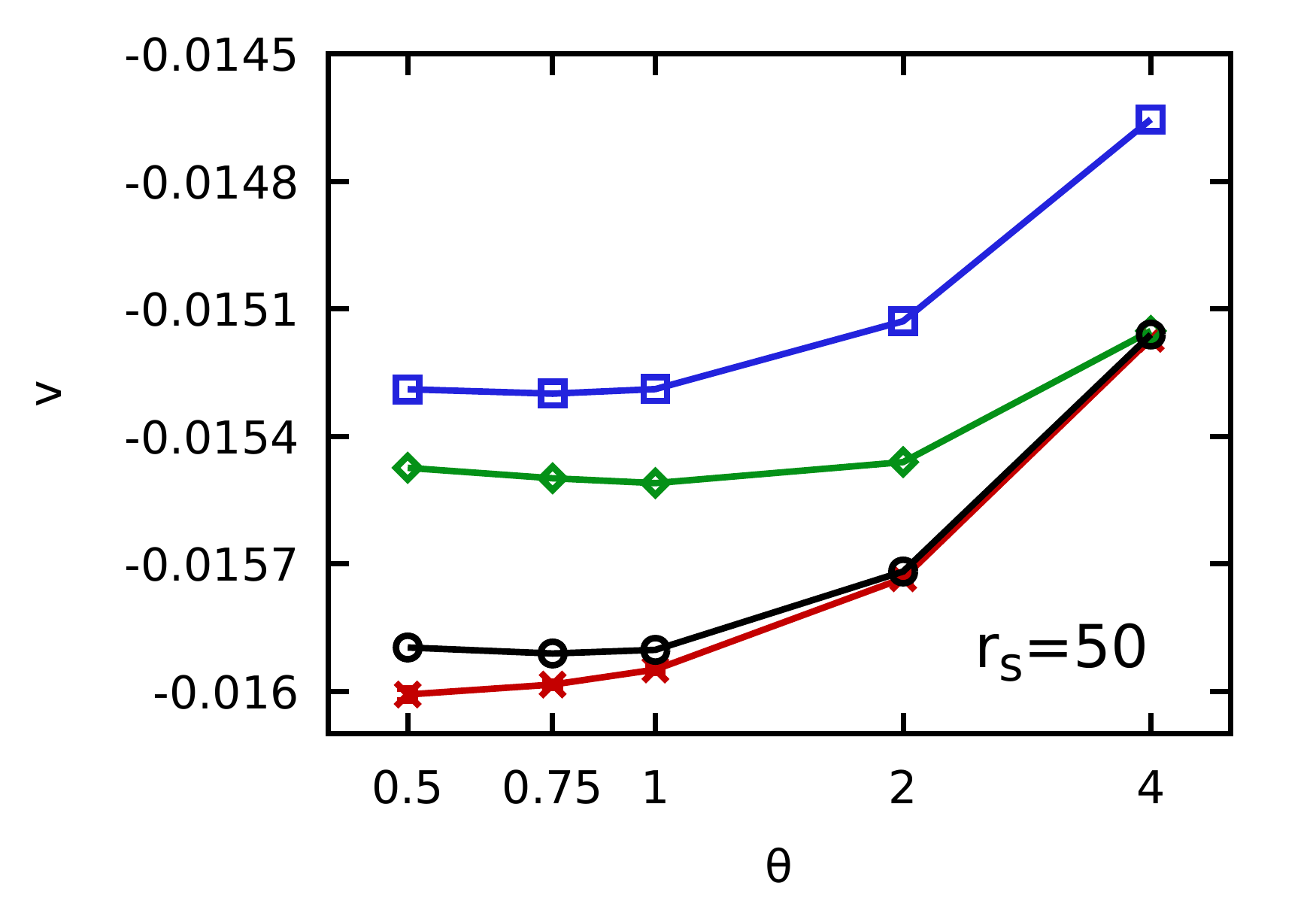}\includegraphics[width=0.448\textwidth]{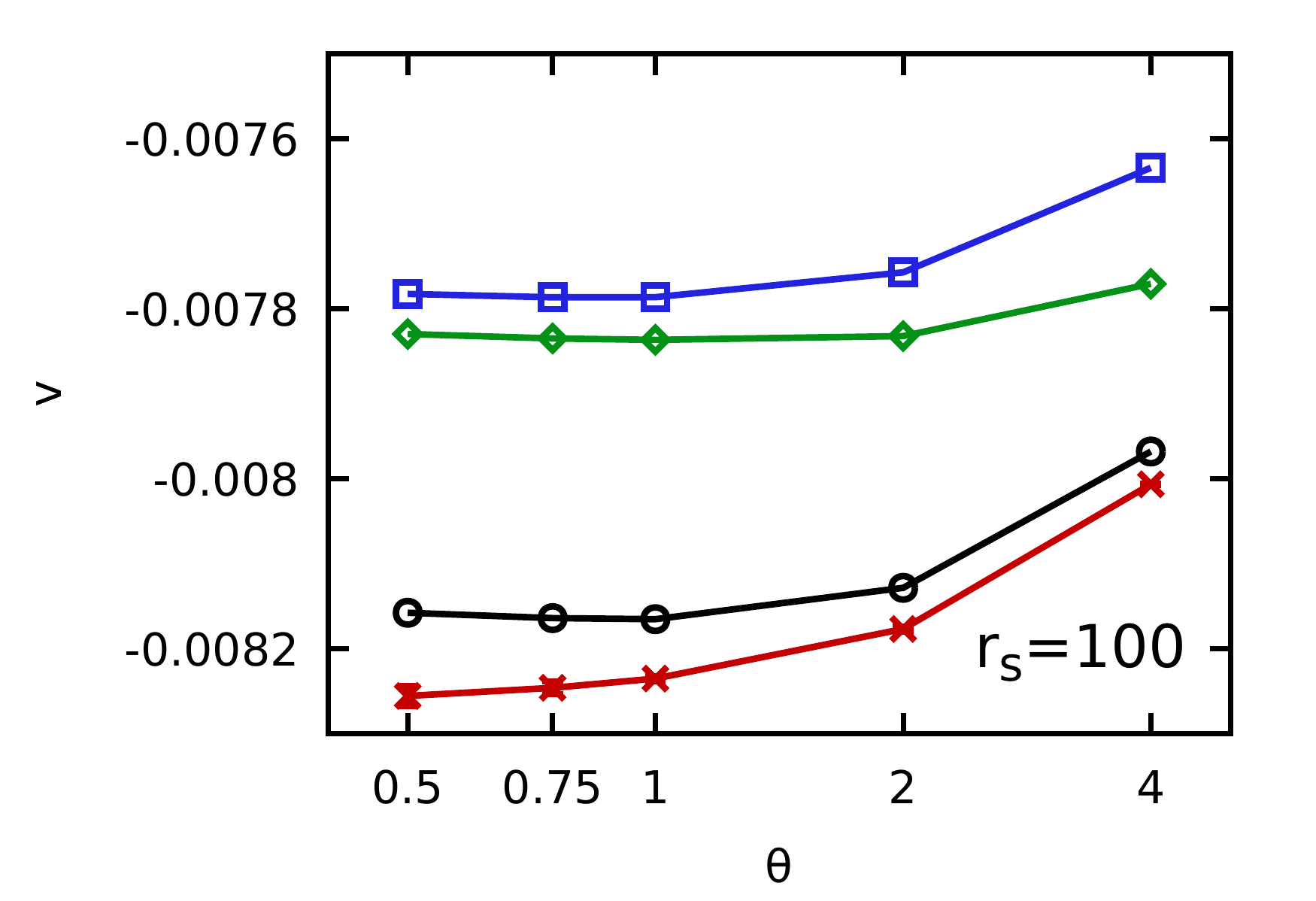}
\caption{\label{fig:v}
Comparison of interaction energies of the unpolarized electron liquid for $r_s=20$ (top left), $r_s=30$ (top right), $r_s=50$ (bottom left), and $r_s=100$ (bottom right). Shown are finite-size corrected PIMC data (red crosses), dielectric results within STLS (blue squares), VS (green diamonds), and HNC (black circles). In addition, we show interaction energies computed from two recent parametrizations of the exchange--correlation free energy $f_\textnormal{xc}$ via Eq.~(\ref{eq:v_derivative}), namly by Groth \textit{et al.}~\cite{groth_prl} (GDSMFB, dash-dotted yellow) for $r_s=20$ and Karasiev \textit{et al.}~\cite{ksdt} (KSDT, black dashed) for $r_s\leq 30$.
}
\end{figure*}

In Fig.~\ref{fig:v}, we show extensive comparisons of the interaction energy per particle $v$ in the thermodynamic limit between different methods and data sets. More specifically, we show the $\theta$-dependence of $v$ for four different values of the density parameter $r_s$ covering the entire relevant electron liquid regime.

Let us start this investigation by considering the top left panel corresponding to $r_s=20$. The red crosses depict our new PIMC data, which have been extrapolated to the TDL as discussed in Sec.~\ref{sec:PIMC_results}, cf.~Fig.~\ref{fig:FSC_v}. Further, the blue squares, green diamonds, and black circles show dielectric results within STLS, VS, and HNC, which have been obtained by numerically integrating over $S(q)$ via Eq.~(\ref{eq:static_structure_factor}). First and foremost, we note that the interaction energy increases towards high temperature, as it is expected. Somewhat surprisingly, we find that STLS overall constitutes the least accurate dielectric theory, which is in contrast to previous findings in the warm dense matter regime~\cite{review}. Furthermore, we find that VS and HNC are almost indistinguishable for the three lowest considered temperatures. In this context, it is important to note that VS does give an accurate description of $S(q)$ at these parameters, whereas the high accuracy of HNC is due to a fortunate cancellation of errors under the integral in Eq.~(\ref{eq:static_structure_factor}). Interestingly, VS provides the least accurate results for $\theta=2$ and $\theta=4$, with a systematic deviation of $\Delta v/v\sim3\%$, and only the HNC scheme gives reliable results over the entire $\theta$-range.
In addition to dielectric theory, we also show the interaction energy computed from two recent parametrizations of the exchange--correlation free energy $f_\textnormal{xc}$ via
\begin{eqnarray}\label{eq:v_derivative}
v(r_s,\theta) = 2 f_\textnormal{xc}(r_s,\theta) + r_s \left. \frac{\partial f_\textnormal{xc}(r_s,\theta)}{\partial r_s}\right|_{\theta} \quad .
\end{eqnarray}
The dash-dotted yellow line depicts the results from Eq.~(\ref{eq:v_derivative}) using the parametrization by Groth \textit{et al.}~\cite{groth_prl} (GDSMFB) that is valid for $0\leq r_s \leq 20$ and based on a combination of different new PIMC methods. More specifically, it uses as input finite-size corrected~\cite{dornheim_prl} interaction energies from the permutation blocking PIMC approach, which relies on a small number of fourth-order factorizations of the thermal density matrix, see Refs.~\cite{dornheim,dornheim2,dornheim_neu} for extensive discussions. Yet, we find perfect agreement to our new, independent PIMC data over the entire $\theta$-range.
Finally, the dashed dark grey curve has been computed by evaluating Eq.~(\ref{eq:v_derivative}) using the parametrization of $f_\textnormal{xc}$ by Karasiev \textit{et al.}~\cite{ksdt} (KSDT), which is based on the restricted PIMC data from Ref.~\cite{brown_ethan}. Although the input data have been shown to be somewhat unreliable due to i) the uncontrolled fixed node approximation (see Ref.~\cite{schoof_prl} for the first systematic quantification of the corresponding nodal errors, and Ref.~\cite{review} for an overview) and ii) an insufficient treatment of finite-size effects~\cite{dornheim_prl}, the KSDT curve is in good qualitative agreement to our PIMC data and is, overall, more accurate than VS and STLS. For completeness, we mention that such discrepancies in the parametrization of $f_\textnormal{xc}$ in the low density regime are not expected to impact a density functional theory calculation in the warm dense matter regime and, thus, are primarily of academic interest, see Ref.~\cite{karasiev_status} for a recent discusion of this point.

Upon increasing the coupling parameter $r_s$, we observe the following major trends: i) VS becomes significantly less accurate, even though it correctly captures the increase in the peak height of $S(q)$, cf.~Fig.~\ref{fig:S}; ii) STLS does constitute the least reliable dielectric method for almost all depicted $r_s$-$\theta$-combinations with a systematic deviation of $\Delta v/v\approx 6\%$ at $r_s=100$ and $\theta=0.5$; iii) HNC provides the most accurate interaction energies for nearly all parameter combinations, although the accuracy somewhat decreases for large $r_s$ and small $\theta$. Still, even at $r_s=100$ and $\theta=0.5$, we find $\Delta v/v\approx 1\%$.

Note that the parametrization by KSDT is available at $r_s=30$ (it was constructed for $0\leq r_s \leq 40$), which is beyond the range of validity of GDSMFB. Again, we find that the dashed grey curve is in good qualitative agreement with the new PIMC data, and is more accurate than STLS and VS, though not HNC.

Finally, we mention that all extrapolated PIMC data and dielectric results are listed in Tab.~\ref{tab:interaction}.

\subsection{Density response and local field correction\label{sec:response_and_LFC}}

\begin{figure*}\centering
\hspace*{-0.4cm}\includegraphics[width=0.36\textwidth]{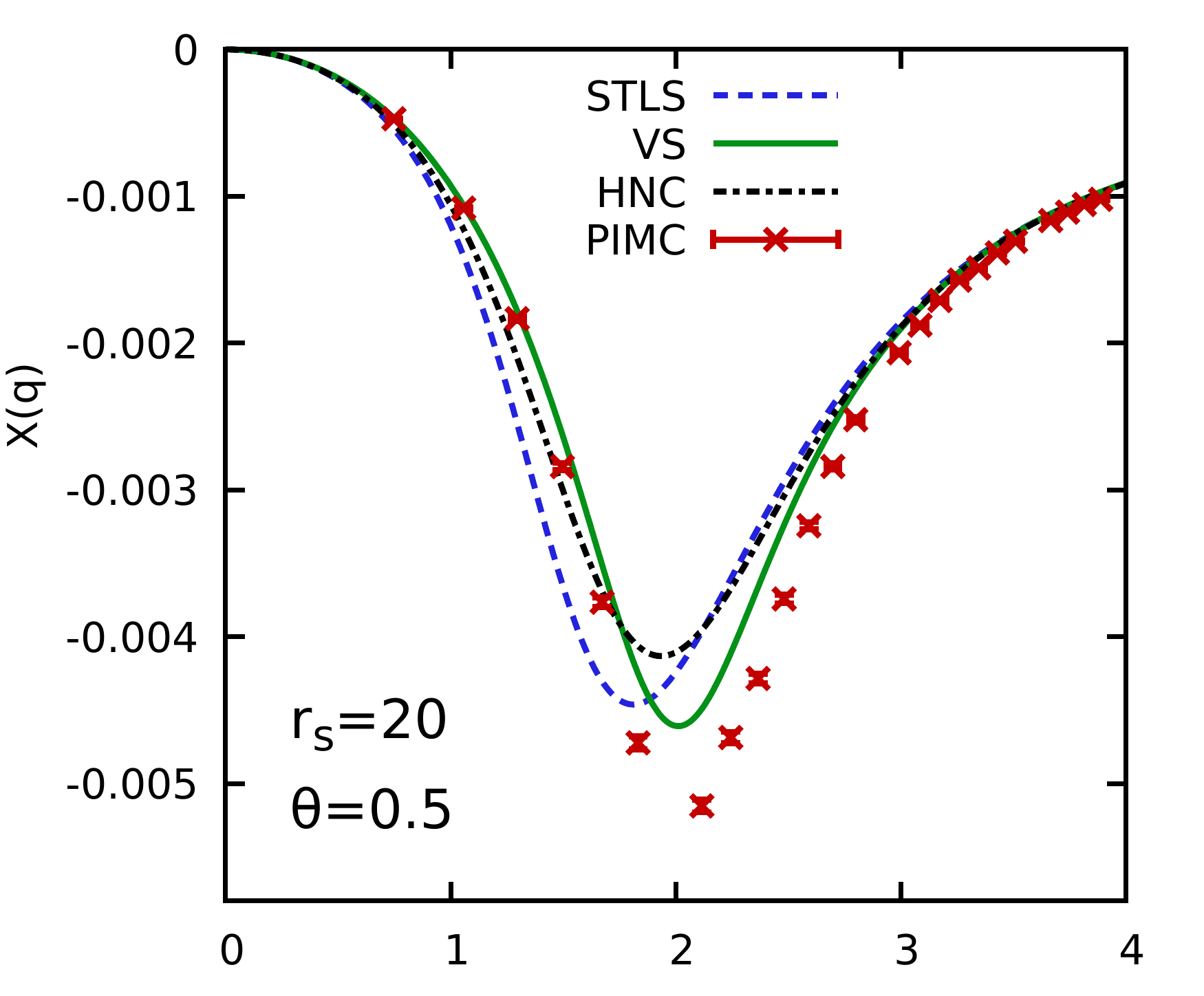}\hspace*{-0.4cm}\includegraphics[width=0.36\textwidth]{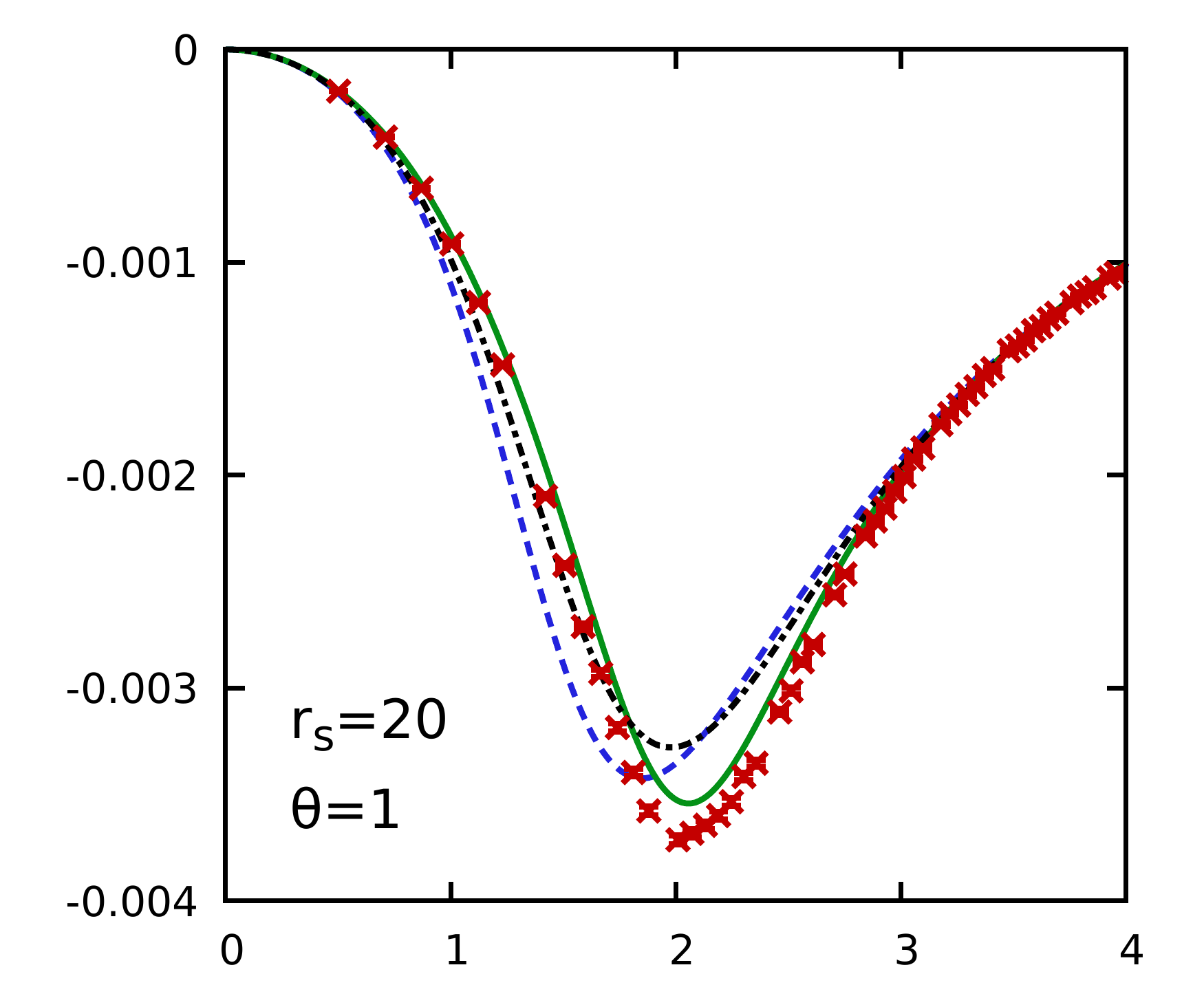}\hspace*{-0.4cm}\includegraphics[width=0.36\textwidth]{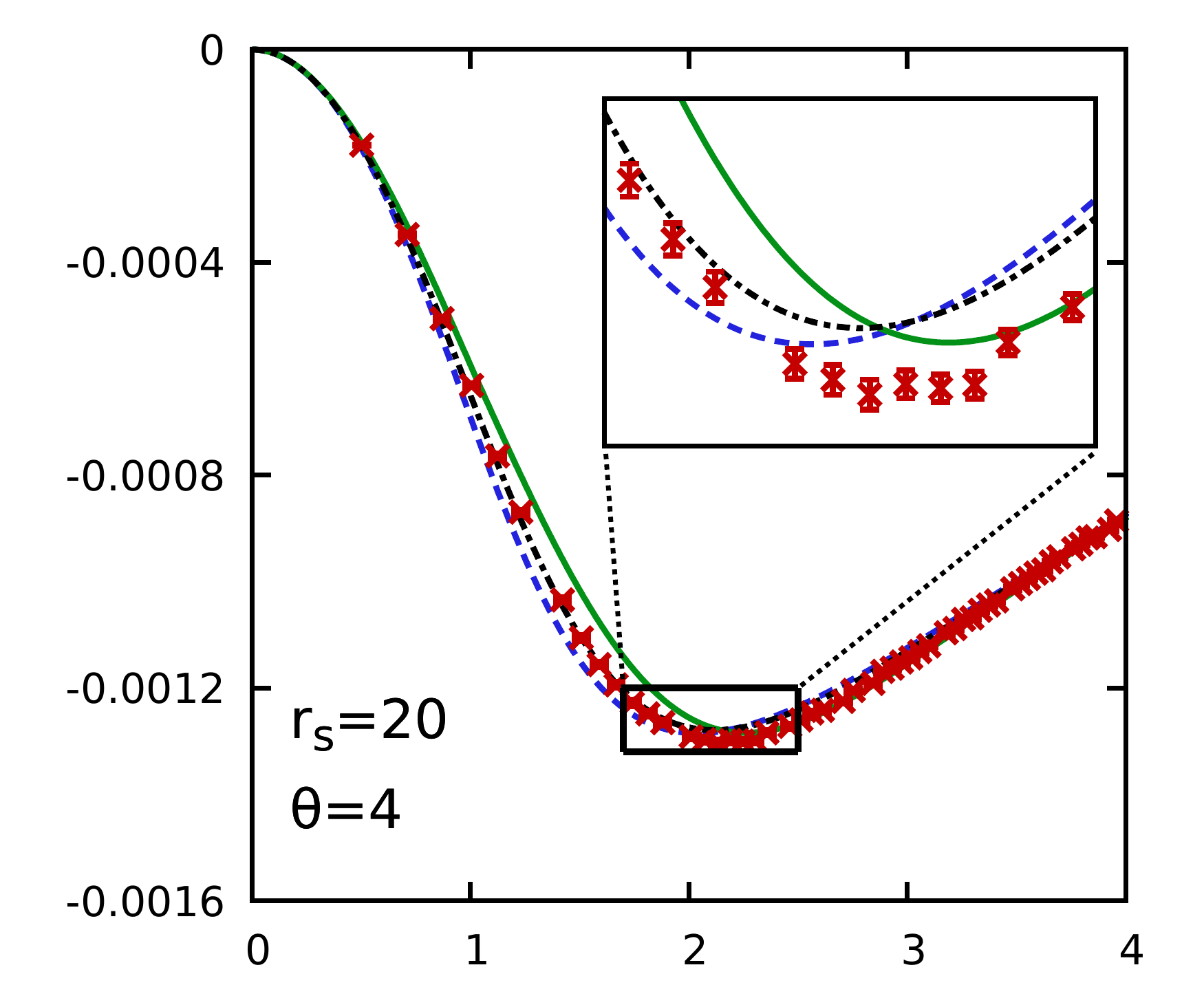}\\ 
\hspace*{-0.4cm}\includegraphics[width=0.36\textwidth]{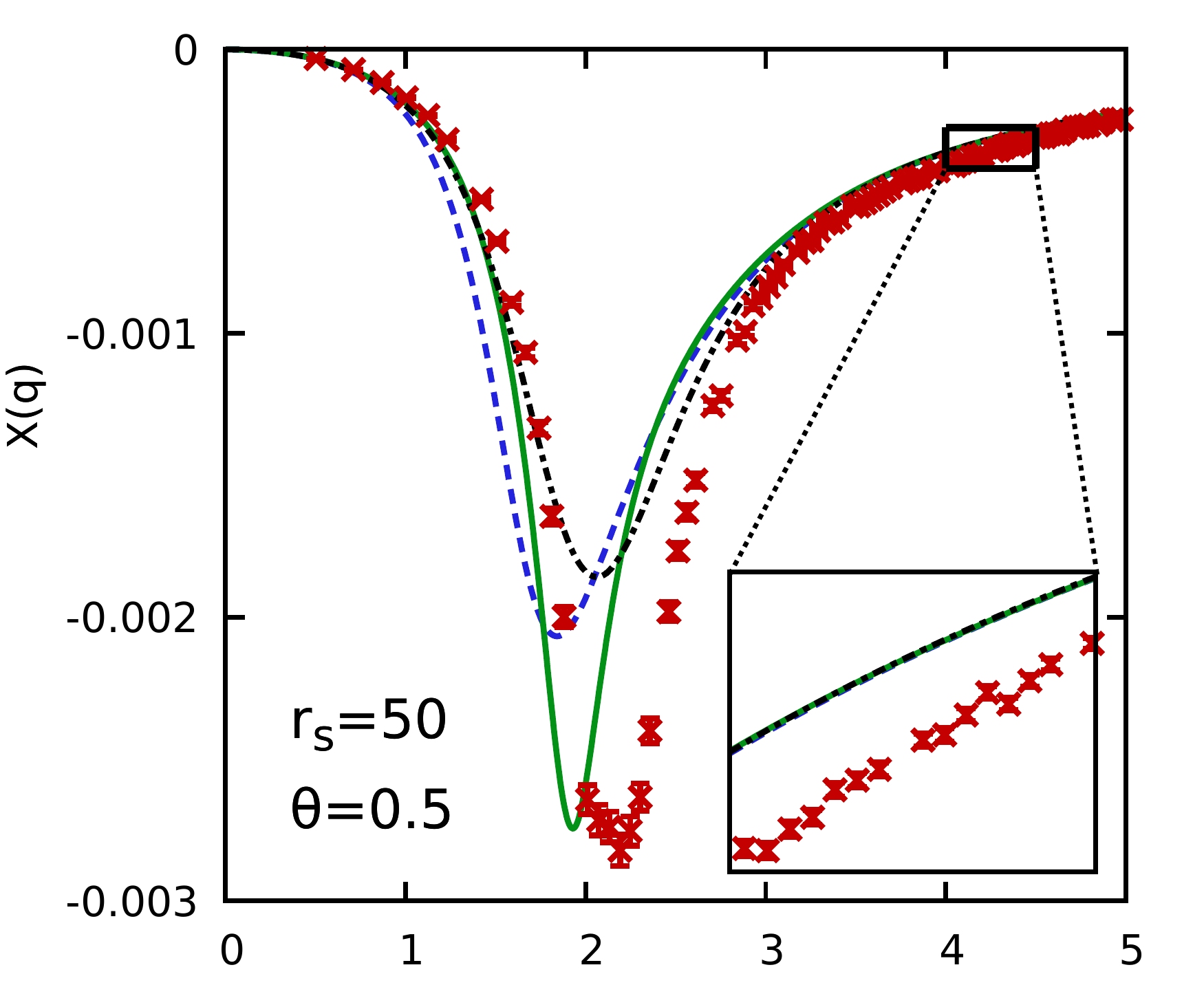}\hspace*{-0.4cm}\includegraphics[width=0.36\textwidth]{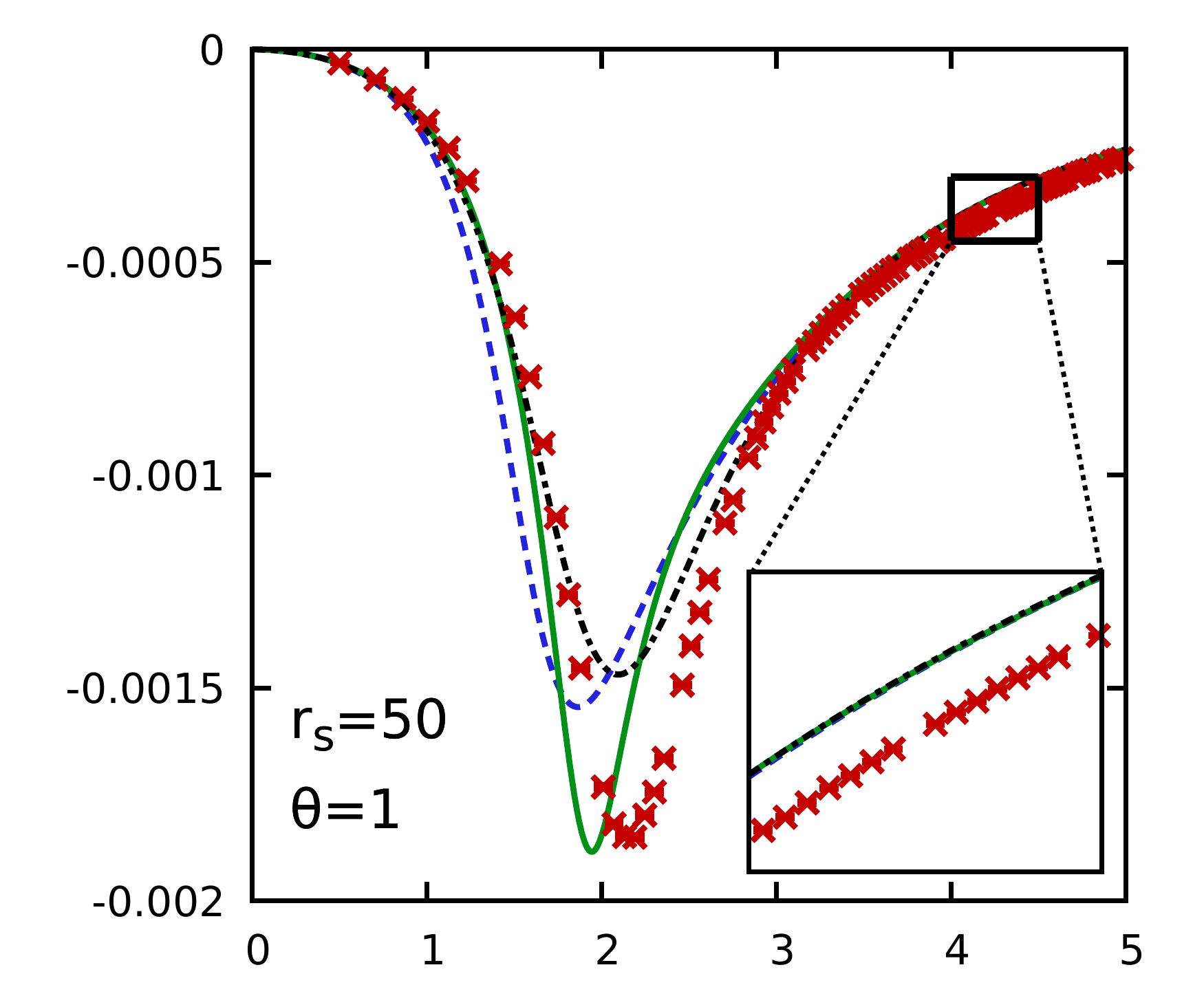}\hspace*{-0.4cm}\includegraphics[width=0.36\textwidth]{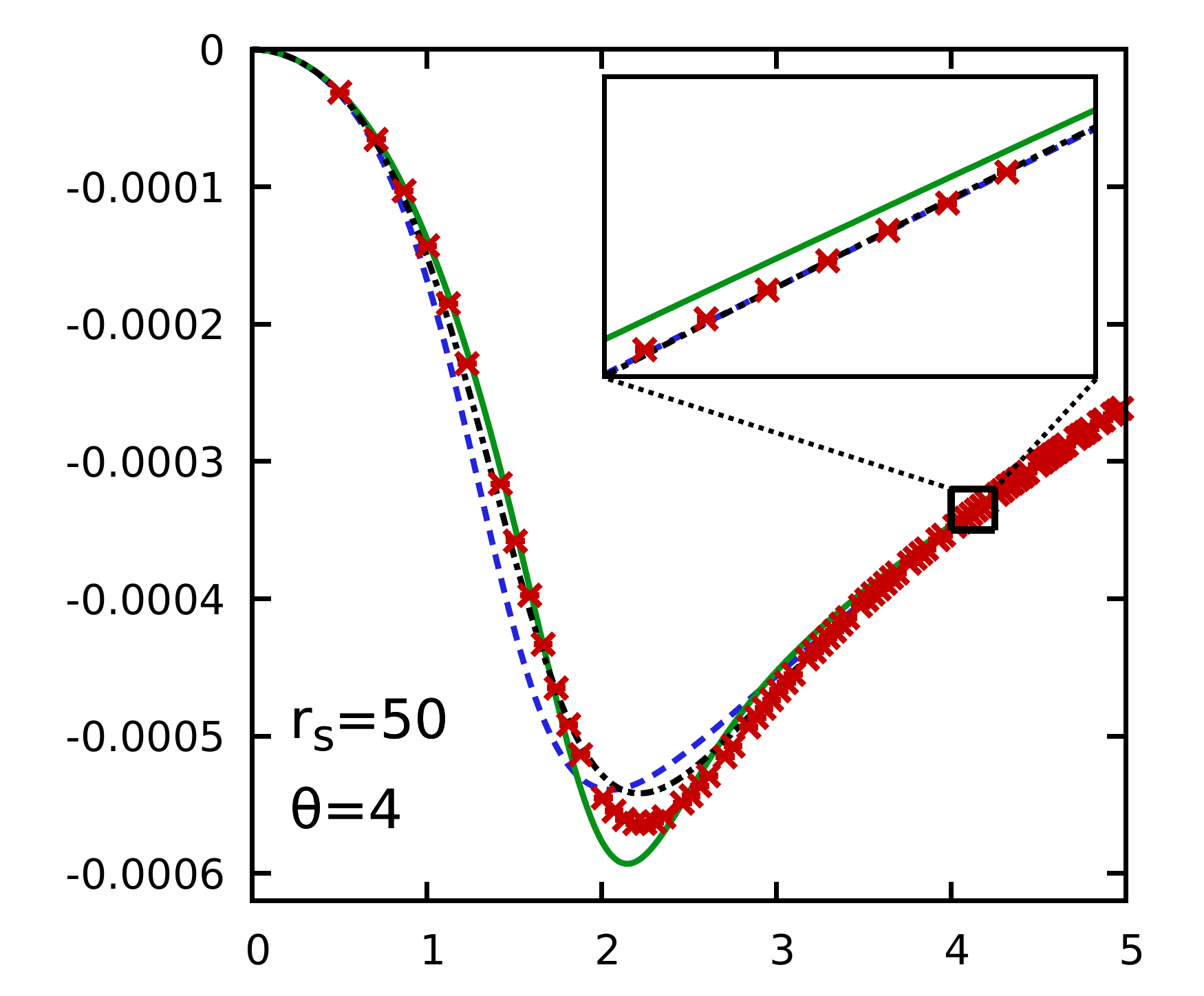}\\ 
\hspace*{-0.4cm}\includegraphics[width=0.36\textwidth]{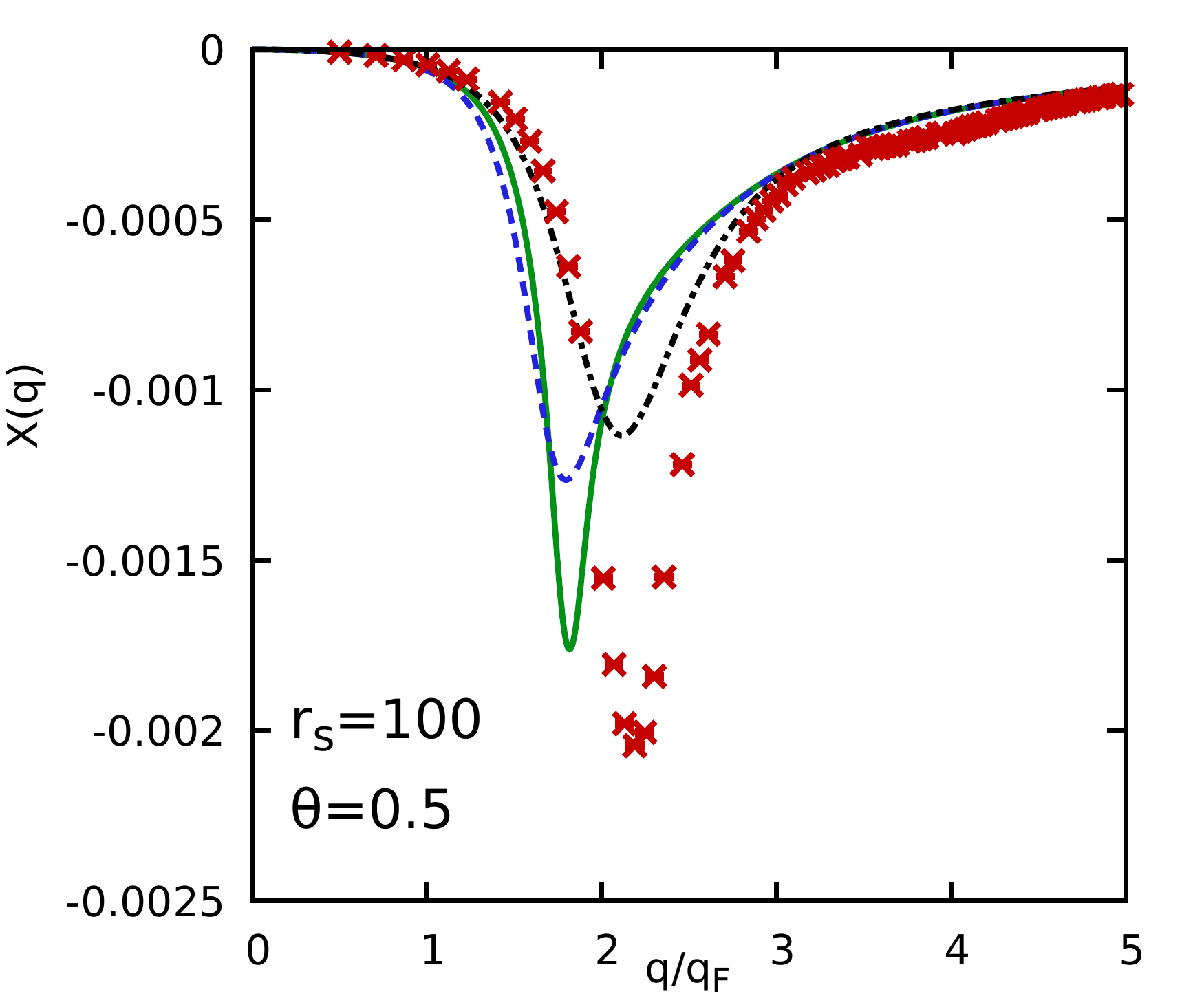}\hspace*{-0.4cm}\includegraphics[width=0.36\textwidth]{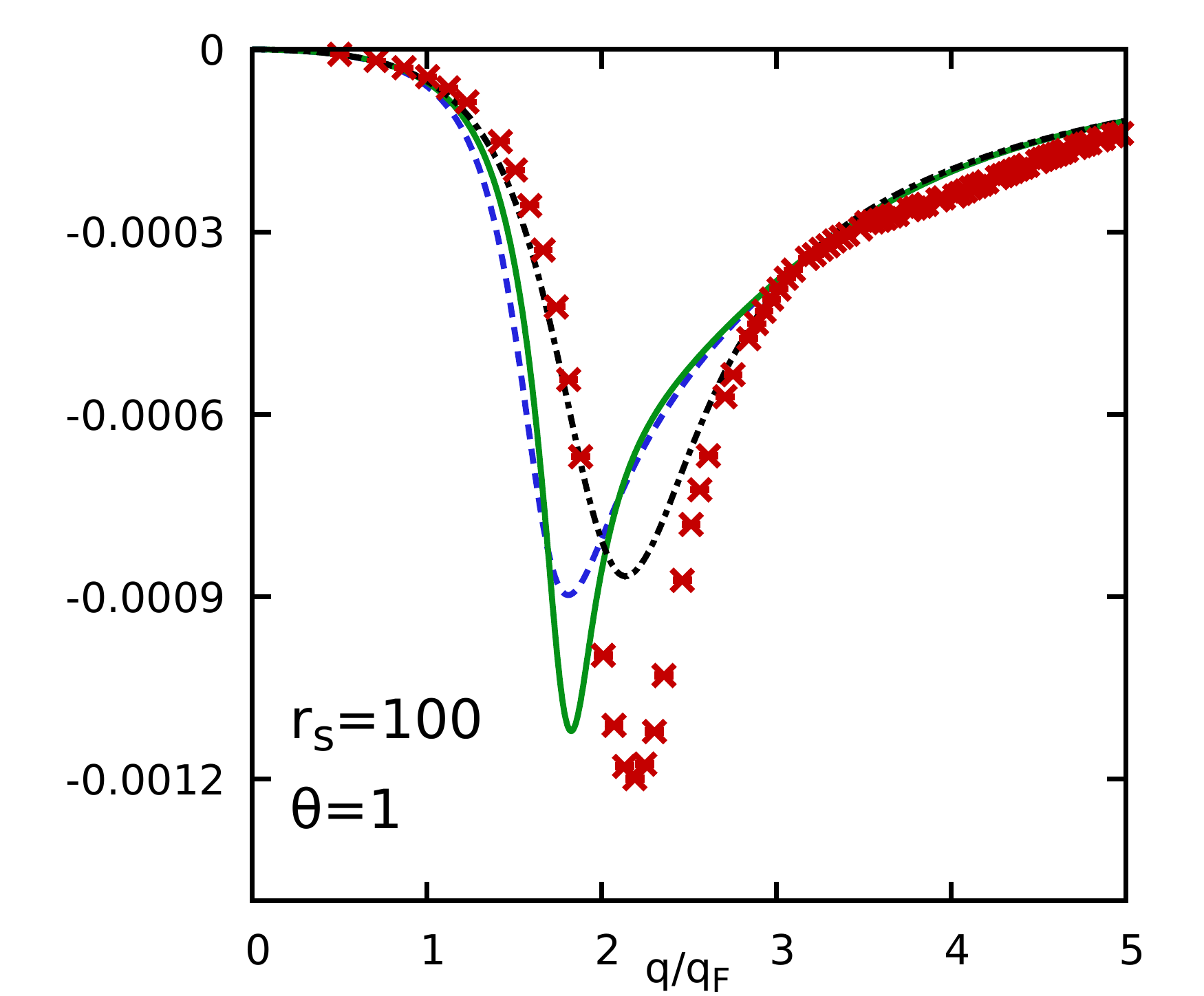}\hspace*{-0.4cm}\includegraphics[width=0.36\textwidth]{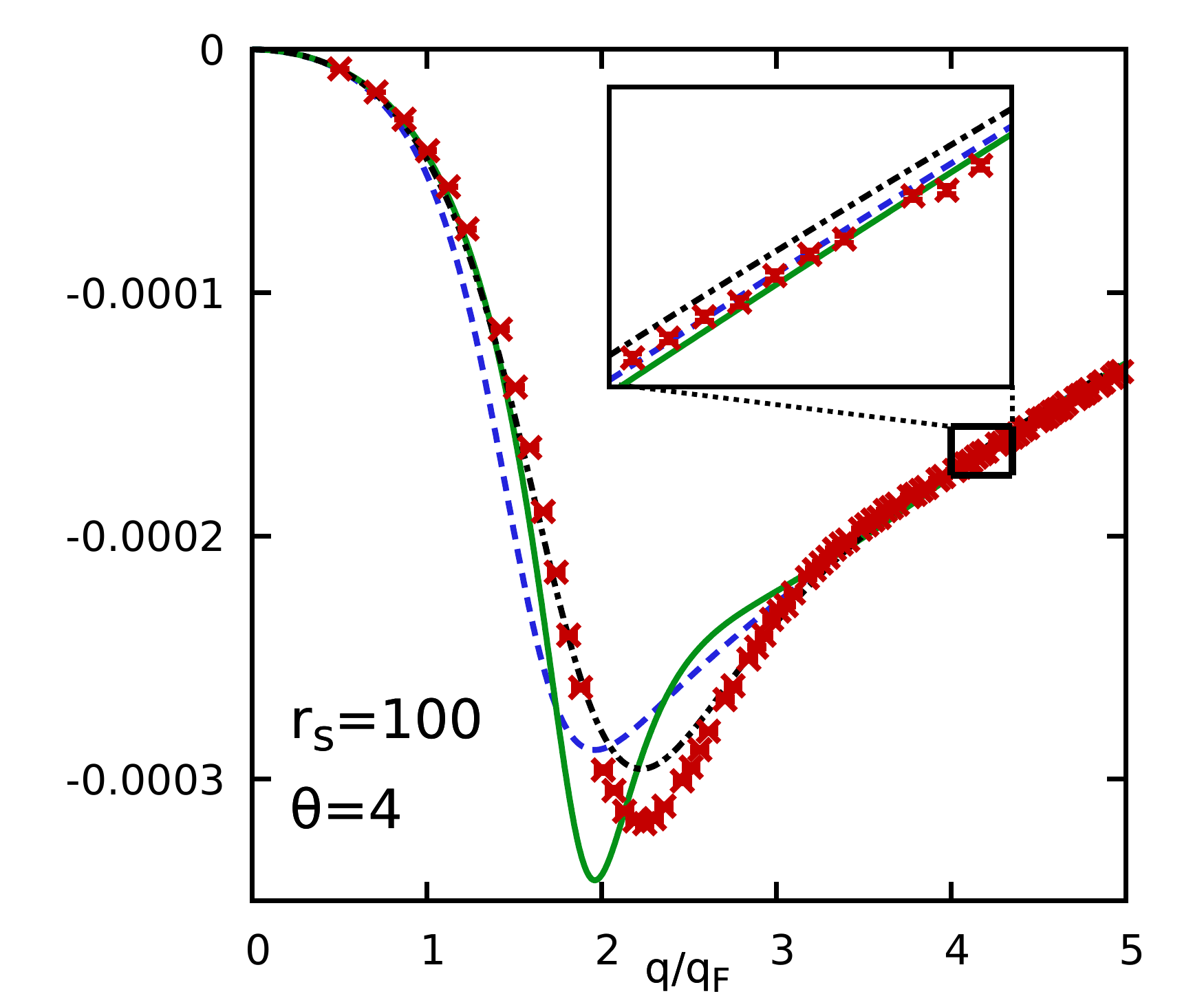}
\caption{\label{fig:Chi}
Comparison of the static density response function of the unpolarized electron liquid. Shown are PIMC data (red crosses, $N=66$ electrons, except for $r_s=20$ and $\theta=0.5$ where we show $N=20$) and the results from various dielectric approximations, in particular STLS (dashed blue), VS (solid green), and HNC (dash-dotted black). All PIMC data are available online~\cite{github_link}.
}
\end{figure*}

In the following section, we turn our attention to the original raison d'\^{e}tre of the dielectric formalism as introduced in Sec.~\ref{sec:dielectric}, which is the description of the response of the electron liquid to an external (static) perturbation.
To this end, we show an extensive comparison of the static density response function $\chi(q)$ between our PIMC data and the dielectric methods in Fig.~\ref{fig:Chi}. Firstly, we mention that both the curves and the crosses exhibit the exact parabolic behavior at small $q$,
\begin{eqnarray}
\lim_{q\to0} \chi(q) = -\frac{q^2}{4\pi} \quad ,
\end{eqnarray}
which is an immediate consequence of screening effects in the electron gas~\cite{kugler_bounds}. Furthermore, we observe a qualitatively similar behavior of $\chi(q)$ for all depicted cases, i.e., a monotonic increase, followed by a pronounced maximum around $q_\textnormal{max}\approx2q_\textnormal{F}$, and a subsequent monotonic decay with increasing $q$, where it converges towards the ideal function $\chi_0(q)$ and eventually decays to zero. Interestingly, the position of the peak does not coincide with the peak in $S(q)$ (or in $G(q)$, see Fig.~\ref{fig:G} below), but occurs for somewhat smaller wave numbers. A further pronounced trend that is manifest in our results for $\chi(q)$ concerns the peak width, which becomes increasingly sharp both towards low temperature and large $r_s$. This trend is connected to the possible emergence of a change-density wave instability~\cite{iyetomi_cdw,dynamic_ii,schweng}, which is discussed in more detail in Sec.~\ref{sec:CDW}.

Let us next examine in detail the accuracy of the different dielectric methods regarding the description of the density response. At $r_s=20$ (top row), VS gives an almost exact description of the peak position, and constitutes also the most accurate approximation for the peak height. This is in contrast to the previously discussed static structure factor $S(q)$, where the peak position was more accurately estimated by HNC, whereas VS was systematically too small. Moreover, we note that HNC provides the best description for small to intermediate wave numbers, $0\leq q \lesssim 1.5q_\textnormal{F}$, and that STLS is the least accurate method for all temperatures. Finally, we note that all dielectric methods become more accurate in the description of $\chi(q)$ with increasing $\theta$, as the impact of the approximate local field correction vanishes.

Upon increasing the coupling strength to $r_s=50$ (center row) and $r_s=100$ (bottom row), all dielectric methods become significantly less accurate, as it is expected. At $r_s=50$, the VS method gives the best qualitative description of the static density response, as the peak height is predicted with remarkable accuracy even for $\theta=0.5$ and $\theta=1$, whereas it is significantly underestimated both by STLS, and even more so by HNC. Interestingly, the latter approach almost exactly predicts the right peak position at $r_s=100$, although here, too, the peak height is captured better by VS. A further detail of interest concerns the large-$q$ behavior of $\chi(q)$ at $\theta=0.5,1$, where the PIMC data exhibit significant deviations to all dielectric theories. This has some interesting consequences for the local field correction, which are discussed below.

In a nutshell, we again find a trade-off regarding the description of peak height and position between VS and HNC, whereas STLS consistently provides the worst description of $\chi(q)$ both qualitatively and quantitatively.

\begin{figure*}\centering
\hspace*{-0.4cm}\includegraphics[width=0.36\textwidth]{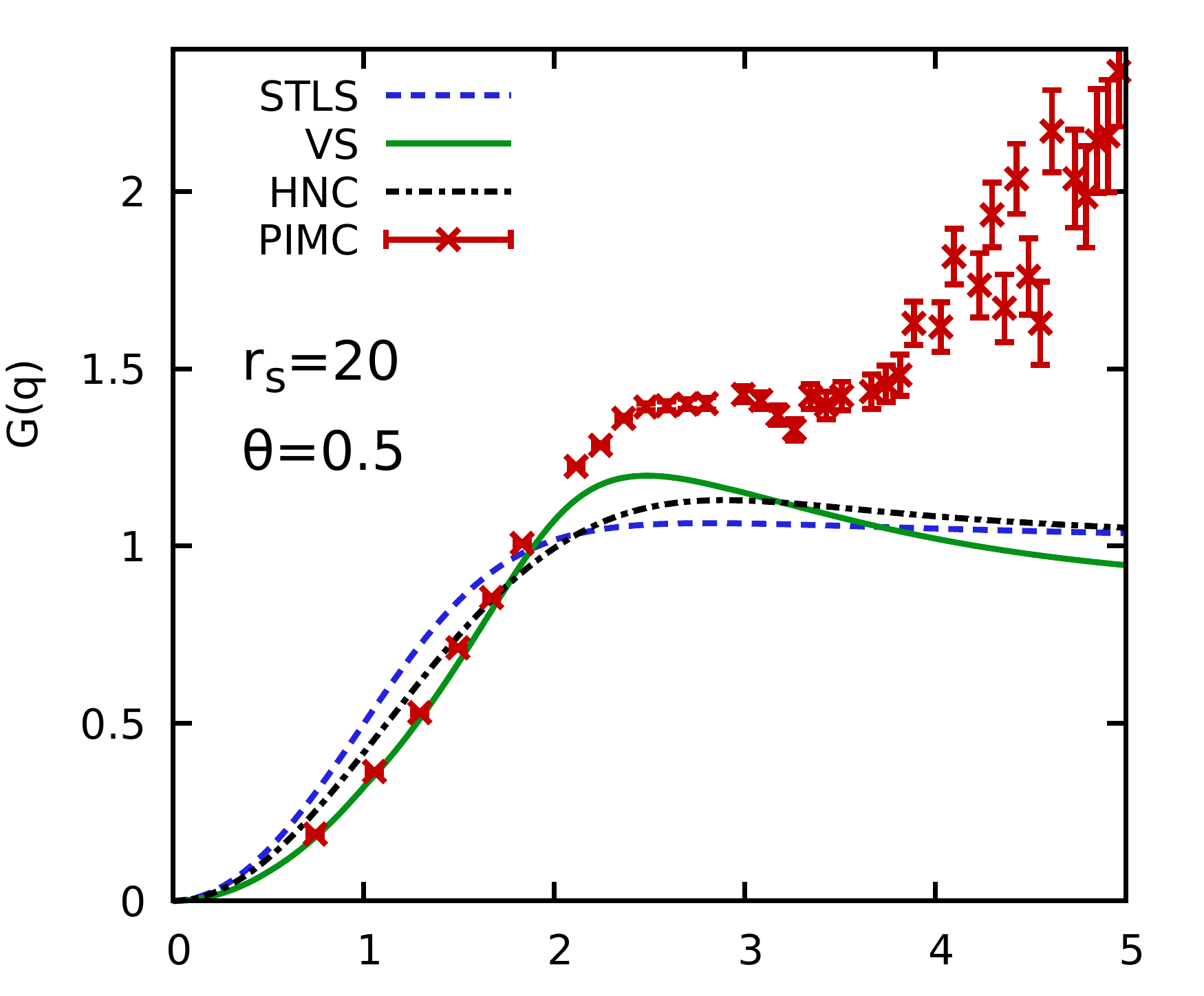}\hspace*{-0.4cm}\includegraphics[width=0.36\textwidth]{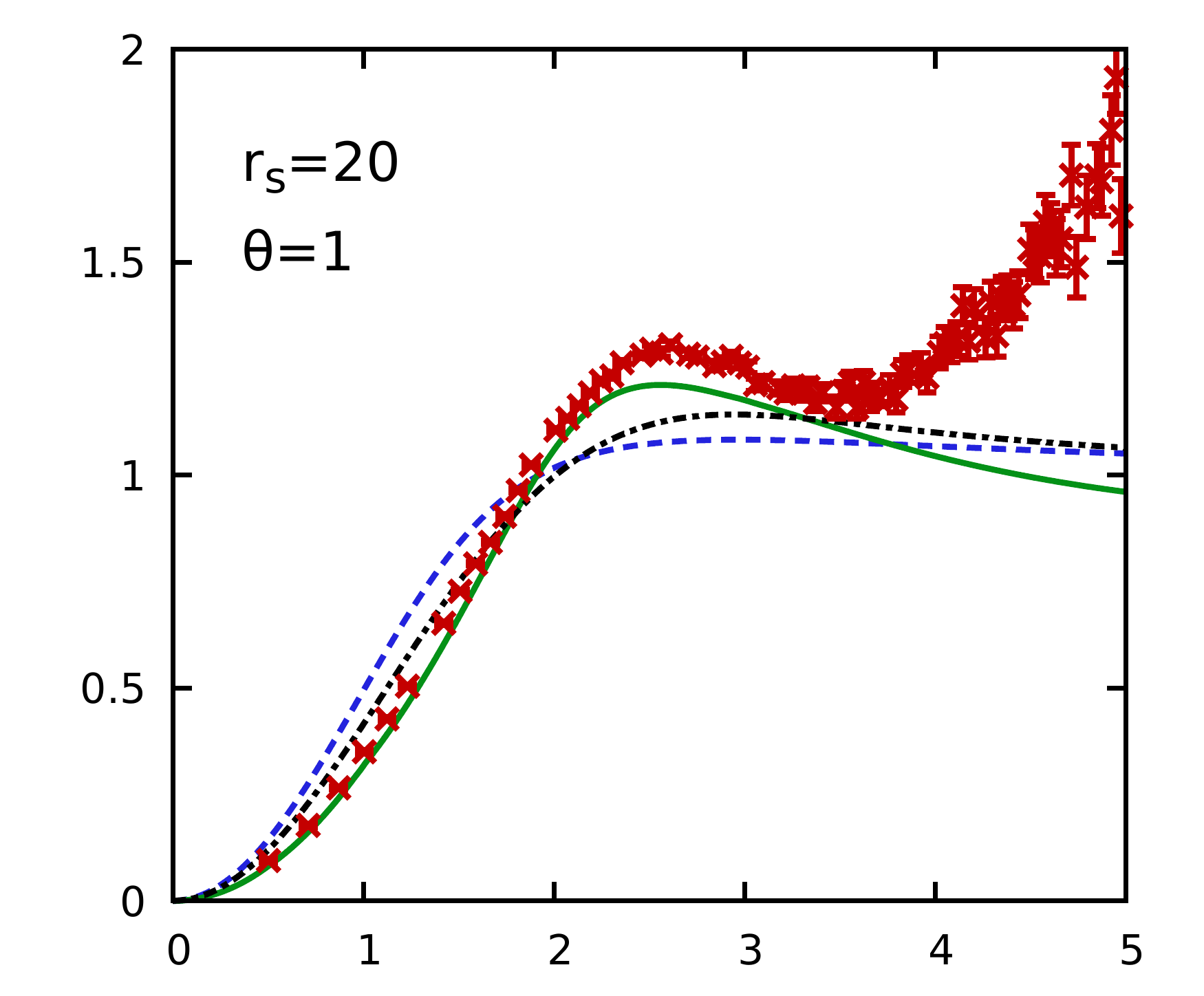}\hspace*{-0.4cm}\includegraphics[width=0.36\textwidth]{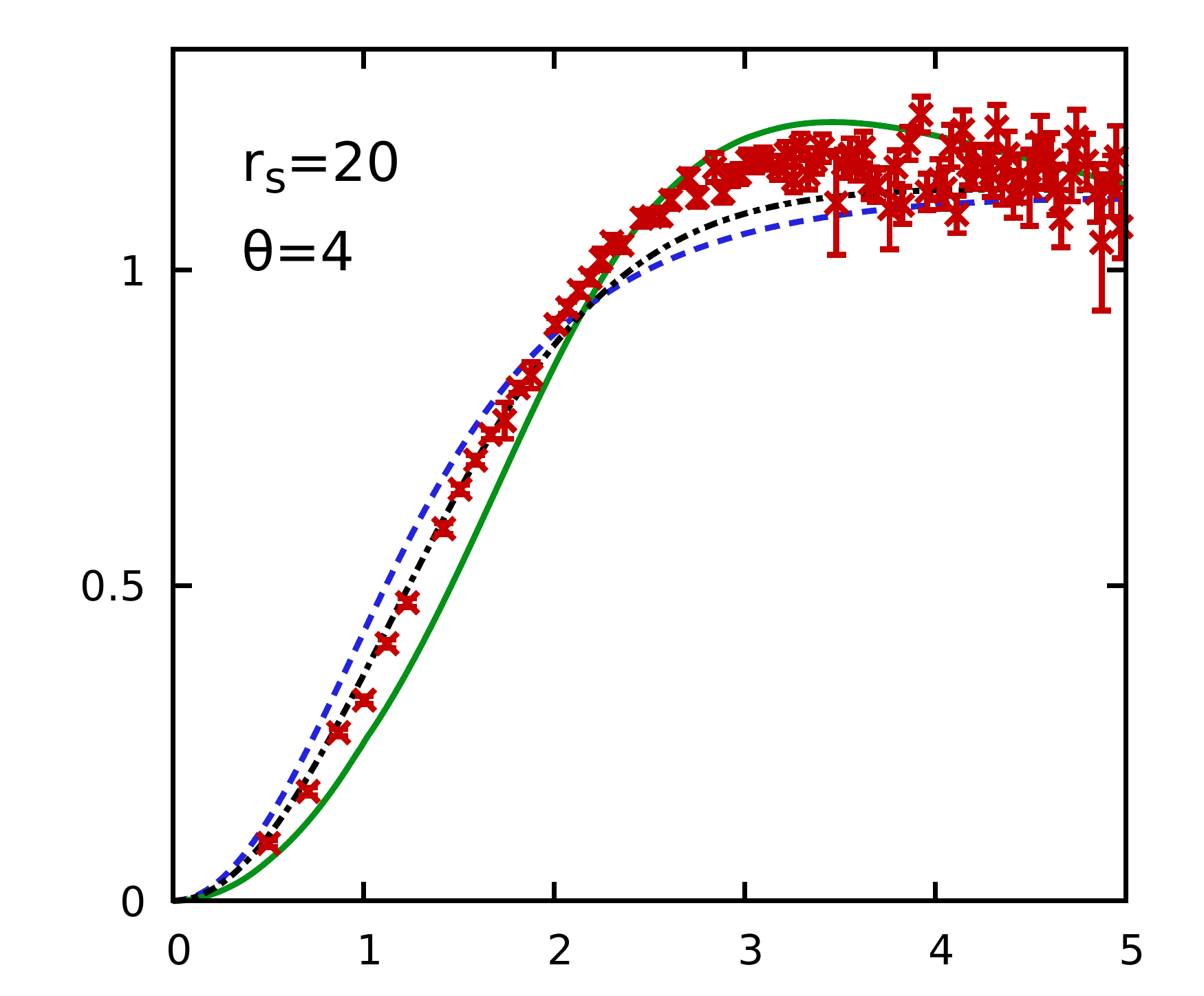}\\ 
\hspace*{-0.4cm}\includegraphics[width=0.36\textwidth]{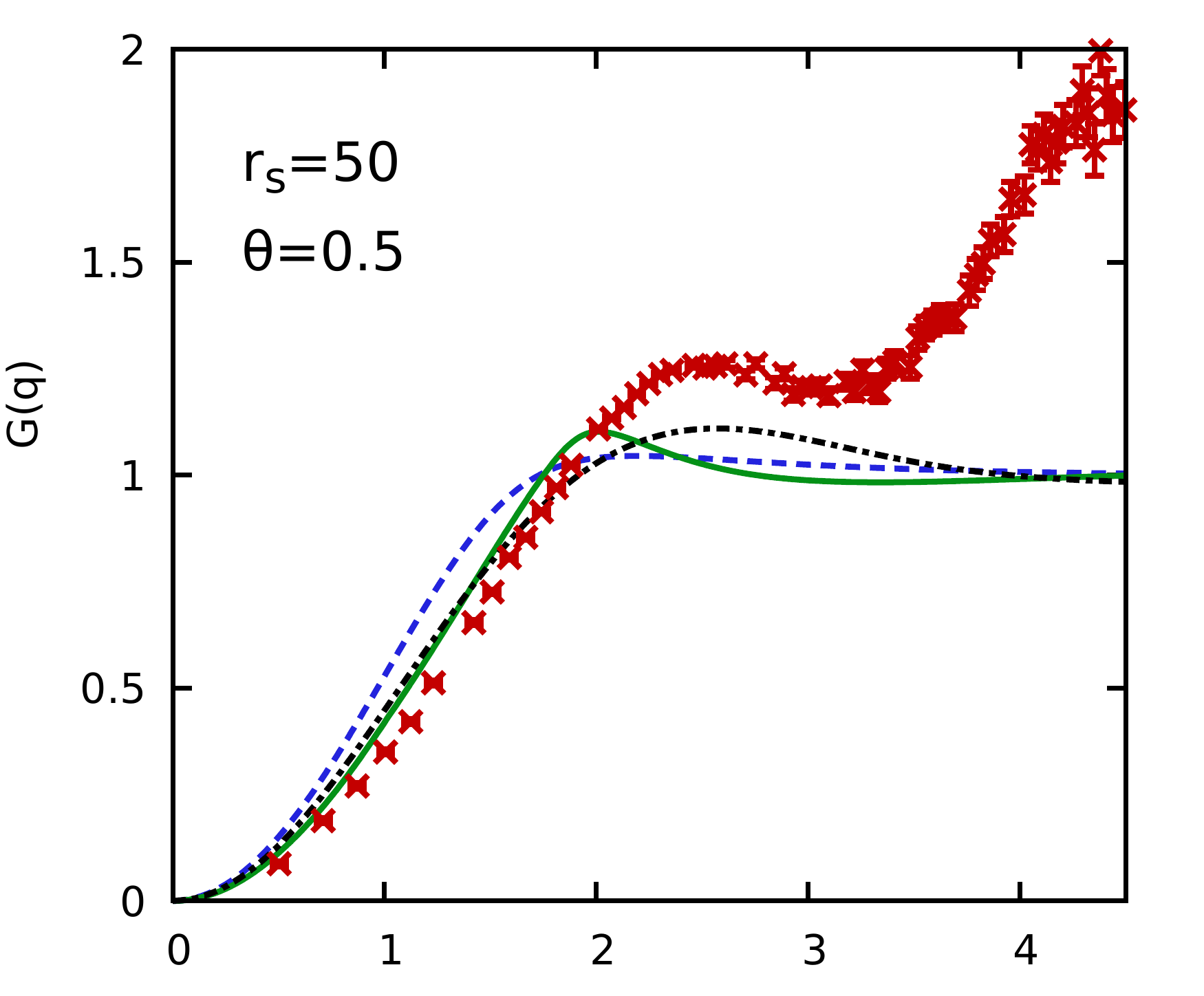}\hspace*{-0.4cm}\includegraphics[width=0.36\textwidth]{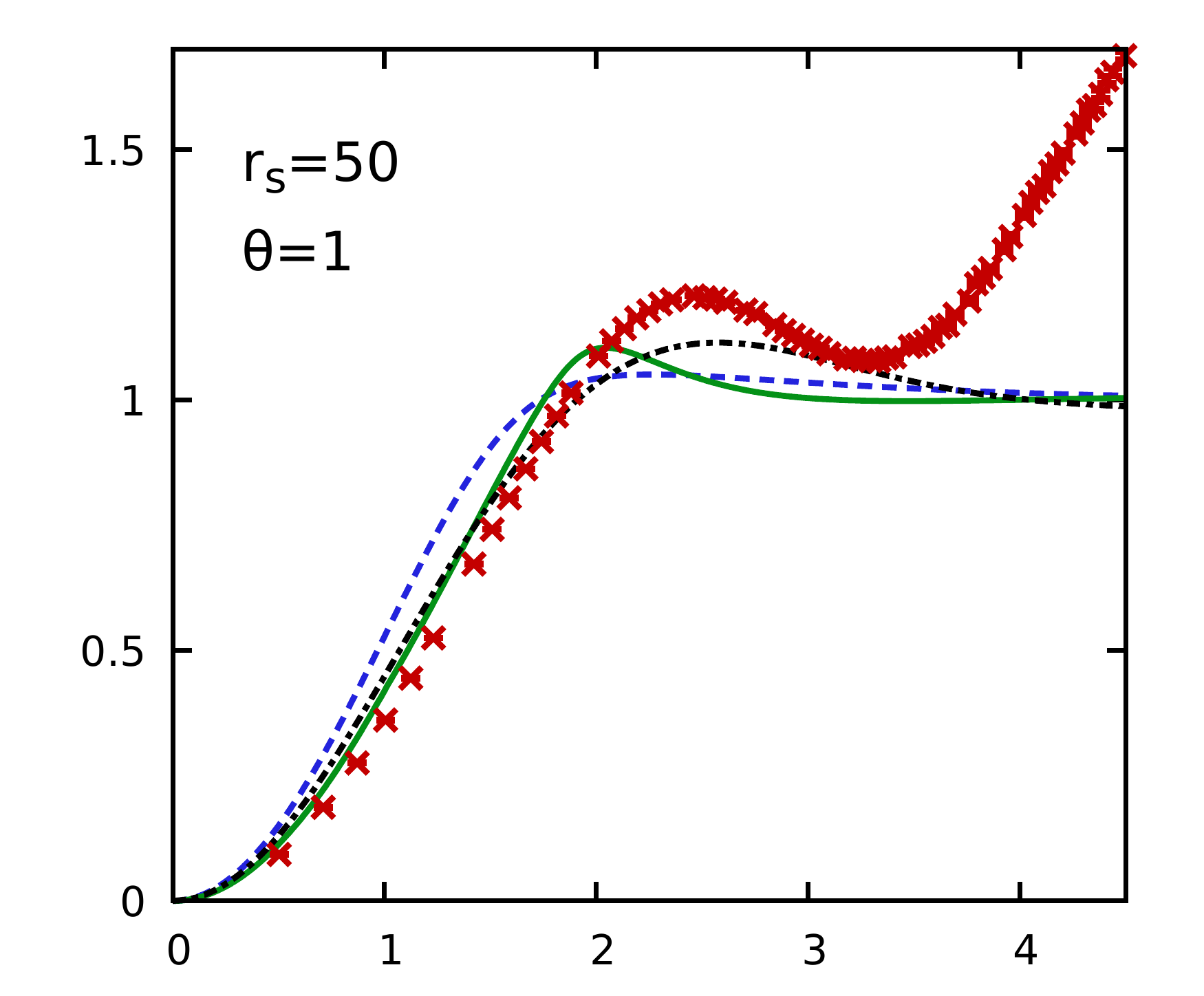}\hspace*{-0.4cm}\includegraphics[width=0.36\textwidth]{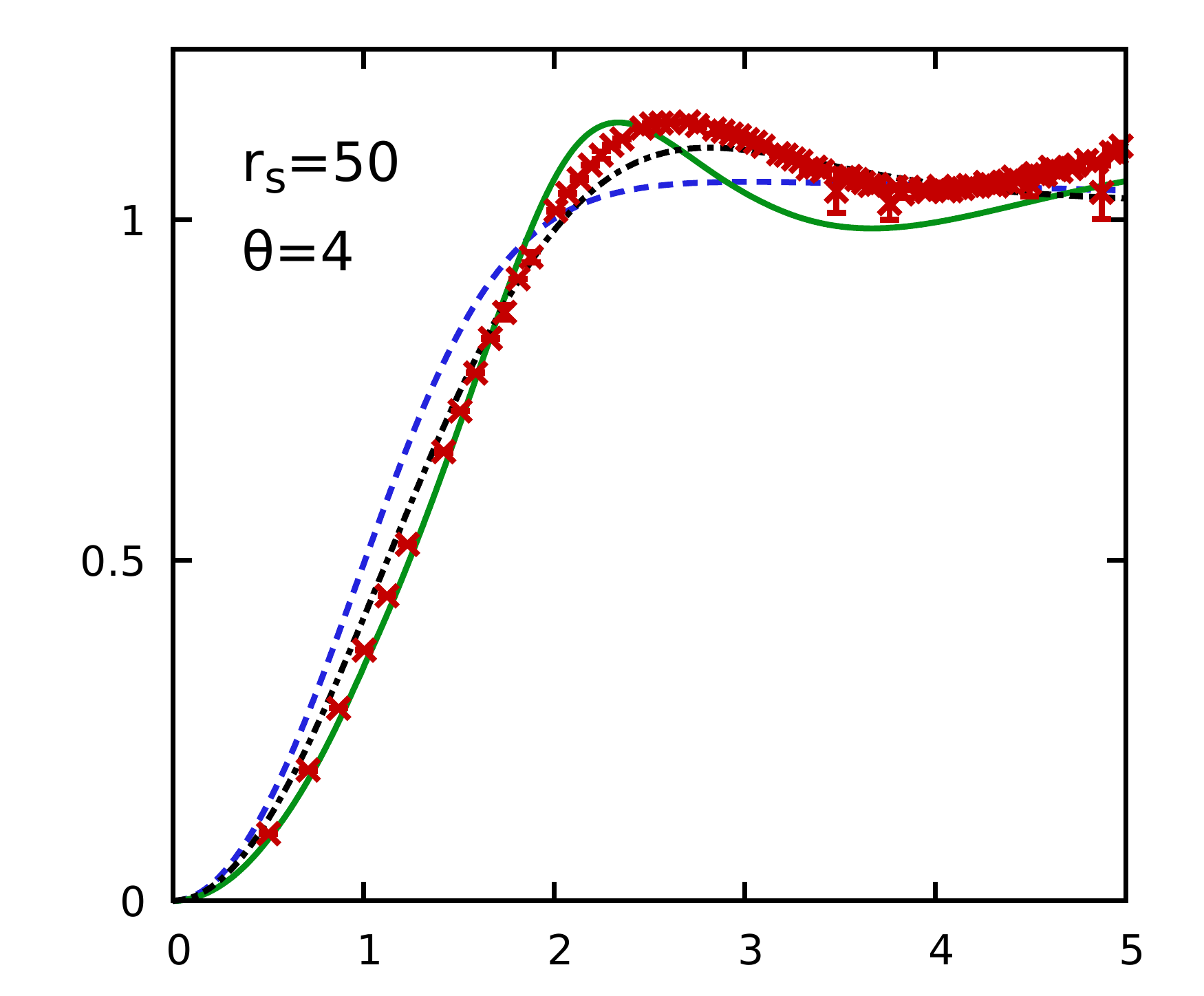}\\ 
\hspace*{-0.4cm}\includegraphics[width=0.36\textwidth]{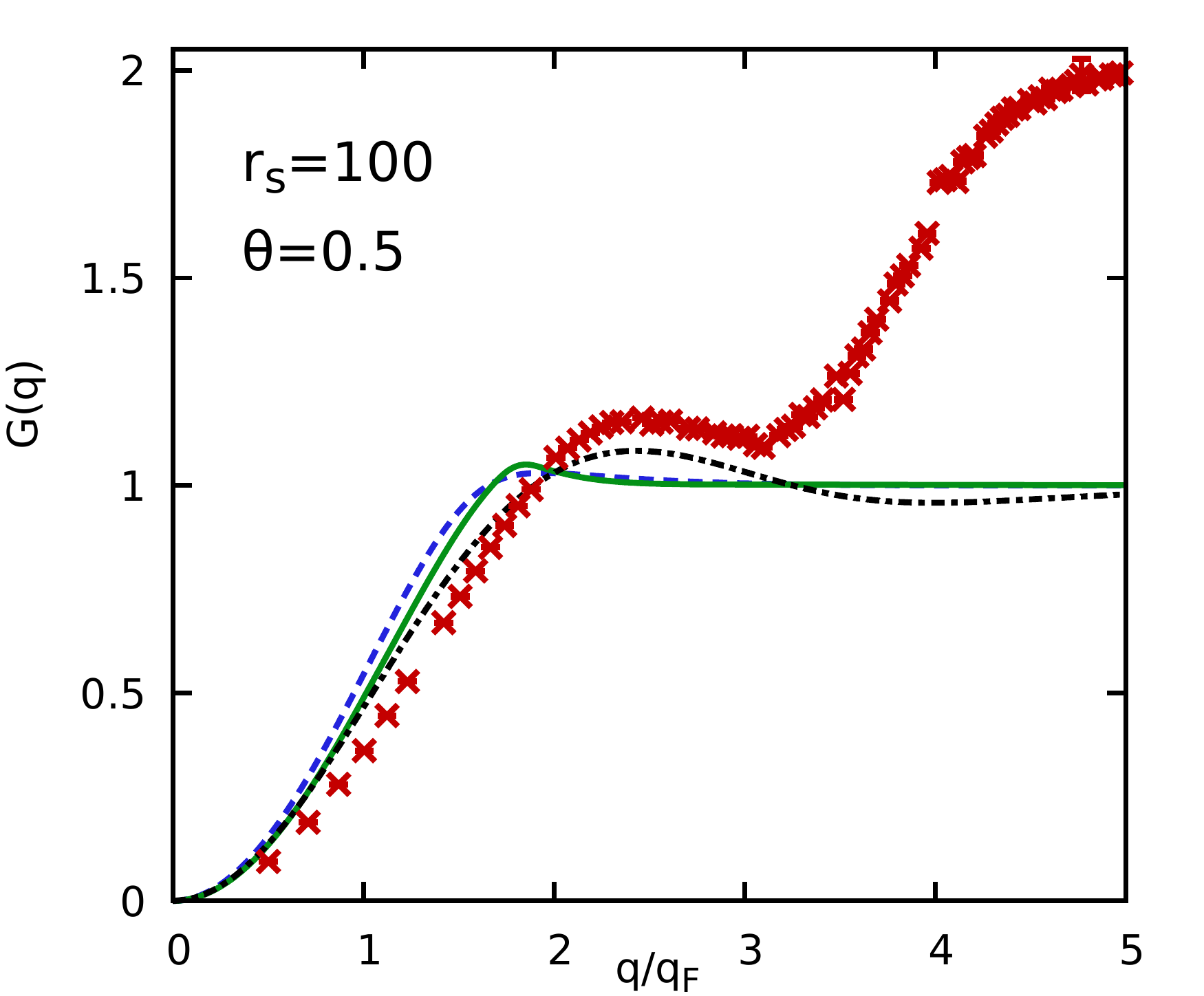}\hspace*{-0.4cm}\includegraphics[width=0.36\textwidth]{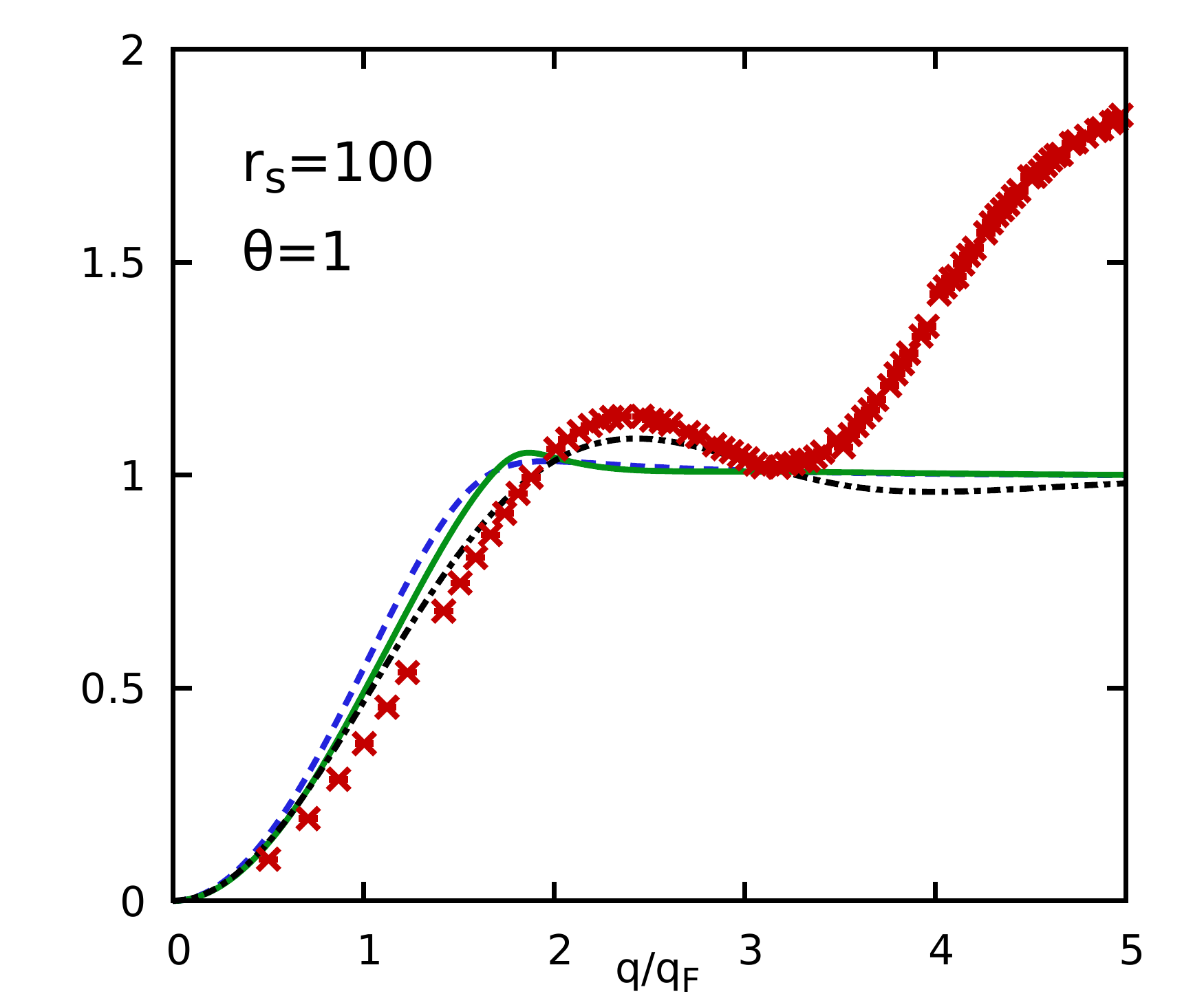}\hspace*{-0.4cm}\includegraphics[width=0.36\textwidth]{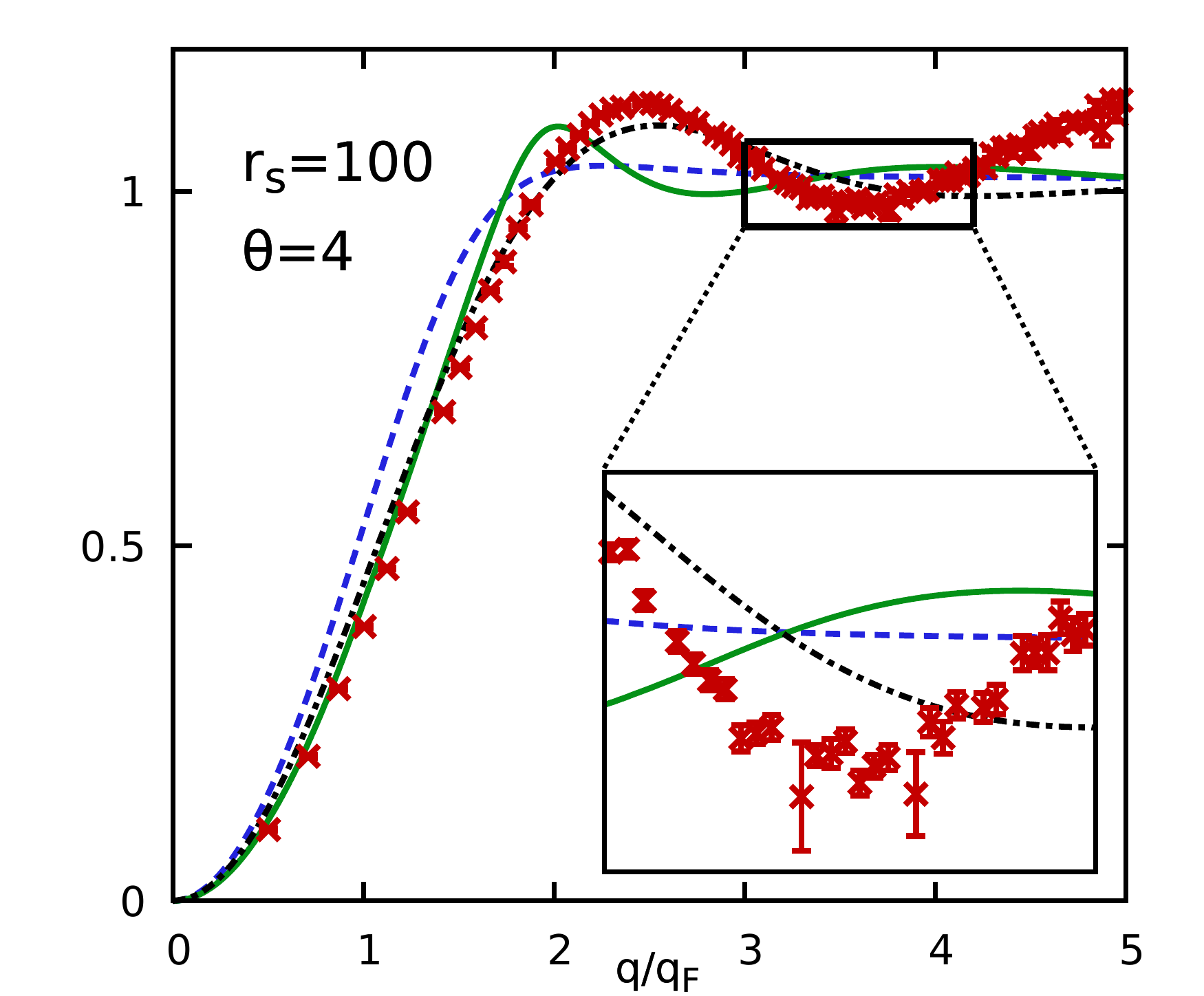}
\caption{\label{fig:G}
Comparison of the static local field correction of the unpolarized electron liquid. Shown are PIMC data (red crosses, $N=66$ electrons, except for $r_s=20$ and $\theta=0.5$ where we show $N=20$) and the results from various dielectric approximations, in particular STLS (dashed blue), VS (solid green), and HNC (dash-dotted black). All PIMC data are available online~\cite{github_link}.
}
\end{figure*}

The second key quantity regarding the description of the density response of the electron liquid is the (static) local field correction $G(q)$, which is shown in Fig.~\ref{fig:G} for the same conditions as in Fig.~\ref{fig:Chi}. For completeness, we note that the PIMC data for $G(q)$ have been finite-size corrected by using CPIMC data~\cite{groth_jcp} for $\chi_0^N(q)$ as described in Sec.~\ref{sec:PIMC_results}, although the effect can hardly be seen with the naked eye. In contrast to the previously shown comparisons for the static structure factor $S(q)$ (Fig.~\ref{fig:S}) and density response function $\chi(q)$ (Fig.~\ref{fig:Chi}), the dielectric theories give a distinctly different qualitative behavior of $G$ than the exact PIMC results, as we will discuss in detail below.

Let us first focus on the PIMC data, which are, as usual, given by the red crosses. Firstly, we note that $G(q)$ exhibits a quadratic behavior for small $q$, which is given by the exact compressibility sum-rule Eq.~(\ref{eq:CSR}). Moreover, $G$ exceeds unity for some wave numbers for all shown cases, which indicates an attractive electron-hole interaction in the description of screening effects~\cite{iyetomi_cdw}. Moreover, this feature is a necessary, though not sufficient condition for a CDW, which is discussed in detail in Sec.~\ref{sec:CDW}. At the Fermi temperature (middle column), $G(q)$ exhibits a pronounced maximum around $q\approx2.5q_\textnormal{F}$ for all $r_s$-values, which is followed by a minimum at $q\approx3.2q_\textnormal{F}$, and a subsequent increase in $G$ for large $q$.
Interestingly, this structure becomes less pronounced both with increasing and decreasing $\theta$, and is absent in our PIMC data at $r_s=20$ for both $\theta=0.5$ and $\theta=4$. To understand the increase of $G(q)$ towards large wave numbers, we might recall the corresponding expansion by Holas~\cite{holas_limit},
\begin{eqnarray}\label{eq:TAIL}
\lim_{q\to\infty}G(q) = B(r_s) + C(r_s) q^2 \ ,
\end{eqnarray}
which states that, in the ground state, the static local field correction does not become constant, but instead approaches a parabola in this limit. More specifically, both constants are known from theory, and the pre-factor $C(r_s)$ is directly proportional to the change in the kinetic energy $K$ due to interaction effects, $K_\textnormal{xc}$. While we again stress that Eq.~(\ref{eq:TAIL}) does not hold at finite temperature~\cite{dornheim_ML}, it still qualitatively explains the observed increase in $G(q)$ as $K_\textnormal{xc}$ is strictly positive under the present conditions~\cite{negative_kxc_note}. A similar behavior was reported recently in Ref.~\cite{dornheim_ML} in the warm dense matter regime. Moreover, we note that $K_\textnormal{xc}$ vanishes in the classical limit, and it is well known that $G(q)$ approaches unity for large $q$ in this regime~\cite{iyetomi_review}.

Let us next discuss the quality of the dielectric methods. In general, we find that VS constitutes the most accurate approximation for small $q$ (the only exception is $r_s=20$ and $\theta=4$, where HNC is superior). This is consistent with previous findings~\cite{review,dynamic_folgepaper} and is most likely a direct consequence of the incorporation of the compressibility sum-rule Eq.~(\ref{eq:CSR}) into the formalism. Still, it is important to note that fulfilling Eq.~(\ref{eq:CSR}) does not guarantee the correct small-$q$ behavior, as the exchange-correlation free energy $f_\textnormal{xc}^\textnormal{VS}$ as computed via VS is not equal to the exact quantity. 
Furthermore, we find that VS provides a surprisingly good description of $G(q)$ in particular for $r_s=20$ and $\theta=0.5,1$, and, in the latter case, agrees with our PIMC data nearly up to $q=3.5q_\textnormal{F}$. In contrast, the HNC scheme constitutes the most accurate dielectric approximation at strong coupling, where it again correctly predicts the position of the peak around $q_\textnormal{max}\approx2.3q_\textnormal{F}$. Finally, we mention that STLS provides the least accurate description of $G(q)$, which is consistent with our previous findings for $S(q)$ and $\chi(q)$.

Let us conclude this section with a discussion of the large-$q$ limit of the dielectric methods, which are all approaching a constant value in this limit. First and foremost, we note that all three schemes constitute \textit{static approximations}, i.e., they treat the dynamic local field correction as frequency independent for all $\omega$, $G_\textnormal{static}(q,\omega)=G(q)$. Therefore, the static local field correction from those theories should be interpreted as some kind of "frequency average", but cannot reproduce the exact behavior in the static limit by design~\cite{farid}. While the PIMC data, too, are only available for $\omega\to0$, they nevertheless have been obtained from a full "dynamic theory". More specifically, the full frequency-dependence is incorporated into the PIMC formalism via a propagation in the imaginary time $\tau$ corresponding to the exact description of the thermodynamic equilibrium.

In fact, it is well known from the electronic cusp condition~\cite{kimball} that neglecting the frequency dependence in $G(q,\omega)$ leads to a constant value for large $q$,
\begin{eqnarray}\label{eq:kim}
\lim_{q\to\infty} G(q) = 1 - g(0) \quad ,
\end{eqnarray}
with $g(0)$ being the pair-correlation function at zero distance. Indeed, Eq.~(\ref{eq:kim}) is fulfilled by all dielectric theories, which explains the observation in Fig.~\ref{fig:G}, see also Ref.~\cite{stls} for a more extensive discussion.

\subsection{Charge-density wave instability\label{sec:CDW}}

\begin{figure}\centering
\includegraphics[width=0.448\textwidth]{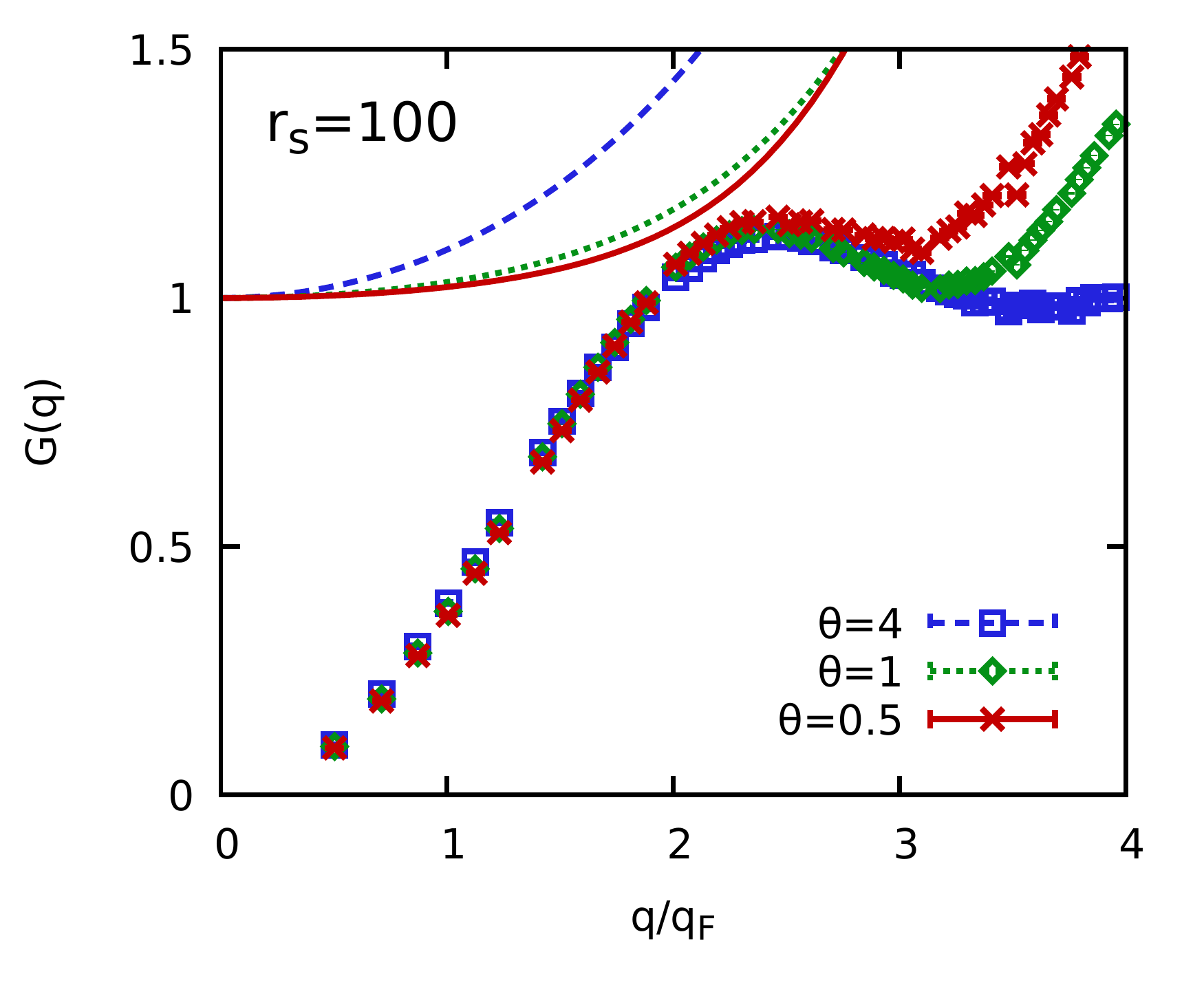}
\includegraphics[width=0.448\textwidth]{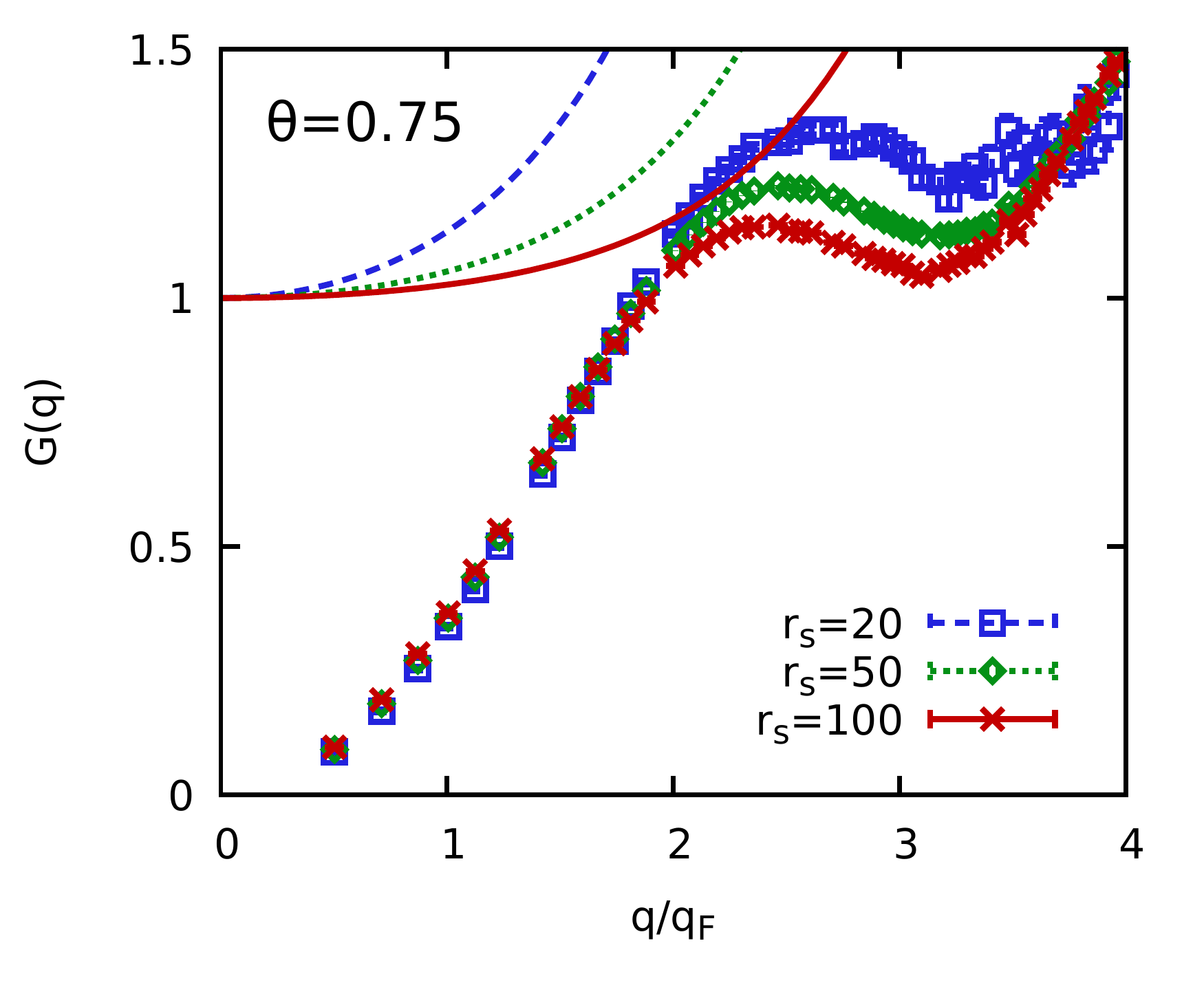}
\caption{\label{fig:CDW}
Excluding the possibility of a charge-density wave: Shown are our PIMC data for $N=66$ unpolarized electron for the static local field correction at $r_s=100$ and $\theta=4$, $\theta=1$, and $\theta=0.5$ (top panel), as well as for $\theta=0.75$ and $r_s=20$, $r_s=50$, and $r_s=100$ (bottom panel). The corresponding lines depict the condition for a CDW [cf.~Eq.~(\ref{eq:CDW})] at the respective parameters.
}
\end{figure}

Is has long been speculated~\cite{iyetomi_cdw,dynamic_ii,schweng} that the electron liquid might become unstable to an infinitesimal density perturbation for some specific wave number at strong coupling. Such a charge-density wave is defined by a divergence of the static density response function,
\begin{eqnarray}
\lim_{q\to q_\textnormal{CDW}}\chi(q) \to \infty \quad ,
\end{eqnarray}
which can be re-formulated in terms of the local field correction as
\begin{eqnarray}\label{eq:CDW}
G_\textnormal{CDW}(q) = 1 - \frac{q^2}{4\pi \chi_0(q)} \quad .
\end{eqnarray}
For the corresponding condition $G(q)=G_\textnormal{CDW}(q)$ to hold, the local field correction must be larger than one, since the denominator in Eq.~(\ref{eq:CDW}) is strictly negative.

\begin{table*}
\caption{\label{tab:summary}Overview of dielectric theories in the electron liquid regime. The first and second rows show the basis for the approximation (cf.~Sec.~\ref{sec:DielectricTheory}) and maximum error in the interaction energy (which appears at $r_s=100$ and $\theta=0.5$, see Fig.~\ref{fig:v} and Tab.~\ref{tab:interaction}). The last three rows summarize trends regarding the comparison to our new \textit{ab intio} PMIC data for the static structure factor $S(q)$ (see Fig.~\ref{fig:S}), density response function $\chi(q)$ (see Fig.~\ref{fig:Chi}) and local field correction $G(q)$ (see Fig.~\ref{fig:G}). The different colours indicate \textcolor{green}{good}, \textcolor{orange}{ok}, and \textcolor{red}{bad} qualitative agreement to the PIMC results regarding a particular feature.
}
\begin{ruledtabular}
\begin{tabular}{r|lll}
 & STLS & VS & HNC \\ \hline 
 approximation{\huge \textcolor{white}{)}} & $f_2(r_1,r_2) \approx f_1(r_1) f_1(r_2) g_\textnormal{eq}(r_1,r_2)$ & STLS + CSR & STLS + HNC \\ \hline 
 maximum $\Delta v/v${\huge \textcolor{white}{R}} & $5.7\%$ & $5.4\%$ & $1.2\%$ \\ \hline 
$S(q)$: peak{\huge \textcolor{white}{)}} & \textcolor{red}{height} / \textcolor{red}{position} & \textcolor{orange}{height} / \textcolor{red}{position} & \textcolor{red}{height} / \textcolor{green}{position} \\ \hline 
$\chi(q)$: peak{\huge \textcolor{white}{)}} & \textcolor{red}{height} / \textcolor{red}{position} & \textcolor{orange}{height} / \textcolor{red}{position} & \textcolor{red}{height} / \textcolor{green}{position} \hspace*{1.5cm} \\ \hline 
$G(q)$: peak{\huge \textcolor{white}{)}} & \textcolor{red}{height} / \textcolor{red}{position} & \textcolor{orange}{height} / \textcolor{red}{position} & \textcolor{orange}{height} / \textcolor{green}{position} \vspace*{-0.2cm} \\ 
large-$q$ tail{\huge \textcolor{white}{)}} & \textcolor{red}{flat} & \textcolor{red}{flat} & \textcolor{red}{flat}
 
\end{tabular}
\end{ruledtabular}
\end{table*}

In Fig.~\ref{fig:CDW}, we investigate this condition for the unpolarized electron liquid  by comparing the above condition to our PIMC data. The top panel corresponds to $r_s=20$, and we plot $G(q)$ for three different temperatures. Firstly, we note that $G(q)$ only weakly depends on temperature for small to intermediate wave numbers, whereas they substantially deviate for large $q$. At $\theta=4$, Eq.~(\ref{eq:CDW}) is rapidly increasing with $q$ and does not come close to our PIMC data for any $q$. With decreasing temperature, the maximum in $G(q)$ slightly increases, and $G_\textnormal{CDW}(q)$ approaches the data points around twice the Fermi wave number. Still, the two curves do not intersect at the present conditions, which means that a CDW is not present in this regime. This is consistent with previous findings from Schweng and B\"ohm~\cite{schweng}, who predicted  the formation of a CDW at $r_s=142$ and $\theta=0.5$ based on the dynamic version of the STLS approximation. Moreover, we mention that the trend observed in the top panel of Fig.~\ref{fig:CDW} does not conclusively point to the formation of a CDW at $r_s=100$ with decreasing temperature, as both the exact static LFC and Eq.~(\ref{eq:CDW}) are not expected to drastically change for $\theta < 0.5$ in the relevant $q$-range.

In the bottom panel, we show the same investigation for $\theta=0.75$ and three different values of the density parameter $r_s$. The most noticeable trend in our PIMC data is the increase in the maximum of $G(q)$ with increasing the density, and the simultaneous decrease in the slope for large $q$. Regarding the possibility of a CDW, we find that Eq.~(\ref{eq:CDW}) approaches the exact LFC for large $r_s$, as it is expected. Still, we cannot conclude that a CDW will actually occur upon further increasing $r_s$ based on this trend: while $G_\textnormal{CDW}(q)$ will indeed get further shifted to the right for increasing $r_s$, the magnitude of $G(q)$ in the relevant $q$-range decreases, and the two might potentially not intersect even then. On the other hand, it is imaginable that the CDW might occur at large $r_s$ for wave numbers not around the maximum, but the increasing large-$q$ tail. This, too, however, cannot be inferred from our PIMC data, as the electron liquid will eventually crystallize, which might significantly change the density response, and renders an extrapolation of the observed trends highly dubious.

\section{Summary and discussion\label{sec:summary}}

In summary, we have studied the uniform electron gas in the strongly coupled electron liquid regime. First and foremost, we have carried out extensive PIMC simulations for $20$ different density--temperature combinations and, to mitigate finite-size effects, different particle numbers $N$. We expect these data to be very valuable in themselves, as they substantially extend our current picture of the UEG beyond the WDM regime. In addition, we have applied dielectric theory within the STLS, VS, and HNC approximations, and have found that different schemes reproduce different features of the exact results, see Tab.~\ref{tab:summary} for a juxtaposition. More specifically, our key findings are that (i) VS is remarkably accurate for $r_s=20$ and almost exactly reproduces the exact data for $S(q)$. This is in contrast to previous findings in the WDM regime~\cite{review}; (ii) the recent HNC scheme provides the most accurate interaction energies, which is the result of a fortunate error cancellation in the integration of $S(q)$; (iii) our new PIMC data for $v$ are in good agreement with parametrizations of $f_\textnormal{xc}$, where they are available; and (iv) no dielectric method provides an accurate description of $G(q)$ [and, to a lesser extend, $\chi(q)$], in particular towards large $q$.
Moreover, we have investigated the possibility of a charge-density wave, which, however, does not manifest at the present conditions.

We expect our new results to be useful to extend parametrizations of the UEG~\cite{ksdt,groth_prl,review,karasiev_status} towards the low-density regime, and to provide a challenging benchmark for new approximations~\cite{panholzer1}. In addition, the comparison to dielectric theory has revealed partly substantial shortcomings of the latter, which can be used as the starting point for the further development of dielectric approximations. Possible future projects include the investigation of the dynamic structure factor (see Refs.~\cite{dornheim_dynamic,dynamic_folgepaper}) and the corresponding incipient excitonic mode~\cite{takada2,higuchi}, and PIMC simulations at even stronger coupling to study the onset of Wigner crystallization at finite temperature~\cite{clark_casula}.

All PIMC data are available online~\cite{github_link}.

\par 
$ $

\section*{Acknowledgments}
We wish to thank Simon Groth for providing the CPIMC data for the ideal density response function $\chi_0^N(q)$ used for the finite-size correction of $G(q)$.

This work was partly funded by the Center of Advanced Systems Understanding (CASUS) which is financed by Germany's Federal Ministry of Education and Research (BMBF) and by the Saxon Ministry for Science and Art (SMWK) with tax funds on the basis of the budget approved by the Saxon State Parliament.

All PIMC calculations were carried out on the clusters \emph{hypnos} and \emph{hemera} at Helmholtz-Zentrum Dresden-Rossendorf (HZDR), and at the Norddeutscher Verbund f\"ur Hoch- und H\"ochstleistungsrechnen (HLRN) under grant shp00015.


\begin{thebibliography}{10}
\bibitem{github_link} A link to a GitHub repository containing the PIMC data will be made available upon publication.
\bibitem{negative_kxc_note} It is well known~\cite{neg1,neg2,stls} that $K_\text{xc}$ becomes negative for weak coupling above a certain temperature. Indeed, the negative large-$q$ asymptote of $G(q)$ predicted by Eq.~(\ref{eq:TAIL}) has recently been reported in this regime, see Ref.~\cite{dornheim_ML}.
\end{thebibliography}
\end{document}